\documentclass[
twocolumn,
superscriptaddress,
nofootinbib,
amsmath,amssymb,
aps,
]{revtex4-2}

\usepackage{amsmath}
\usepackage{amsthm}
\usepackage{mathptmx}
\usepackage{newtxmath}
\usepackage[hidelinks]{hyperref}
\hypersetup{pdftitle={Adaptive cut reveals multiscale complexity in networks},pdfauthor={Louis Boucherie, Yong-Yeol Ahn, Sune Lehmann}}
\usepackage{stmaryrd}
\usepackage{graphicx}
\usepackage{bm}
\usepackage{booktabs}
\usepackage{longtable}

\graphicspath{{figures/}}

\renewcommand{\min}{\mathrm{min}}
\renewcommand{\max}{\mathrm{max}}

\theoremstyle{plain}
\newtheorem{lemma}{Lemma}[section]
\newtheorem{theorem}[lemma]{Theorem}
\newtheorem{corollary}[lemma]{Corollary}

\begin{document}

\title{Adaptive cut reveals multiscale complexity in networks}

\author{Louis Boucherie}
\email{louibo@dtu.dk}
\affiliation{Department of Applied Mathematics and Computer Science, Technical University of Denmark, Lyngby, Denmark}
\affiliation{Copenhagen Center for Social Data Science (SODAS), University of Copenhagen, Copenhagen, Denmark}

\author{Yong-Yeol Ahn}
\affiliation{School of Data Science, University of Virginia, Charlottesville, Virginia, United States}

\author{Sune Lehmann}
\affiliation{Department of Applied Mathematics and Computer Science, Technical University of Denmark, Lyngby, Denmark}
\affiliation{Copenhagen Center for Social Data Science (SODAS), University of Copenhagen, Copenhagen, Denmark}

\date{Published in \textit{PNAS Nexus} (2026), \href{https://doi.org/10.1093/pnasnexus/pgag344}{doi:10.1093/pnasnexus/pgag344}}

\begin{abstract}
Hierarchical clustering and community detection are important problems in machine learning and complex network analysis. A common approach to identify clusters is to simply cut dendrograms at some threshold. However, single-level cuts are often suboptimal in terms of capturing underlying structure in the data, especially when the dendrogram is unbalanced. In this paper, we present the adaptive cut, a method that uses the hierarchical structure of dendrograms through multi-level cuts to overcome the limitations of single-level approaches. The adaptive cut optimizes an objective function using a Markov chain Monte Carlo with simulated annealing, resulting in better partitions. We demonstrate the adaptive cut through applications to link clustering and modularity optimization, but note that the method is applicable to any clustering task that relies on a dendrogram and an objective function. Beyond the adaptive cut, we introduce the balancedness score, an information-theoretic metric that quantifies how balanced a dendrogram is. Balancedness predicts the potential benefits of using multi-level cuts. For the community detection examples, we evaluate our method on more than 200 real-world networks and multiple synthetic datasets, demonstrating improvements in partition density and modularity over traditional single-cut approaches. We also show the generality of the adaptive cut by applying it across hierarchical clustering techniques and objective functions. The adaptive cut improves clustering outcomes across a range of hierarchical clustering tasks.
\end{abstract}


\maketitle

\section{Introduction}

Identifying groups of objects that are closely related, which is commonly referred to as `clustering' or `community detection' in the case of networks, is an important problem in many scientific fields. A plethora of clustering methods have been proposed in the literature \cite{johnson1967hierarchical, lancichinetti2008benchmark, heller2005bayesian, xu2005survey, radicchi2004defining}. Community detection and cluster analysis methods frequently produce hierarchies, reflecting the multi-scale nature found in many datasets \cite{louvain, infomap, ex_clustering}. The end product of hierarchical clustering is a merge tree (or dendrogram if binary) originating from leaves containing data elements and culminating at the root, which encompasses the entire dataset. Dendrograms are a common feature of both clustering and community detection methods. In this paper, we focus on hierarchical clustering and community detection in complex networks, in particular the Link Clustering algorithm \cite{ahn2010link} and the Louvain optimization of modularity \cite{louvain}. However, our approach can be applied to any clustering process that produces a dendrogram and an independent function of clustering quality.

The most common method for obtaining clusters (or communities) is to cut the dendrogram at a single level or a constant height cutoff value \cite{campello2013framework, milligan1985examination, jain1988algorithms}. There are several methods for selecting this value, including optimizing an objective function \cite{dunn1974well, ahn2010link}, obtaining a specific number of clusters (through the elbow method or the silhouette method) \cite{milligan1985examination}, or optimizing for clusters with high intra-cluster similarity and high inter-cluster differences \cite{davies1979cluster}. However, single-level cuts may not always identify the most useful clusters. In particular, many dendrograms are unbalanced (e.g.\ due to different levels of data density across a dataset), and a single-level cut cannot resolve the trade-off between over-aggregation in one part of the dendrogram and providing an excessively granular view in another part. Consequently, single-level cuts are incapable of optimally separating distinct clusters.

In this sense, single-level cuts are unable to use all of the information contained within the hierarchical structure of the dendrogram. In unbalanced dendrograms, choosing a single cut point can result in the formation of clusters with very different sizes. This can have a range of undesired downstream consequences. For example, statistical tests that assume equal cluster sizes may be unsuitable when employed in the analysis of unbalanced clusters \cite{yen2009cluster, lin2017clustering}. Similarly, machine learning algorithms that rely on balanced training sets may underperform \cite{he2009learning, branco2016survey}.

Here, we propose the \emph{adaptive cut}, a method for cutting dendrograms using a multi-level cut. Our approach improves a range of tasks that involve hierarchical tree structures. To demonstrate the robustness of our results, we evaluate the method on over 200 real networks. The adaptive cut optimizes an objective function along the dendrogram using a Markov chain Monte Carlo with a simulated annealing scheme. We demonstrate the generality of the method across two distinct use cases of community detection for network nodes and edges. The previous key approach to examine this has been based on a visualization tool for exploring clustering scenarios at different cut levels \cite{visual_cut}. However, this approach depends on expert knowledge and does not provide a quantitative measure. Another approach \cite{dynamic_cut} proposes a heuristic based on dendrogram shape, akin to the silhouette coefficient \cite{rousseeuw1987silhouettes}.

We also present a new measure of dendrogram balance. Numerous tree balance indices have been proposed in the literature \cite{tree_balance_old, tree_balance_review, tree_balance_axiom, fernandez_gomez_2020}. Classical indices such as the Colless and Sackin indices measure the topological shape of the tree, while more recent information-theoretic indices \cite{tree_balance_axiom} quantify the entropy of subtree sizes. The entropy-based normalisation closest to ours is the normalised tree balance of \cite{fernandez_gomez_2020}, which averages a local Shannon entropy over the internal nodes of the dendrogram. Our index, named \textit{balancedness}, differs from these approaches in that it is defined at each \emph{similarity level} of the dendrogram rather than at each branching event. This level-wise formulation captures how balanced the partition is at each hierarchical scale, making it directly relevant as a predictor of adaptive cut improvement (a head-to-head comparison with the index of \cite{fernandez_gomez_2020} on the 202-network catalogue is reported in Supplementary Section~S6.4). Our index is computationally efficient and satisfies the axioms required for membership in the class of robust, universal tree balance indices \cite{tree_balance_axiom} (see Methods, \emph{Balancedness Axioms}, for the proof).

Existing work on community detection also connects to our method. Markov chain Monte Carlo with simulated annealing has been employed for community detection in networks to optimize modularity \cite{sune_mcmc,Guimera_MCMC}, description length \cite{infomap}, or fit stochastic block models \cite{peixoto2014efficient}. These approaches optimize community splits directly and as such do not use the tree structure to optimize the objective function. They operate on a large state space, making convergence to the optimal partition challenging and computationally expensive. By using an underlying tree structure, adaptive cut reduces the size of the state space and enables a more efficient optimization of the objective function and faster convergence to an optimal solution.

\begin{figure*}[ht]
    \centering
    \includegraphics[width=\textwidth]{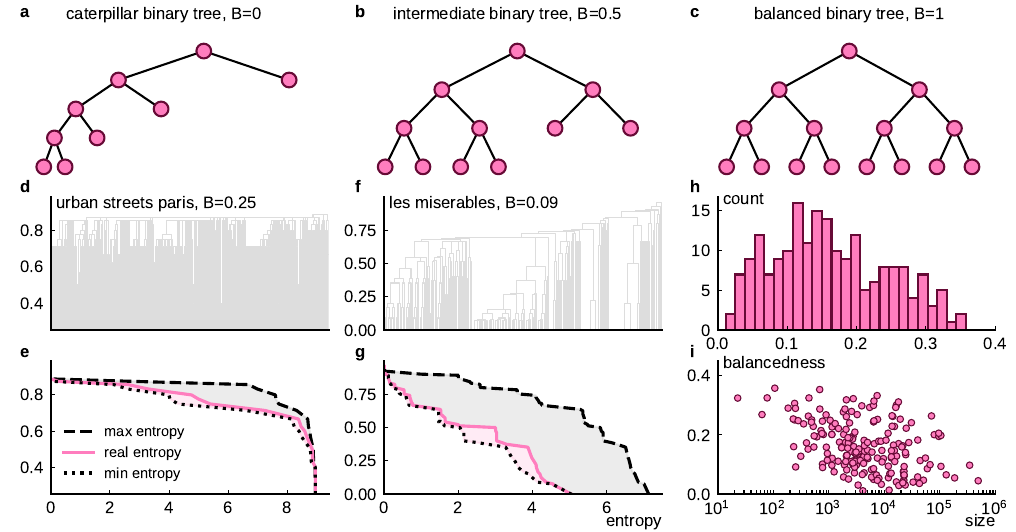}
    \caption{\textbf{Explanation of the balancedness Measure.}
\textbf{(a, b, c)} Illustrations of different tree structures: (a) an unbalanced caterpillar tree, (b) an intermediate tree, and (c) a balanced tree.
\textbf{(d)} Dendrogram of the urban street network of Paris based on link similarities \cite{ahn2010link}. This network sits in the top quartile of our 202-network catalogue ($B = 0.25$, see (h)); see (e) for the per-level entropy curves.
\textbf{(e)} The Shannon entropy $H(\pi_l)$ of the partition at each dendrogram level (pink curve), alongside the theoretical upper bound $H_{\max}$ and lower bound $H_{\min}$ (black curves). The y-axis is shared with the dendrogram in (d). The balancedness score (Eq.~\ref{eq:balancedness}) corresponds to the average normalized position of the pink curve between the two bounds.
\textbf{(f)} Dendrogram of the ``Les Miserables'' character network based on link similarities \cite{ahn2010link}. The dendrogram is highly unbalanced, as shown in (g).
\textbf{(g)} The entropy curves for the Les Miserables network, computed as in (e).
\textbf{(h)} The distribution of balancedness scores for 202 real networks.
\textbf{(i)} The balancedness metric plotted against network size (number of edges $M$). There is a negative correlation between $B$ and size (Pearson $r = -0.21$), indicating that larger networks tend to produce more unbalanced dendrograms under link clustering.}
\label{fig:balancedness}
\end{figure*}

\section{Balancedness}

\subsection{Definition}

We begin by introducing the \textit{balancedness} score, an information theory based metric that quantifies the balance of a dendrogram. The balancedness metric compares the actual branching structure of a dendrogram (real entropy) with both a perfectly balanced scenario (maximal entropy) and a highly skewed one (minimal entropy) to quantify how balanced a particular dendrogram is. At each level $l$ of the dendrogram, the partition of the $n$ leaves in each of the $k$ branches is noted as follows:
\begin{align}
    \pi_l = \{ B_1,B_2,\dots,B_k\},
\end{align}
where $B_i$ represents the set of leaves having branch $B_i$ as an ancestor. To normalize the entropy at each level, we compute the upper and lower bounds on Shannon entropy for any partition of $n$ elements into $k$ non-empty groups. The \emph{maximal entropy} is attained by the most balanced partition, where $n \bmod k$ groups have size $\lceil n/k \rceil$ and the remaining groups have size $\lfloor n/k \rfloor$:
\begin{align}
H_{\max}(k,n) = - \sum_{i=1}^{k} \frac{n_i}{n}\log_{2}\Big(\frac{n_i}{n}\Big)\,,
\label{eq:max-entropy}
\end{align}
where $n_i = \lceil n/k \rceil$ for $i \leq n \bmod k$ and $n_i = \lfloor n/k \rfloor$ otherwise. Note that $H_{\max}(k,n) = \log_2 k$ only when $k$ divides $n$ evenly. The \emph{minimal entropy} is attained by the most skewed partition, where one group contains $n-(k-1)$ elements and the remaining $k-1$ groups each contain one element:
\begin{align}
H_{\min}(k,n) = & - \sum_{i=1}^{k-1}\frac{1}{n}\log_{2}\Big(\frac{1}{n}\Big) \notag \\
& - \frac{n-(k-1)}{n}\log_{2}\Big(\frac{n-(k-1)}{n}\Big).
\end{align}
We also determine the realized entropy of the leaf partition $\pi_l$ at level $l$ as follows:
\begin{align}
H_{}(\pi_l) = - \sum_{i=1}^{|\pi_l|}p(B_{i})\log_{2}p(B_{i})\,,
\end{align}
where $p(B_{i})=|B_i|/n$. The balancedness score is the average, over all levels, of the ratio between the realized entropy minus the minimal entropy and the maximum entropy minus the minimal entropy across all levels (we write $H_{\max}(\pi_l)$ and $H_{\min}(\pi_l)$ for $H_{\max}(k,n)$ and $H_{\min}(k,n)$ evaluated at the level-$l$ partition, with $k=|\pi_l|$ branches),
\begin{align}
B = \frac{1}{L}\sum_{i=1}^{L} \frac{H_{}(\pi_l)-H_{min}(\pi_l)}{H_{max}(\pi_l)-H_{min}(\pi_l)}\,.
\label{eq:balancedness}
\end{align}
The balancedness metric ranges from 0 (unbalanced dendrogram, Fig.~\ref{fig:balancedness}a) to 1 (perfectly balanced dendrogram, Fig.~\ref{fig:balancedness}c). The balancedness score also satisfies the axioms of a robust and universal measure of tree balance \cite{tree_balance_axiom}. Balancedness can be computed in $\mathcal{O}(n)$ with a single pass over levels.

\subsection{Examples}

To illustrate our metric, we examine the balancedness score of two real networks. Figure~\ref{fig:balancedness}d,f show the dendrograms obtained by link clustering of the street network of Paris and the character network of Les Miserables, respectively. The Paris network sits in the top quartile of our catalogue ($B = 0.25$, ranked 36 of 202, see Fig.~\ref{fig:balancedness}h), whereas the Les Miserables network is highly unbalanced. To determine the balancedness of each dendrogram, we compute the maximum, minimum, and realized entropy values at each level as defined in Eqs.~\ref{eq:max-entropy}--\ref{eq:balancedness}. These values are displayed alongside each dendrogram in Fig.~\ref{fig:balancedness}e,g. The balancedness score corresponds to the average normalized position of the realized entropy curve between the upper and lower bounds.

A majority of real-world networks yield unbalanced dendrograms, as illustrated in Figure~\ref{fig:balancedness}h. Figure~\ref{fig:balancedness}i shows a negative correlation between balancedness and network size (Pearson $r = -0.21$), indicating that larger networks tend to produce more unbalanced dendrograms under link clustering. This further motivates the adaptive cut, as larger networks are those that benefit most from multi-level cuts.

\section{Adaptive cut}

\begin{figure*}[t]
    \centering
    \includegraphics[width=\textwidth]{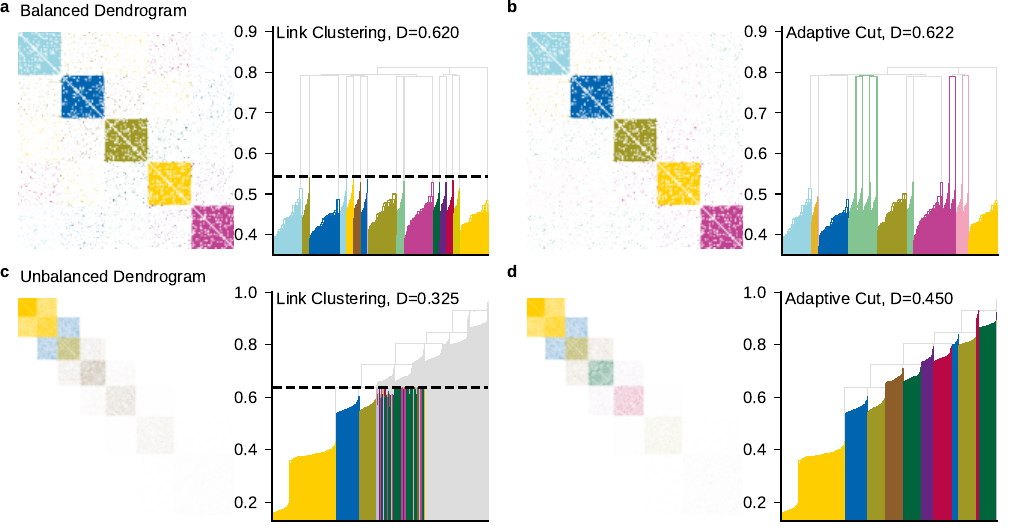}
    \caption{\textbf{Comparison between Link Clustering and adaptive cut.}
    \textbf{(a)} The adjacency matrix of a stochastic block model network, with nodes colored according to edge communities identified by the link clustering method \cite{ahn2010link}. \textbf{(b)} The corresponding dendrogram for the same network, with partitions or communities defined by the single-level link clustering cut \cite{ahn2010link}. The similarity level at which the cut is made is indicated by the dashed line. \textbf{(c, d)} Similar network to (a) and (b), but using an adaptive cut method instead of link clustering.}
    \label{fig:adaptive-cut}
\end{figure*}

To identify the multi-levels of the adaptive cut, we optimize an objective function over the partitions (or cluster membership). We employ a Markov chain Monte Carlo (MCMC) approach to optimize the objective function $f$ within a finite search space $X$, using the softmax distribution
\begin{align}
\label{eq:pi_star}
\pi^{\star}(x) = \frac{e^{f(x)/T}}{Z}\,,
\end{align}
where $x$ represents a state within the search space $X$, $T$ denotes the temperature parameter, and $Z$ is the partition function.

Because moves between adaptive cuts change the number of clusters, the sampler is a reversible-jump MCMC (RJMCMC)~\cite{green1995reversible} with an asymmetric proposal. The Markov chain walks up or down the dendrogram: an up-move merges a selected cluster with its dendrogram sibling (collapsing the sibling's descendants when needed), and a down-move splits a selected cluster by sampling an adaptive cut of its subtree. Each move is accepted or rejected with a Metropolis--Hastings probability that includes the proposal-density ratio; this guarantees detailed balance with respect to $\pi^\star$ even though up- and down-moves are not symmetric. The chain is ergodic because any partition can be reached from any other via a sequence of up- and down-moves. The full proposal mechanism, transition probabilities $Q_{x\to y}$, Metropolis--Hastings acceptance ratio, and simulated-annealing cooling schedule are given in Methods; the formal proofs of detailed balance and ergodicity are in Supplementary Section~S2. We initialize the chain at the single-level cut returned by the underlying clustering algorithm: starting at a known local maximum dramatically shortens mixing time relative to a random partition~\cite{neal1993probabilistic}.

\section{Results}

\subsection{Link Clustering with toy network models}

The Link Clustering method \cite{ahn2010link} uses hierarchical clustering based on link similarity (Jaccard index of neighbors) to create a dendrogram, where each link is a leaf, and branches represent link communities. Communities are extracted by cutting the dendrogram at a specific similarity threshold, allowing nodes to belong to multiple overlapping communities (see Methods, \emph{Link Similarity and Dendrogram Construction}). To identify the most relevant communities, the method introduces partition density $D$, which measures link density within communities and does not suffer from resolution limits \cite{ahn2010link, fortunato2007resolution, lancichinetti2011limits}. The optimal cut is determined by maximizing $D$ along the levels of the dendrogram (see Methods, \emph{Partition Density}).

However, determining the best cut level can be challenging. The adaptive cut method optimizes $D$ further, potentially revealing more accurate community structures, especially in networks with communities of varying sizes and densities. This approach addresses the limitations of a single-level cut by better capturing complex network structures (Fig.~\ref{fig:adaptive-cut}c).

\subsubsection{Varying Density Stochastic Block Model}

To illustrate the limitations of the single-level cut, we focus on two toy networks with a given community structure, a stochastic block model \cite{holland1983stochastic, karrer2011stochastic} and a varying density stochastic block model that we introduce. We simplify the stochastic block model (SBM) by assuming that the probability of an edge between two nodes depends only on whether the nodes belong to the same community or different communities (see the adjacency matrix Fig.~\ref{fig:adaptive-cut}a). The model is described by the following equation:
\begin{align}
P(A_{ij} = 1 \mid z_i = z_j) = \theta_{\text{intra}}, \quad P(A_{ij} = 1 \mid z_i \neq z_j) = \theta_{\text{inter}},
\end{align}
where $A_{ij}$ is the adjacency matrix entry for nodes $i$ and $j$, where $A_{ij} = 1$ indicates the presence of an edge between these nodes, and $A_{ij} = 0$ otherwise. The variables $z_i$ and $z_j$ denote the community assignments of nodes $i$ and $j$ respectively. The term $\theta_{\text{intra}}$ represents the probability of an edge between any two nodes in the same community, while $\theta_{\text{inter}}$ represents the probability of an edge between nodes in different communities. In this model, $\theta_{\text{intra}}$ is the same for all communities, and $\theta_{\text{inter}}$ is the same for all pairs of different communities. Our varying density stochastic block model is the same, except that the $\theta_{\text{intra}}$ decreases and the $\theta_{\text{inter}}$ also decreases. In the Varying Density Stochastic Block Model, we generalize the traditional stochastic block model by varying the intra-community and inter-community densities. This approach allows the creation of communities with different internal structures and varying degrees of connectivity between them. In particular, the intra- and inter-community densities decrease across communities (see Methods, \emph{Varying Density Stochastic Block Model}).

The stochastic block model (adjacency matrix in Fig.~\ref{fig:adaptive-cut}a) exhibits clear communities of identical size and density, exhibiting identical similarities with respect to link clustering. The single-level cut is capable of distinguishing between the communities (see Fig.~\ref{fig:adaptive-cut}a), while the adaptive cut does not offer a superior partitioning (see Fig.~\ref{fig:adaptive-cut}b). The partition densities of the two cuts are close, $D = 0.620$ for link clustering and $D = 0.622$ for the adaptive cut. Conversely, although the Varying Density Stochastic Block Model still exhibits a clear community structure, the single-level cut is unable to provide a partition that reflects these communities. This is due to the inability of a single-cut to capture communities with varying densities. The adaptive cut improves performance, with a value of $D = 0.45$, in comparison to the single-level cut, which has a value of $D = 0.325$ (Fig.~\ref{fig:adaptive-cut}c,d). The communities are aligned with the blocks of the adjacency matrix, as illustrated in Figure~\ref{fig:adaptive-cut}d. Finally, it can be observed that the stochastic block model dendrogram is more balanced than the dendrogram of the varying density stochastic block model. This is evidenced by the respective balancedness values of $B=0.23$ and $B=0.03$.

\subsubsection{Community Structure Evaluation}

\begin{figure}[ht]
    \centering
    \includegraphics[width=\columnwidth]{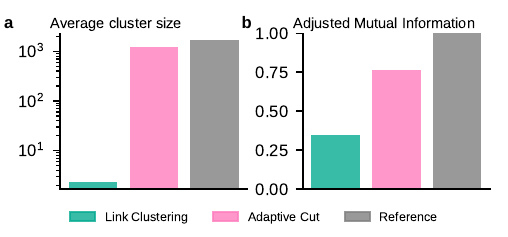}
    \caption{Varying Density Stochastic Block Model Network: Comparison of average cluster size and recovery accuracy.  
    \textbf{(a)} Comparison of the average cluster size (log-scale) for Link Clustering, the adaptive cut, and the ground truth. Link Clustering has many singleton communities that yield a smaller average size, whereas the adaptive cut merges the singletons into clusters, so the average cluster size matches the ground truth better.  
    \textbf{(b)} Adjusted Mutual information between the detected partitions and the ground truth assignments. The adaptive cut has higher mutual information with ground truth than Link Clustering, indicating a better community detection.}
    \label{fig:distribution_plots}
\end{figure}

In Figure~\ref{fig:distribution_plots}a, we compare the log-scale average cluster size produced by each method with the true cluster size. The Link Clustering method has a smaller average community size because it leaves many links isolated in singleton communities (21,254). This observation is due to the tendency for the single-level cuts of unbalanced networks to `over-partition'. In contrast, the adaptive cut method successfully merges these small communities into larger communities -- shifting its mean cluster size much closer to the ground truth, reflecting a more accurate representation of the structure of the network, as can be seen in the adjacency matrix (see Fig.~\ref{fig:adaptive-cut}d).

Figure~\ref{fig:distribution_plots}b shows the mutual information between each detected partition and the true cluster assignments. The adaptive cut has a higher mutual information with the ground truth than Link Clustering, demonstrating that its multi-level cutting scheme recovers the underlying community structure more accurately than a single-level cut.

\subsection{Link Clustering with real networks examples}

\begin{figure}[ht]
    \centering
    \includegraphics[width=\columnwidth]{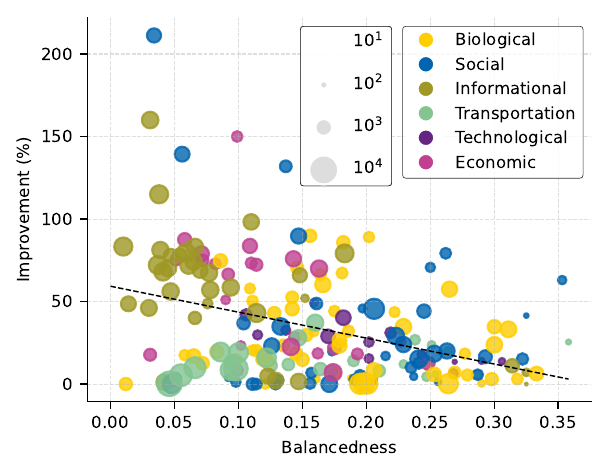}
    \caption{Improvement (in \%) of the partition density between the single-level cut (Link Clustering \cite{ahn2010link}) and the multi-level adaptive cut as a function of the dendrogram balancedness. The color of the symbols indicates the domain of the network, and the size indicates the number of nodes in the network.}
    \label{fig:link_clustering_results}
\end{figure}

We compare the Link Clustering and adaptive cut across 202 real-world networks to evaluate a) the typical improvement and b) the influence of balancedness on the outcome of the community detection. Figure~\ref{fig:link_clustering_results} presents the extent to which adaptive cut enhances partition density (expressed as a percentage) relative to the balancedness of each dendrogram. Our findings indicate that as the balancedness of a network decreases, the likelihood of achieving significant improvements with adaptive cut increases. Across the 202 networks the adaptive cut improves $D$ over the single-level cut on $195$ networks (mean $+34\%$, median $+21\%$), with the largest gains concentrated at low $B$: networks with $B < 0.15$ have a $\geq 75\%$ chance of $\geq 10\%$ improvement under adaptive cut (Supplementary Section~S6). A deterministic greedy hill-climber on the same dendrogram and the same objective recovers about $21$ of those $34$ percentage points; the remaining $\sim 13$ points come from the RJMCMC's ability to escape greedy local optima via the geometric-penalty multi-level proposal (Supplementary Section~S3). This relationship holds across network domains, including economic, transportation, informational, biological, and social networks (the full per-network results, with domain, size, $D_{\text{lc}}$, $D_{\text{ac}}$ and source, are listed in Supplementary Table~S11 (Section~S7)). The robustness of this finding to the choice of linkage method (single, average, weighted, Ward) is examined in Supplementary Section~S4, where we show that the adaptive cut improves partition density on the large majority of the eighty network--linkage combinations from a sample of twenty catalogue networks (the only exceptions are two near-degenerate runs), confirming that the method is agnostic to the dendrogram construction method.

\subsection{Louvain with real network examples}

The Louvain method \cite{louvain} is a popular algorithm for community detection that optimizes modularity to identify community structures in networks. It operates iteratively, initially assigning each node to its own community and then repeatedly merging communities to maximize modularity gains. However, this greedy optimization can lead to the algorithm getting trapped in local maxima, potentially missing the globally optimal community structure. The standard Louvain method also does not produce a complete dendrogram that represents the full hierarchy of community merges, as it stops merging when no further modularity gain is possible.

To integrate the adaptive cut into the Louvain method, we modify the algorithm to construct a full dendrogram, extending beyond the point where modularity gain ceases. This is achieved by allowing merges that result in a modularity decrease, thereby capturing the entire hierarchical structure of the network from individual nodes up to the entire network as a single community. The MCMC then optimises modularity over the resulting dendrogram using the sampler defined in Methods (\emph{Reversible-Jump MCMC Sampler}).

\begin{figure}[ht]
    \centering
    \includegraphics[width=\columnwidth]{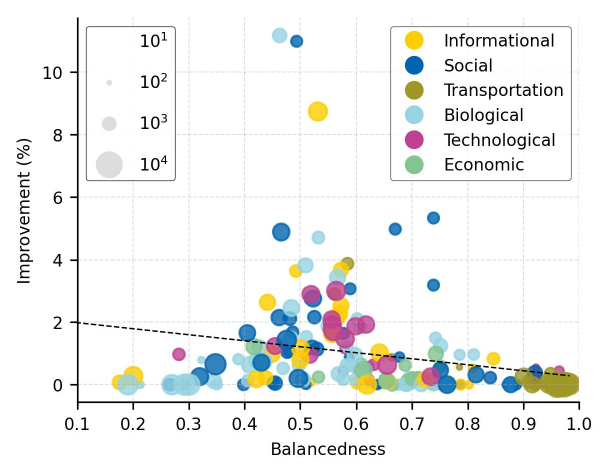}
    \caption{Improvement (in \%) of the modularity between the single-level cut (Louvain) and the multi-level cut (Louvain + adaptive cut) as a function of balancedness. The symbol colors indicate the domain of the network, and the size the number of nodes in the network.}
    \label{fig:louvain_results}
\end{figure}

Our experiments on real-world networks demonstrate improvements in modularity when applying the adaptive cut to the Louvain method. For instance, in networks with unbalanced dendrograms, the adaptive cut effectively adjusts for imbalance by selecting cuts at different levels of the dendrogram, leading to partitions with higher modularity. Figure~\ref{fig:louvain_results} shows the percentage improvement in modularity achieved by the adaptive cut compared to the original Louvain partition, plotted against the balancedness score of each network's dendrogram. The results indicate that networks with lower balancedness scores benefit more from the adaptive cut, validating our approach.

\subsection{Generalizing to any hierarchical clustering}

The adaptive cut framework extends beyond network community detection and is applicable to any clustering task that produces a dendrogram. This generalization requires two main components: a method for constructing the dendrogram and an objective function to optimize.

For dendrogram construction, any hierarchical clustering algorithm can be used, such as single linkage \cite{johnson1967hierarchical}, average linkage \cite{sokal1958statistical}, or Ward's method \cite{ward1963hierarchical}. These methods build a hierarchical tree by iteratively merging clusters based on a chosen linkage criterion.

An appropriate objective function is necessary to evaluate the quality of partitions at different levels of the dendrogram. Common choices include the Davis-Bouldin index \cite{davies1979cluster}, silhouette score \cite{rousseeuw1987silhouettes}, within-cluster sum of squares as in K-means clustering \cite{macqueen1967some}, or the ratio of intra-cluster to inter-cluster distances.

Once the dendrogram and objective function are defined, the adaptive cut is applied by optimizing the objective function using the reversible-jump Markov chain Monte Carlo sampler described in Methods (\emph{Reversible-Jump MCMC Sampler}). The chain explores partitions by moving up and down the dendrogram to find the partition that optimizes the objective function, with detailed balance and ergodicity guaranteed by the matched-support construction proved in Supplementary Section~S2.

\begin{figure}[ht]
    \centering
    \includegraphics[width=\columnwidth]{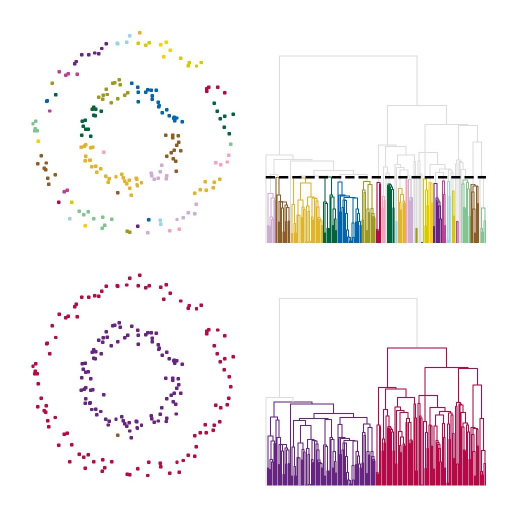}
    \caption{Comparison of clustering results on the two-circle dataset using a single-level cut and the adaptive cut. The adaptive cut identifies the two concentric clusters, which the single-level cut fails to separate.}
    \label{fig:clustering_circle}
\end{figure}

To illustrate the generality of our method, we apply the adaptive cut to the classic two-circle toy dataset \cite{pedregosa2011scikit}, which presents a challenging clustering problem due to its non-convex cluster shapes. We construct a dendrogram using hierarchical clustering and use the within-cluster sum of squares as the objective function. Figure~\ref{fig:clustering_circle} compares the clustering results of the traditional single-level cut and the adaptive cut. While the single-level cut fails to separate the concentric circles, the adaptive cut successfully identifies the inherent cluster structure through multi-level cuts.

\section{Methods}

\subsection{Link Similarity and Dendrogram Construction}
\label{methods:link_similarity}

To construct a dendrogram representing link communities~\cite{ahn2010link}, we first define the similarity between pairs of links. For an undirected, unweighted network, let $n(i)$ denote the set of neighbors of node $i$. The similarity between two links $e_{ik}$ and $e_{jk}$ that share a common node $k$ is the Jaccard index of the endpoint neighborhoods,
\begin{align}
S(e_{ik}, e_{jk}) = \frac{|n(i) \cap n(j)|}{|n(i) \cup n(j)|}.
\label{eq:jaccard}
\end{align}
For weighted and directed networks this generalizes to the Tanimoto coefficient~\cite{tanimoto1958elementary}. Single-linkage hierarchical clustering on the resulting similarity matrix yields the dendrogram: each link starts as its own cluster and clusters are merged iteratively in decreasing order of similarity until a single cluster remains.

\subsection{Partition Density}
\label{methods:partition_density}

To identify a meaningful level of clustering within a dendrogram we use the partition density $D$~\cite{ahn2010link}. Given a network with $M$ links partitioned into $C$ subsets, where subset $c$ contains $m_c$ links spanning $n_c$ nodes, the link density of community $c$ is
\begin{align}
D_c = \frac{m_c - (n_c-1)}{n_c(n_c-1)/2 - (n_c-1)}.
\label{eq:Dc}
\end{align}
The overall partition density is the link-weighted average,
\begin{align}
D = \frac{2}{M} \sum_c m_c \,\frac{m_c - (n_c-1)}{(n_c-2)(n_c-1)}.
\label{eq:D}
\end{align}
$D$ is free from the resolution limit~\cite{fortunato2007resolution,lancichinetti2011limits} and serves as the objective function maximized by the adaptive cut.

\subsection{Varying Density Stochastic Block Model}
\label{methods:vdsbm}

The varying density stochastic block model (VDSBM) generalizes the standard SBM by allowing the intra-community edge probability to vary across communities. For community $c$ of size $k_c$ with per-node edge probabilities $\{p_i\}_{i\in c}$, the intra-community density is
\begin{align}
\theta_{\text{intra}}^{(c)} = \frac{1}{k_c} \sum_{i \in c} p_i,
\label{eq:theta-intra}
\end{align}
and $\theta_{\text{intra}}^{(c)}$ decreases (linearly or non-linearly) as the community index $c$ grows. The inter-community density between adjacent communities $c$ and $c+1$ scales with the smaller of the two,
\begin{align}
\theta_{\text{inter}}^{(c,c+1)} = \big(\theta_{\text{intra}}^{(c+1)}\big)^2,
\label{eq:theta-inter}
\end{align}
producing networks with both highly dense and very sparse communities, a regime where single-level cuts cannot resolve the structure.

\subsection{State Space}
\label{methods:state_space}

The MCMC state space is the set of all valid adaptive cuts of the dendrogram; equivalently, sets of internal nodes whose subtrees partition all leaves. The number of valid cuts of a binary dendrogram rooted at $v$ is given by the recursion
\begin{align}
a(v) = \begin{cases} 1 & \text{if } v \text{ is a leaf,} \\ 1 + a(\ell)\cdot a(r) & \text{otherwise,} \end{cases}
\label{eq:state-space-recursion}
\end{align}
where $\ell$ and $r$ are the children of $v$ and the leading $1$ counts the trivial cut $\{v\}$. For a perfectly balanced binary tree on $n = 2^d$ leaves this gives $a_d = 1 + a_{d-1}^2$ (OEIS A003095), which grows doubly exponentially in depth $d$, giving $a_d = 2^{\Theta(n)}$ (single-exponential in $n$, since $\log_2 a_d \approx 0.59\,n$). This is dramatically smaller than the Bell number $B(n)$ that counts all partitions of $n$ elements, and is what makes the adaptive-cut MCMC tractable.

\subsection{Reversible-Jump MCMC Sampler}
\label{methods:rjmcmc}

Because moves between adaptive cuts change the number of clusters, the sampler is a reversible-jump MCMC (RJMCMC)~\cite{green1995reversible}. Given a partition with $k$ clusters, each proposal step proceeds as follows.
\begin{enumerate}
\item Select a cluster $v_i$ uniformly at random with probability $1/k$.
\item Choose a direction: forced up ($P_{\text{up}}=1$) if $v_i$ is a dendrogram leaf, forced down ($P_{\text{down}}=1$) if $v_i$ is the root, otherwise $P_{\text{up}} = P_{\text{down}} = 1/2$.
\item Execute the move. \textbf{Up:} merge $v_i$ with its dendrogram sibling into their parent; if the sibling has been split into smaller descendants, collapse all of them. \textbf{Down:} sample a random adaptive cut $\sigma$ of $v_i$'s subtree from the geometric-penalty proposal
\begin{align}
\rho(\sigma \mid v_i) = \frac{2^{-|\sigma|}}{Z_c(v_i)},
\label{eq:rho}
\end{align}
where the constrained partition function $Z_c(v_i)$ restricts $\sigma$ so that at least one direct child of $v_i$ remains whole. This guarantees that the reverse of any down-move is an up-move from a whole direct child; when both direct children remain whole, either can trigger the reverse and the reverse proposal sums their contributions (Eq.~\ref{eq:proposal}).
\end{enumerate}

The normalization constants are precomputed bottom-up in $\mathcal{O}(n)$:
\begin{align}
Z(v) &= \begin{cases} 1/2 & \text{if } v \text{ is a leaf,} \\ 1/2 + Z(\ell)\,Z(r) & \text{otherwise,} \end{cases} \label{eq:Z}\\
Z_c(v) &= \tfrac{1}{2}Z(\ell) + \tfrac{1}{2}Z(r) - \tfrac{1}{4}. \label{eq:Zc}
\end{align}
Values are stored in log-space for large subtrees.

The resulting proposal probability $Q_{x\to y}$ for a transition from state $x$ (with $k$ clusters) to state $y$ is asymmetric:
\begin{equation}
Q_{x\to y} = \left\{
\begin{array}{@{}ll@{}}
\frac{1}{k}\sum_{v\in S_{\text{up}}} P_{\text{up}}(v), & \text{if } y \text{ is an up-move from } x, \\[6pt]
\frac{1}{k}\,P_{\text{down}}(v_i)\,\rho(\sigma\mid v_i), & \text{if } y \text{ is a down-move from } x, \\[4pt]
0 & \text{otherwise,}
\end{array}\right.
\label{eq:proposal}
\end{equation}
where $S_{\text{up}}$ is the set of clusters in $x$ whose selection and up-move yields $y$. Although $Q$ is not symmetric, the Metropolis--Hastings rule~\cite{metropolis1953equation,hastings1970monte} preserves detailed balance for any ergodic proposal by incorporating the proposal-density ratio. The acceptance probability is
\begin{align}
\alpha(x_i \to x_\star) &= \min\Bigl\{1,\; \frac{\pi^{\star}(x_\star)\,Q_{x_\star\to x_i}}{\pi^{\star}(x_i)\,Q_{x_i\to x_\star}}\Bigr\} \notag\\
                       &= \min\Bigl\{1,\; e^{\Delta f / T}\cdot\frac{Q_{x_\star\to x_i}}{Q_{x_i\to x_\star}}\Bigr\},
\label{eq:alpha_ratio}
\end{align}
where $\Delta f = f(x_\star) - f(x_i)$. With this acceptance probability the stationary distribution is $\pi^\star(x) \propto e^{f(x)/T}$ on the space of adaptive cuts. The matched-support property ($Q_{x\to y}>0 \iff Q_{y\to x}>0$), detailed balance, and ergodicity are proved formally in Supplementary Section~S2.

To prevent the chain from getting trapped in local maxima we use the logarithmic simulated-annealing~\cite{kirkpatrick1983optimization} cooling schedule
\begin{align}
T_k = \frac{T_0}{\log(k - k_{\text{SA}} + 2)}, \qquad k > k_{\text{SA}},
\label{eq:annealing}
\end{align}
with $T_k = T_0$ for $k \leq k_{\text{SA}}$. We use $k_{\text{SA}} = 2 N_{\text{steps}}/3$ by default, which lets the chain explore at high temperature before converging.

\subsection{Balancedness Axioms}
\label{methods:balancedness_axioms}

The balancedness score $B$ (Eq.~\ref{eq:balancedness}) satisfies the three universal tree-balance axioms of~\cite{tree_balance_axiom}.
\paragraph*{Maximum-value axiom.} For every level $l$, $H_{\min}(\pi_l) \leq H(\pi_l) \leq H_{\max}(\pi_l)$, so each summand of $B$ lies in $[0,1]$ and $0 \leq B \leq 1$. Equality $B = 1$ requires $H(\pi_l) = H_{\max}(\pi_l)$ at every level, i.e., every internal node splits its leaves into branches of identical size.
\paragraph*{Minimum-value axiom.} In a caterpillar (fully unbalanced) dendrogram every internal node separates a single leaf from the rest, so $H(\pi_l) = H_{\min}(\pi_l)$ at every level and $B = 0$. Conversely, $B = 0$ implies every split ejects a single leaf, i.e., the dendrogram is a caterpillar.
\paragraph*{Insensitivity to vanishing subtrees.} Replacing a leaf by a subtree with $\varepsilon n$ leaves ($\varepsilon \to 0$) perturbs each entropy by $\mathcal{O}(\varepsilon\log(1/\varepsilon))$. Since the denominators $H_{\max}-H_{\min}$ are independent of $\varepsilon$, each summand of $B$ perturbs by at most $C\,\varepsilon\log(1/\varepsilon)$, so $|B(T_\varepsilon) - B(T)| \to 0$.

\subsection{Computational Complexity}

The sampler performs a one-off $\mathcal{O}(M)$ precomputation of the geometric-cut normalisers $\log Z(v),\log Z_c(v)$, the per-node leaf sets, the down-move probability tables, and the per-subtree partition-density contributions $D(v) = m_v D_c(m_v,n_v)$. The per-step cost of the main loop is then $\mathcal{O}(k)$ for an up-move (where $k = |\sigma|$ is the current number of communities and the dominant cost is a sibling-leaf disjointness scan for the multi-collapse) and $\mathcal{O}(|\sigma'|)$ for a down-move that samples a cut of size $|\sigma'|$ from $\rho(\sigma\mid v) = 2^{-|\sigma|}/Z_c(v)$. Accepting a move updates the total partition density incrementally from the cached $D(v)$ values in $\mathcal{O}(|\sigma' - \sigma|)$. The total optimisation cost is $\mathcal{O}(M) + \mathcal{O}(N_{\text{steps}}\cdot \bar c)$ where $\bar c$ is the average move cost; in practice $\bar c$ stays small because the chain spends most of its time near a low-$k$ partition.

We benchmarked the sampler at $2\times 10^5$ steps, $T = 10^{-4}$, on a single CPU process across ten networks spanning $78 \le M \le 18{,}083$ edges (Supplementary Table~S6). Wall-clock spans $1$--$10\,\mathrm{s}$ for $M < 1{,}000$ and tens of seconds to a few minutes for $1{,}000 \le M < 10{,}000$, with the high-improvement networks at the slow end (e.g.\ \texttt{bible\_nouns}, $M=9{,}059$, $184\,\mathrm{s}$); networks with $M \ge 10{,}000$ take $1$--$10\,\mathrm{min}$. Dendrogram construction (Jaccard similarities and single-linkage) overtakes the MCMC only for the largest networks (e.g.\ \texttt{us\_congress-H93}, $M=18{,}083$). Convergence, initialisation-robustness, temperature-sensitivity, and reproducibility diagnostics are reported in Supplementary Section~S5. The full catalogue of networks used to evaluate the method (with balancedness, $D_{\text{lc}}$, $D_{\text{ac}}$, and provenance) is provided as Supplementary Table~S11 (Section~S7).

\section{Discussion}

In this paper, we introduced the adaptive cut, a new method designed to overcome the limitations of traditional single-level cuts in hierarchical clustering and community detection. Using multi-level cuts and optimizing an objective function using a Markov chain Monte Carlo algorithm with simulated annealing, the adaptive cut uses the full hierarchical structure of dendrograms. This approach is best suited to datasets with unbalanced dendrograms, where single-level cuts often fail to capture the true underlying structure. We also introduce the balancedness score, an information-theoretic metric that quantifies the balance of a dendrogram and predicts the potential benefits of the adaptive cut.

First, we demonstrated that traditional single-level cuts can be inadequate for accurately identifying clusters in complex networks, especially when dealing with unbalanced dendrograms, which can arise when the density of data is uneven across a dataset. Single-level cuts may lead to over-aggregation in some parts of the dendrogram and overly granular partitions in others, resulting in suboptimal clustering. This issue was evident in our experiments with stochastic block models exhibiting varying densities, where the adaptive cut outperformed the single-level approach.

Second, we introduced the balancedness score, an information-theoretic metric that quantifies the balance of a dendrogram. Our analysis of over 200 real-world networks revealed that most dendrograms are unbalanced. The balancedness score not only measures dendrogram balance but also serves as a predictor of the potential improvement achievable with the adaptive cut method. Networks with lower balancedness scores tended to benefit more from the adaptive cut, as it adjusts for the imbalance by employing multi-level cuts.

Applied to both synthetic and real-world networks, the adaptive cut outperforms single-level cuts. In synthetic networks, the adaptive cut produced partitions that more accurately reflected the ground truth community structures, especially in models with varying densities. In real-world networks across diverse domains (including social, economic, biological, and transportation networks), the adaptive cut improved performance metrics such as partition density and modularity. These improvements were most pronounced in networks with unbalanced dendrograms.

We showed the generality of the adaptive cut method. It can be applied across different hierarchical clustering techniques (e.g., single linkage, average linkage, Ward's method) and optimized for objective functions such as modularity, silhouette score, and partition density. This flexibility lets the adaptive cut apply to many clustering tasks beyond network analysis, including machine learning tasks that use hierarchical clustering.

We distinguish adaptive cut from other multiscale community detection approaches. Methods based on multiresolution modularity \cite{reichardt2006statistical} scan a resolution parameter to identify communities at different scales, but operate on the full partition space rather than leveraging dendrogram structure. Stability-based methods \cite{delvenne2010stability} analyze community persistence across Markov time scales, requiring diffusion processes rather than an existing dendrogram. Dynamic tree cut \cite{dynamic_cut} is the most closely related approach, as it also produces multi-level cuts of dendrograms, but relies on shape-based heuristics rather than objective function optimization. The key distinction is that adaptive cut is not a standalone community detection method but a complementary post-processing step: it takes any given dendrogram as input and optimizes the cut, regardless of how the dendrogram was constructed.

The balancedness index, beyond its role as an improvement predictor, serves as a structural descriptor of community heterogeneity. The quantity $B$ reads the spread of merge heights in the dendrogram: a dendrogram whose merges are concentrated near a single similarity gives $B$ near $1$, while a dendrogram with widely spread merge heights gives $B$ near $0$. A controlled $2\times 2$ design that crosses uniform vs varying intra-community density against flat vs nested community structure (SI Sec.~S6.3) shows that the spread is driven by density heterogeneity across communities, not by the depth of nesting: $B$ drops by $0.09$--$0.12$ along the density-variation axis and changes by less than $0.04$ along the nesting axis. The density-axis ordering also holds under average and complete linkage; under single linkage on larger graphs the chaining effect compresses $B$ into a narrow band, so both the absolute scale of $B$ and the resolvability of the density signal depend on the linkage choice (SI Fig.~S8). Balancedness is related to, but distinct from, multiscale notions arising from the resolution limit of modularity optimisation \cite{fortunato2007resolution} and from multi-resolution \cite{reichardt2006statistical} and Markov-stability \cite{delvenne2010stability} community-detection schemes, which target community size relative to total network scale. Density heterogeneity is one ingredient of multiscale structure: low $B$ identifies networks where some communities are tight and others are loose, the regime in which a single-level cut over- or under-fragments and where adaptive cut helps most.

While the adaptive cut offers advantages, it has limitations. One key limitation is its reliance on the dendrogram structure, which confines the set of possible partitions to those represented within the dendrogram. Although this restriction reduces the state space and enhances computational efficiency, it may exclude some potentially optimal partitions not present in the dendrogram.

The optimization process within the adaptive cut uses a Markov chain Monte Carlo method with a simulated annealing schedule. The choice of cooling schedule can impact convergence rates and computational efficiency. Future work could investigate adaptive cooling schemes, where the cooling rate is adjusted dynamically based on the acceptance rate of new solutions. This approach could provide a more flexible optimization strategy, enhancing convergence and allowing the algorithm to balance exploration and exploitation.

Another promising direction is to explore alternative optimization methods, such as dynamic programming, which can optimize objective functions on tree structures with lower computational complexity \cite{bellman1966dynamic, marti2017cut}.

In conclusion, the adaptive cut method, coupled with the balancedness score, represents an advance in hierarchical clustering and community detection. By using the hierarchical information in dendrograms and addressing the shortcomings of single-level cuts, it provides a more accurate clustering method suitable for complex and unbalanced datasets.

\begin{acknowledgments}
We thank Morten Mørup, Nikolaos Nakis and Abdulkadir Çelikkanat for stimulating discussions. This work was supported in part by the Villum Foundation Grant Nation-Scale Social Networks [grant number: 00034288] (S.L.) and the Danish Council for Independent Research [grant number: 0136-00315B] (S.L.). The funders had no role in study design, data collection and analysis, decision to publish or preparation of the manuscript.
\end{acknowledgments}

\section*{Author contributions}

L.B. and S.L. conceived the method. L.B. implemented the algorithms and conducted the experiments. Y.-Y.A. contributed to the theoretical framework. L.B., Y.-Y.A., and S.L. analyzed the results and wrote the manuscript.

\section*{Competing interests}

The authors declare no competing interest.

\section*{Data availability}

The code and data underlying this article are available at \href{https://github.com/LCB0B/adaptive\_cut}{github.com/LCB0B/adaptive\_cut}.

\section*{Supplementary Information}

The Supplementary Information after the main text contains: (S1)~code availability; (S2)~the complete proof of detailed balance and ergodicity for the reversible-jump MCMC sampler, organised as Lemmas~S2.1--S2.5, Theorem~S2.6 (Detailed Balance), and Corollary~S2.7 (verification of the Metropolis--Hastings sufficient conditions); (S3)~a deterministic greedy baseline on the same dendrogram and the same objective, separating the multi-cut gain from the MCMC contribution; (S4)~an adaptive-cut comparison across four linkage methods (single, average, weighted, Ward) on twenty representative networks; (S5)~convergence and sensitivity diagnostics (chain length, initialisation, temperature, complexity, wall-clock); (S6)~balancedness as a structural descriptor (decision criterion, partition-disagreement diagnostic, controlled $2\times 2$ design, NTB comparison); and (S7)~the complete table of all 202 networks used in Figures~\ref{fig:link_clustering_results} and~\ref{fig:louvain_results}, with domain, size, balancedness, $D_{\text{lc}}$, $D_{\text{ac}}$, source, and reference for each.

\clearpage
\onecolumngrid
\begin{center}
{\large\textbf{Supplementary Information}}
\end{center}

\setcounter{section}{0}
\setcounter{figure}{0}
\setcounter{table}{0}
\setcounter{equation}{0}
\renewcommand{\thesection}{S\arabic{section}}
\renewcommand{\thesubsection}{\thesection.\arabic{subsection}}
\renewcommand{\thesubsubsection}{\thesubsection.\arabic{subsubsection}}
\renewcommand{\thefigure}{S\arabic{figure}}
\renewcommand{\thetable}{S\arabic{table}}
\numberwithin{equation}{section}
\renewcommand{\theequation}{S\arabic{section}.\arabic{equation}}
\renewcommand{\thelemma}{S\arabic{section}.\arabic{lemma}}
\makeatletter
\renewcommand{\p@subsection}{}
\renewcommand{\p@subsubsection}{}
\makeatother

\section{Code availability}

Code is available on GitHub \url{https://github.com/LCB0B/adaptive_cut}.

\section{Reversible-Jump MCMC: Detailed Balance and Ergodicity}
\label{sec:rjmcmc}

The adaptive-cut sampler defined in Methods is a reversible-jump MCMC~\cite{green1995reversible} on the space $\mathcal{X}$ of adaptive cuts of a binary dendrogram. Every transition is either a single-cluster up-move (collapse the selected cluster with its sibling, possibly absorbing the sibling's descendants) or a down-move that samples a constrained adaptive cut of the selected cluster's subtree from the geometric-penalty proposal $\rho(\sigma\mid v) = 2^{-|\sigma|}/Z_c(v)$ (Eq.~\ref{eq:rho} of Methods). In this section we (i) verify that the unconstrained and constrained proposals are bona-fide probability distributions over cuts, (ii) prove that up- and down-moves come in reverse pairs with $Q_{x\to y}>0 \iff Q_{y\to x}>0$ (matched support), and (iii) conclude that the Metropolis--Hastings acceptance rule preserves the target $\pi^{\star}(x) \propto e^{f(x)/T}$ and produces an ergodic chain. Throughout, $\Sigma(v)$ denotes the set of adaptive cuts of the subtree rooted at $v$ and $\Sigma_c(v)\subset\Sigma(v)$ the constrained cuts in which at least one direct child of $v$ remains whole, so that the state space is $\mathcal{X}=\Sigma(\mathrm{root})$.

\subsection{Sampler correctness}

\begin{lemma}[Single-cut proposal normalises]
\label{lem:sampler-correctness}
For every node $v$ of the dendrogram, the unconstrained geometric-penalty proposal $\rho_u(\sigma\mid v) = 2^{-|\sigma|}/Z(v)$ defines a probability distribution over the set of adaptive cuts of the subtree rooted at $v$, with
\begin{align}
Z(v) = \begin{cases} 1/2 & \text{if } v \text{ is a leaf,} \\ 1/2 + Z(\ell)\,Z(r) & \text{if } v \text{ has children } \ell, r. \end{cases}
\label{eq:Z-recursion}
\end{align}
\end{lemma}

\begin{proof}
By structural induction on $v$. If $v$ is a leaf, the only cut is $\{v\}$ with $|\sigma|=1$, so $\sum_\sigma 2^{-|\sigma|} = 1/2 = Z(v)$. If $v$ has children $\ell, r$, every cut either equals $\{v\}$ (contributing $2^{-1}=1/2$) or descends past $v$, in which case it decomposes uniquely as $\sigma = \sigma_\ell \sqcup \sigma_r$ with $\sigma_\ell\in\Sigma(\ell)$, $\sigma_r\in\Sigma(r)$ and $|\sigma|=|\sigma_\ell|+|\sigma_r|$: every block of an adaptive cut is a subtree root, so no block straddles the two children, and $\sigma\mapsto(\sigma_\ell,\sigma_r)$ is a bijection onto $\Sigma(\ell)\times\Sigma(r)$. These cuts contribute $\sum_{\sigma_\ell} 2^{-|\sigma_\ell|} \cdot \sum_{\sigma_r} 2^{-|\sigma_r|} = Z(\ell)\,Z(r)$ by the inductive hypothesis. Summing gives $Z(v) = 1/2 + Z(\ell)\,Z(r)$.
\end{proof}

\begin{lemma}[Constrained-cut proposal normalises]
\label{lem:sampler-correctness-constrained}
The constrained proposal $\rho_c(\sigma\mid v) = 2^{-|\sigma|}/Z_c(v)$ over adaptive cuts $\sigma$ of $v$'s subtree in which at least one direct child of $v$ remains whole is a probability distribution, with
\begin{align}
Z_c(v) = \tfrac{1}{2}Z(\ell) + \tfrac{1}{2}Z(r) - \tfrac{1}{4}.
\label{eq:Zc-recursion}
\end{align}
\end{lemma}

\begin{proof}
Let $A$ be the adaptive cuts of $v$'s subtree in which $\ell$ remains whole and $B$ those in which $r$ remains whole; the constrained set is $A\cup B$. Summing the geometric weight over $A$ gives $\sum_{\sigma_r} 2^{-1-|\sigma_r|} = Z(r)/2$ (with $\ell$ whole, the rest of the cut is any cut of $r$'s subtree), and over $B$ gives $Z(\ell)/2$. The overlap $A\cap B$ contains only $\sigma = \{\ell, r\}$: once both children are whole they already cover every leaf of $v$, so no further node can be added without re-covering a leaf or re-introducing $v$; its weight is $2^{-2} = 1/4$. Inclusion--exclusion yields $Z_c(v) = Z(\ell)/2 + Z(r)/2 - 1/4$.
\end{proof}

\begin{lemma}[Bottom-up linear computability]
\label{lem:Zc-recursion}
Both $Z(v)$ and $Z_c(v)$ for every node $v$ can be computed in $\mathcal{O}(n)$ time and stored in log-space, where $n$ is the number of leaves of the dendrogram (the number of edges $M$ in the link-clustering application).
\end{lemma}

\begin{proof}
A single post-order traversal computes $Z(v)$ from its children using Eq.~\ref{eq:Z-recursion}; a second post-order traversal computes $Z_c(v)$ using Eq.~\ref{eq:Zc-recursion}. Each node is visited a constant number of times.
\end{proof}

\subsection{Matched support of proposals (UP \texorpdfstring{$\leftrightarrow$}{<->} DOWN pairing)}

\begin{lemma}[UP $\to$ DOWN pairing]
\label{lem:up-down}
Let $y$ be the state obtained from $x$ by an up-move at cluster $v_i \in x$. Then $Q_{x\to y} > 0$ and there exists a unique reverse down-move from $y$ to $x$, with $Q_{y\to x} > 0$.
\end{lemma}

\begin{proof}
Let $p$ be the parent of $v_i$ in the dendrogram and let $v_s$ be the dendrogram sibling of $v_i$. The up-move removes from $x$ the cluster $v_i$ together with every cluster of $x$ that is a descendant of $v_s$ (a set $R \subseteq x$), and adds the cluster $p$, producing $y$. Since $x$ is a cut, the clusters of $x$ lying under $v_s$ partition $v_s$'s leaves, so together with $\{v_i\}$ (which covers $v_i$'s subtree) the set $R$ covers $p$'s subtree exactly once, i.e.\ $R\in\Sigma(p)$. Two cases arise. If $v_s\in x$ (the sibling is whole), then $R=\{v_i,v_s\}$ and the same $y$ is reached by selecting either $v_i$ or $v_s$, so $S_{\text{up}}=\{v_i,v_s\}$. If $v_s\notin x$ (the sibling has been split), then $|R|\geq 3$ (covering the split subtree of $v_s$ needs at least two clusters, plus $v_i$) and only the selection of $v_i$ yields $y$, so $S_{\text{up}}=\{v_i\}$. In both cases $Q_{x\to y}=(1/|x|)\sum_{v\in S_{\text{up}}}P_{\text{up}}(v)>0$. For the reverse, $v_i$ is a direct child of $p$ by construction; selecting $p$ in $y$ (probability $1/|y|$) and proposing the constrained cut $\sigma=R$ of $p$'s subtree, in which the direct child $v_i$ of $p$ remains whole ($v_i\in R$, so $R\in\Sigma_c(p)$), inverts the up-move. Its probability is $Q_{y\to x}=(1/|y|)\,P_{\text{down}}(p)\,2^{-|R|}/Z_c(p)>0$.
\end{proof}

\begin{lemma}[DOWN $\to$ UP pairing]
\label{lem:down-up}
Let $y$ be the state obtained from $x$ by a down-move at cluster $v_i \in x$ producing a constrained cut $\sigma$. Then $Q_{x\to y} > 0$, and the reverse transition $y\to x$ has $Q_{y\to x} > 0$, triggered by every whole direct child of $v_i$ in $\sigma$.
\end{lemma}

\begin{proof}
Since the only size-one cut of $v_i$'s subtree is $\{v_i\}$ itself, which keeps no direct child whole, every $\sigma\in\Sigma_c(v_i)$ has $|\sigma|\ge 2$, so a down-move strictly increases the cluster count and $y\ne x$. Because $\sigma$ is a constrained cut, the set $W$ of direct children of $v_i$ that are whole in $\sigma$ is non-empty. For each $w\in W$, selecting $w$ in $y$ (probability $1/|y|$) and proposing an up-move collapses $w$ together with the clusters of $y$ in the subtree of $w$'s sibling (the sibling itself if whole, or its descendants if split; in either case exactly $\sigma\setminus\{w\}$) and restores $v_i$. Conversely, any one-step move from $y$ that yields $x$ must increase the cluster count, so it is an up-move, and must reinsert $v_i$, which forces the selected cluster to be a direct child of $v_i$ present in $\sigma$, i.e.\ an element of $W$. The reverse selections are exactly $W$, and $Q_{y\to x} = (1/|y|)\sum_{w\in W} P_{\text{up}}(w) > 0$. When only one child is whole, $|W|=1$; when both children are whole, $|W|=2$ and the two contributions add.
\end{proof}

\subsection{Detailed balance and ergodicity}

\begin{theorem}[Detailed balance]
\label{thm:detailed-balance}
For the proposal $Q$ defined in Eq.~\ref{eq:proposal} of Methods and the Metropolis--Hastings acceptance probability $\alpha(x\to y) = \min\{1,\, \pi^{\star}(y)Q_{y\to x}/[\pi^{\star}(x)Q_{x\to y}]\}$ (Eq.~\ref{eq:alpha_ratio} of Methods), the resulting transition kernel $P(x\to y) = Q_{x\to y}\,\alpha(x\to y)$ (for $y\neq x$) satisfies
\begin{align}
\pi^{\star}(x)\,P(x\to y) = \pi^{\star}(y)\,P(y\to x) \qquad \text{for all } x,y\in\mathcal{X}.
\label{eq:detailed-balance}
\end{align}
\end{theorem}

\begin{proof}
An up-move strictly decreases the cluster count (it removes $|R|\ge 2$ clusters and adds their parent) while a down-move strictly increases it (it removes one cluster and adds $|\sigma|\ge 2$), so for any ordered pair $(x,y)$ at most one case of Eq.~\ref{eq:proposal} of Methods applies and $Q_{x\to y}$ is single-valued. Lemmas~\ref{lem:up-down} and~\ref{lem:down-up} show that $Q_{x\to y}>0 \iff Q_{y\to x}>0$ for every pair $(x,y) \in \mathcal{X}^2$. For such a pair,
\begin{align*}
\pi^{\star}(x)\,Q_{x\to y}\,\alpha(x\to y) &= \min\{\pi^{\star}(x)\,Q_{x\to y},\;\pi^{\star}(y)\,Q_{y\to x}\}\\
&= \pi^{\star}(y)\,Q_{y\to x}\,\alpha(y\to x),
\end{align*}
by the symmetry of the $\min$. For pairs with $Q_{x\to y} = Q_{y\to x} = 0$, both sides of Eq.~\ref{eq:detailed-balance} vanish.
\end{proof}

\begin{corollary}[Metropolis--Hastings applicability]
\label{cor:mh-applicability}
The chain defined by Eqs.~\ref{eq:proposal} and~\ref{eq:alpha_ratio} of Methods satisfies the sufficient conditions for a convergent Metropolis--Hastings sampler. (i)~\emph{Irreducibility.} From any state $x \in \mathcal{X}$, repeated up-moves reach the root state $\{r\}$ in at most $|x|-1$ steps, since each up-move removes at least two clusters and adds one, strictly decreasing $|x|$. From $\{r\}$, any target $y$ is reached by at most $|y|-1$ down-moves: whenever a current cluster $v$ strictly contains more than one cluster of $y$, propose the minimal constrained cut consisting of the two direct children of $v$ (both whole, always in $\Sigma_c(v)$, with probability $1/[4Z_c(v)]>0$), which refines the partition by one cluster; iterating reaches $y$, so $P^{(N)}(x,y) > 0$ for some $N \leq |x|+|y|-2 \leq 2(M-1)$. (ii)~\emph{Matched support}: $Q_{x\to y}>0 \iff Q_{y\to x}>0$ by Lemmas~\ref{lem:up-down}--\ref{lem:down-up}. (iii)~\emph{Evaluable positive target}: $\pi^{\star}(x) = e^{f(x)/T}/Z>0$ enters $\alpha$ only through the ratio $\pi^{\star}(y)/\pi^{\star}(x) = e^{(f(y)-f(x))/T}$, so the normaliser $Z$ is never needed. (iv)~\emph{Detailed balance} holds by Theorem~\ref{thm:detailed-balance}. (v)~\emph{Aperiodicity}: the proposal is asymmetric and does not by itself satisfy detailed balance with respect to $\pi^{\star}$, so some move is accepted with probability $\alpha<1$; the corresponding rejection gives a positive self-transition probability $P(x\to x)>0$, so $x$ has period $1$ and, with irreducibility, the chain is aperiodic. By (i)--(v) and the ergodic theorem for finite irreducible aperiodic Markov chains~\cite{levin2017markov}, $\pi^{\star}$ is the unique stationary distribution and $P^{n}(x,\cdot)\to\pi^{\star}$ in total variation from every initial state.

These guarantees hold at each fixed temperature $T$, where the kernel has stationary law $\pi^{\star}_T \propto e^{f/T}$. As $T\to 0$ the law $\pi^{\star}_T$ concentrates on the maximisers of $f$, but the time-inhomogeneous chain run under the cooling schedule $T_k$ (Eq.~\ref{eq:annealing} of Methods) inherits this concentration only under a sufficiently slow schedule. For $T_k = c/\log k$, Hajek's theorem~\cite{hajek1988cooling} gives convergence in probability to a global maximiser of $f$ exactly when $c \ge d^{\star}$, where $d^{\star}$ is the largest depth (measured in $f$) of a basin surrounding a non-global local maximum. The deployed numerator $T_0$ is small (of order $0.1/M$) and likely below $d^{\star}$, so we use the schedule as a simulated-annealing heuristic that biases the chain toward high-$f$ states rather than as a proof of global optimality; convergence behaviour is assessed empirically in Section~\ref{sec:diagnostics}.
\end{corollary}

\section{Greedy Baseline}
\label{sec:greedy-baseline}

To isolate the contribution of the stochastic, multi-level Metropolis--Hastings sampler from the multi-cut principle itself, we compared the adaptive cut against a deterministic greedy baseline operating on the same single-linkage dendrogram and optimising the same objective ($D$). Starting from $\sigma_{\text{lc}}$, the greedy procedure scans the current communities and applies any UP move (merge with dendrogram sibling, absorbing descendants when needed) or DOWN move (split into a sampled adaptive cut of the subtree) that strictly increases $D$, iterating until no improving move is found. The greedy baseline shares the move set of the RJMCMC sampler but removes all stochasticity and all sub-optimal acceptances.

We ran the greedy baseline on all 202 networks in the catalogue (Section~\ref{sec:network_table}). Greedy alone improves over $\sigma_{\text{lc}}$ on $156/202$ networks ($77\%$, mean improvement $+21\%$, median $+12\%$), with the largest gains on highly unbalanced dendrograms (\texttt{Erdos} $+135.7\%$, \texttt{eu\_procurements\_alt-AT\_2008} $+131.1\%$, \texttt{STEM concept networks Researchers} $+126.0\%$, \texttt{foursquare-NYC\_restaurant\_checkin} $+83.0\%$, \texttt{bitcoin\_alpha} $+70.1\%$). The multi-cut principle, separated from the MCMC, already produces large gains over the single-level cut.

The full RJMCMC sampler matches or exceeds the greedy baseline on every network. Across all 202 networks, $D_{\text{ac}} \geq D_{\text{greedy}}$ with zero violations; $D_{\text{ac}} > D_{\text{greedy}}$ strictly on $196/202$ networks ($97\%$). The RJMCMC mean improvement is $+34\%$ ($+21\%$ median), which leaves a $\sim 13$-percentage-point mean gap between RJMCMC and greedy. This gap quantifies the contribution of the geometric-penalty multi-level DOWN proposal (Eq.~\ref{eq:rho} of Methods) combined with simulated annealing: the chain reaches dendrogram-consistent multi-level cuts that the deterministic greedy cannot escape to from a local optimum.

\begin{figure}[h]
\centering
\includegraphics[width=0.95\textwidth]{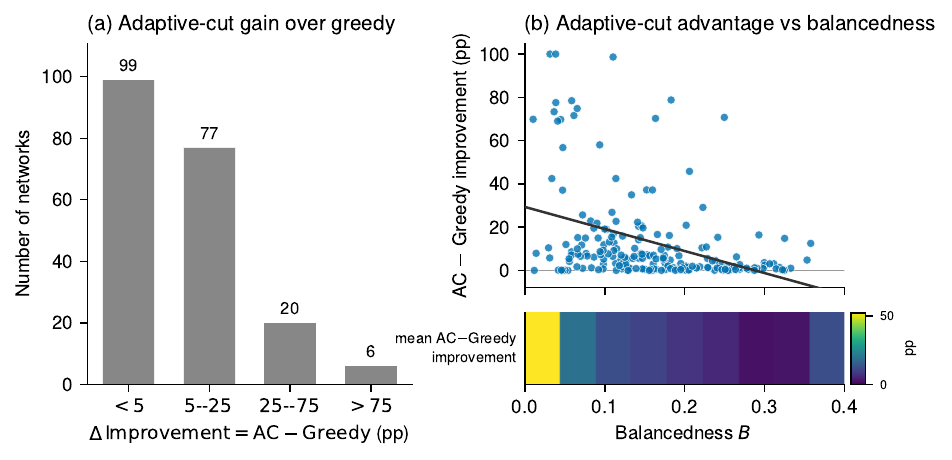}
\caption{Greedy baseline vs adaptive-cut RJMCMC, both operating on the same single-linkage dendrogram and optimising partition density $D$, on all 202 networks in the catalogue (Section~\ref{sec:network_table}). \textbf{(a)} Distribution of the per-network gain of the adaptive cut over greedy, $\Delta\,\mathrm{Improvement} = \mathrm{Improv.}_{\text{AC}} - \mathrm{Improv.}_{\text{greedy}}$, in percentage points of improvement over the single-level cut $D_{\text{lc}}$, bucketed as $<5$, $5$--$25$, $25$--$75$, and $>75$ pp. Greedy improves on $156/202$ networks (mean $+21\%$, median $+12\%$); the full RJMCMC sampler improves on $195/202$ (mean $+34\%$, median $+21\%$) and matches or exceeds greedy on every network. \textbf{(b)} The same advantage $\mathrm{AC}-\mathrm{Greedy}$ as a function of dendrogram balancedness $B$. Top: one point per network with a linear fit (Pearson $r = -0.37$); two networks above $100$ pp are shown at the cap. Bottom: the mean advantage in each $B$ bin as a heat strip. The adaptive cut's gain over greedy concentrates at low $B$, where communities are spread across dendrogram levels and the deterministic greedy is trapped in a local optimum.}
\label{fig:greedy-vs-mcmc}
\end{figure}

Panel~(a): the adaptive cut's gain over greedy exceeds $5$ pp on $103/202$ networks, with $6$ of them above $75$ pp. Panel~(b): this gain concentrates at low balancedness (Pearson $r = -0.37$), the regime where multiscale structure is spread across dendrogram levels and the deterministic greedy cannot escape its local optimum.

This separates three effects: (i)~the multi-cut principle itself accounts for about $21$ of the $34$ percentage points of mean gain over $\sigma_{\text{lc}}$ and is recoverable by a deterministic algorithm; (ii)~the RJMCMC's specific design (geometric-penalty multi-level proposal $\rho(\sigma\mid v) = 2^{-|\sigma|}/Z_c(v)$ plus simulated annealing) adds the remaining $\sim 13$ percentage points by escaping greedy local optima; (iii)~the exploration is not a generic resolution sweep, since both methods stay within the family of dendrogram-consistent cuts.

\paragraph{Dynamic Tree Cut comparison.} As a published-baseline check we also ran Langfelder et al.'s \texttt{dynamicTreeCut} \cite{dynamic_cut} on 30 catalogue networks spanning $M = 23$ to $44{,}000$ and $B \in [0.034, 0.358]$, sweeping the hyperparameter grid \texttt{deepSplit} $\in \{0,\dots,4\}$, \texttt{minClusterSize} $\in \{2, 5, 10, M/1000\}$, and both standard handlings of unassigned leaves. The best DTC partition per network gives a mean change of $-41.5\%$ relative to $D_{\text{lc}}$ (median $-45.1\%$) and never beats $\sigma_{\text{lc}}$ on any of the 30 networks; adaptive cut delivers a higher $D$ than DTC on all 30 (\texttt{results/dtc\_comparison\_30.csv}). This is expected: DTC's selection rules are objective-agnostic, using branch-shape statistics rather than partition density, and since $\sigma_{\text{lc}}$ is by construction the $D$-optimal horizontal cut, any $D$-agnostic horizontal-cut rule is upper-bounded by $D_{\text{lc}}$. AC beating DTC on $D$ is forced; the comparison shows instead that adaptive cut combines DTC's multi-level idea with objective-aware optimization. The greedy baseline above is the cleaner ablation since it shares both the move set and the objective $D$ with adaptive cut.

\section{Adaptive Cut Across Linkage Methods}
\label{sec:linkage-comparison}

To verify that the adaptive cut is agnostic to the dendrogram-construction method, we repeat the link-clustering experiment using four standard linkages: single, average (UPGMA), weighted (WPGMA), and Ward.

For each network and each linkage method, we (i) compute the link similarity matrix as in the main text (Methods, Eq.~\ref{eq:jaccard}), (ii) build a dendrogram using the corresponding linkage criterion, (iii) compute the optimal single-level cut by maximising partition density $D$ along the dendrogram, and (iv) run the adaptive cut MCMC initialised at the single-level optimum. We report the partition density of the single-level cut $D_{\text{lc}}$, the partition density of the adaptive cut $D_{\text{ac}}$, the relative improvement, and the balancedness $B$ of the dendrogram.

We evaluate twenty networks with $78 \le M \le 1{,}954$ links spanning social, biological, technological, transportation, and informational domains, drawn from our benchmark catalogue (Section~\ref{sec:network_table}). For each network we build the dendrogram with each of four linkage methods (single, average, weighted, Ward) applied to the same Jaccard edge-similarity matrix, then run the adaptive cut on each dendrogram. Table~\ref{tab:linkage-comparison} reports the results across all eighty network--linkage combinations.

\begin{table}[!htbp]
\centering
\tiny
\setlength{\tabcolsep}{3pt}
\renewcommand{\arraystretch}{0.95}
\caption{Partition density of the single-level cut ($D_{\text{lc}}$) and the adaptive cut ($D_{\text{ac}}$), relative improvement, and balancedness $B$ across four linkage methods and twenty networks drawn uniformly at random (seed $42$) from the catalogue with $50 \le M < 2000$. Networks are sorted by edge count.}
\label{tab:linkage-comparison}
\begin{tabular}{llrrrr}
\toprule
Network & Linkage & $D_{\text{lc}}$ & $D_{\text{ac}}$ & Improv. & $B$ \\
\midrule
high-school-contacts & single   & 0.446 & 0.476 &  +6.8\% & 0.16 \\
                     & average  & 0.268 & 0.300 & +11.9\% & 0.65 \\
                     & weighted & 0.265 & 0.275 &  +3.9\% & 0.67 \\
                     & Ward     & 0.242 & 0.242 &  +0.0\% & 0.67 \\
\midrule
human-brain-AAL      & single   & 0.435 & 0.465 &  +6.8\% & 0.18 \\
                     & average  & 0.331 & 0.338 &  +2.0\% & 0.64 \\
                     & weighted & 0.265 & 0.274 &  +3.2\% & 0.64 \\
                     & Ward     & 0.221 & 0.221 &  +0.0\% & 0.65 \\
\midrule
workplace-contacts   & single   & 0.325 & 0.333 &  +2.4\% & 0.20 \\
                     & average  & 0.162 & 0.162 &  +0.0\% & 0.71 \\
                     & weighted & 0.166 & 0.190 & +14.4\% & 0.73 \\
                     & Ward     & 0.232 & 0.237 &  +2.0\% & 0.71 \\
\midrule
Zachary              & single   & 0.285 & 0.403 & +41.5\% & 0.33 \\
                     & average  & 0.244 & 0.260 &  +6.7\% & 0.48 \\
                     & weighted & 0.214 & 0.259 & +21.0\% & 0.44 \\
                     & Ward     & 0.205 & 0.278 & +35.3\% & 0.62 \\
\midrule
star-wars            & single   & 0.321 & 0.384 & +19.6\% & 0.17 \\
                     & average  & 0.230 & 0.292 & +27.3\% & 0.64 \\
                     & weighted & 0.233 & 0.294 & +26.0\% & 0.64 \\
                     & Ward     & 0.193 & 0.260 & +34.9\% & 0.65 \\
\midrule
software-colt        & single   & 0.138 & 0.173 & +25.5\% & 0.20 \\
                     & average  & 0.058 & 0.083 & +42.4\% & 0.38 \\
                     & weighted & 0.068 & 0.090 & +32.3\% & 0.43 \\
                     & Ward     & 0.063 & 0.099 & +57.2\% & 0.50 \\
\midrule
human-brain-CPAC     & single   & 0.249 & 0.295 & +18.4\% & 0.28 \\
                     & average  & 0.167 & 0.224 & +34.3\% & 0.69 \\
                     & weighted & 0.170 & 0.214 & +26.3\% & 0.71 \\
                     & Ward     & 0.121 & 0.215 & +77.8\% & 0.70 \\
\midrule
software-jmail       & single   & 0.224 & 0.265 & +17.9\% & 0.24 \\
                     & average  & 0.136 & 0.181 & +33.2\% & 0.60 \\
                     & weighted & 0.161 & 0.196 & +22.2\% & 0.62 \\
                     & Ward     & 0.115 & 0.170 & +47.8\% & 0.69 \\
\midrule
internet-TataNld     & single   & 0.152 & 0.173 & +14.1\% & 0.29 \\
                     & average  & 0.129 & 0.147 & +14.2\% & 0.51 \\
                     & weighted & 0.129 & 0.141 &  +9.6\% & 0.58 \\
                     & Ward     & 0.129 & 0.148 & +14.3\% & 0.61 \\
\midrule
facebook-friends     & single   & 0.414 & 0.437 &  +5.7\% & 0.29 \\
                     & average  & 0.354 & 0.386 &  +9.1\% & 0.60 \\
                     & weighted & 0.358 & 0.401 & +12.1\% & 0.61 \\
                     & Ward     & 0.270 & 0.416 & +54.0\% & 0.65 \\
\midrule
london-transport     & single   & 0.092 & 0.111 & +20.2\% & 0.26 \\
                     & average  & 0.085 & 0.088 &  +4.2\% & 0.44 \\
                     & weighted & 0.085 & 0.089 &  +5.3\% & 0.44 \\
                     & Ward     & 0.091 & 0.093 &  +1.3\% & 0.46 \\
\midrule
Les Miserables       & single   & 0.577 & 0.598 &  +3.8\% & 0.09 \\
                     & average  & 0.525 & 0.546 &  +4.0\% & 0.34 \\
                     & weighted & 0.552 & 0.571 &  +3.4\% & 0.31 \\
                     & Ward     & 0.473 & 0.583 & +23.3\% & 0.37 \\
\midrule
internet-Kdl         & single   & 0.071 & 0.076 &  +6.5\% & 0.27 \\
                     & average  & 0.055 & 0.065 & +17.4\% & 0.39 \\
                     & weighted & 0.055 & 0.063 & +14.7\% & 0.32 \\
                     & Ward     & 0.055 & 0.060 &  +8.9\% & 0.40 \\
\midrule
political-books      & single   & 0.287 & 0.307 &  +7.1\% & 0.32 \\
                     & average  & 0.233 & 0.297 & +27.3\% & 0.69 \\
                     & weighted & 0.214 & 0.260 & +21.5\% & 0.71 \\
                     & Ward     & 0.204 & 0.266 & +30.1\% & 0.75 \\
\midrule
streets-bologna      & single   & 0.124 & 0.130 &  +4.4\% & 0.25 \\
                     & average  & 0.089 & 0.120 & +35.2\% & 0.41 \\
                     & weighted & 0.087 & 0.111 & +26.2\% & 0.41 \\
                     & Ward     & 0.104 & 0.110 &  +5.3\% & 0.40 \\
\midrule
unicodelang          & single   & 0.016 & 0.024 & +48.8\% & 0.08 \\
                     & average  & 0.010 & 0.013 & +33.9\% & 0.38 \\
                     & weighted & 0.009 & 0.012 & +26.5\% & 0.38 \\
                     & Ward     & 0.007 & 0.011 & +52.6\% & 0.46 \\
\midrule
software-jung        & single   & 0.163 & 0.211 & +29.8\% & 0.11 \\
                     & average  & 0.143 & 0.177 & +23.7\% & 0.59 \\
                     & weighted & 0.153 & 0.179 & +17.2\% & 0.57 \\
                     & Ward     & 0.122 & 0.195 & +59.2\% & 0.65 \\
\midrule
eu-proc-EE           & single   & 0.006 & 0.010 & +49.5\% & 0.16 \\
                     & average  & 0.005 & 0.007 & +44.9\% & 0.29 \\
                     & weighted & 0.004 & 0.005 & +24.8\% & 0.34 \\
                     & Ward     & 0.002 & 0.002 & +28.5\% & 0.32 \\
\midrule
eu-proc-PT           & single   & 0.004 & 0.006 & +52.6\% & 0.09 \\
                     & average  & 0.003 & 0.004 &  +9.1\% & 0.26 \\
                     & weighted & 0.002 & 0.002 & +17.2\% & 0.34 \\
                     & Ward     & 0.002 & 0.002 & +19.9\% & 0.35 \\
\midrule
eu-proc-AT           & single   & 0.002 & 0.004 & +142.0\% & 0.10 \\
                     & average  & 0.002 & 0.000 & $-85.0$\% & 0.29 \\
                     & weighted & 0.001 & 0.001 &  $-9.7$\% & 0.28 \\
                     & Ward     & 0.002 & 0.003 & +20.8\% & 0.26 \\
\bottomrule
\end{tabular}
\end{table}

Two patterns emerge. First, average, weighted, and Ward linkage produce more balanced dendrograms than single linkage. Across the twenty networks, single linkage yields balancedness scores in the range $B = 0.08$--$0.33$, whereas the other three methods yield $B = 0.26$--$0.75$. This is consistent with the chaining behaviour of single linkage, which tends to absorb leaves one at a time into a growing cluster and produce caterpillar-like trees.

Second, the adaptive cut improves partition density on most network--linkage combinations. Single-linkage gains span $+2.4\%$ to $+142\%$ with a median of $+18\%$. Gains for the other three linkages span $-85\%$ to $+77.8\%$, where the two negative cases (average and weighted linkage on eu-proc-AT) are near-degenerate runs with $D_{\text{lc}} \approx 0.001$--$0.002$ and the chain drifts off the single-level optimum within a single annealing window. As predicted by the balancedness analysis (Fig.~\ref{fig:link_clustering_results}), the largest single-linkage gains are concentrated in the more unbalanced dendrograms, but meaningful improvements persist on the more balanced trees produced by average, weighted, and Ward linkage.

The adaptive cut is agnostic to the choice of linkage method: it improves partition density across all four methods on the bulk of the test set, and the magnitude of the improvement is well predicted by the balancedness score of the underlying dendrogram. They also show that, on the part of the catalogue where single linkage is competitive on the adaptive-cut objective, the gain from the adaptive cut is largest for single linkage.

\section{Convergence and Sensitivity Diagnostics}
\label{sec:diagnostics}

This section quantifies how the adaptive-cut MCMC behaves as a function of chain length, initialisation, and temperature, and gives a complexity analysis together with a wall-clock benchmark. The supporting scripts and result files are listed at the end of each subsection.

\subsection{Convergence vs.\ chain length}
\label{sec:diag-convergence}

\begin{figure}[h]
    \centering
    \includegraphics[width=\linewidth]{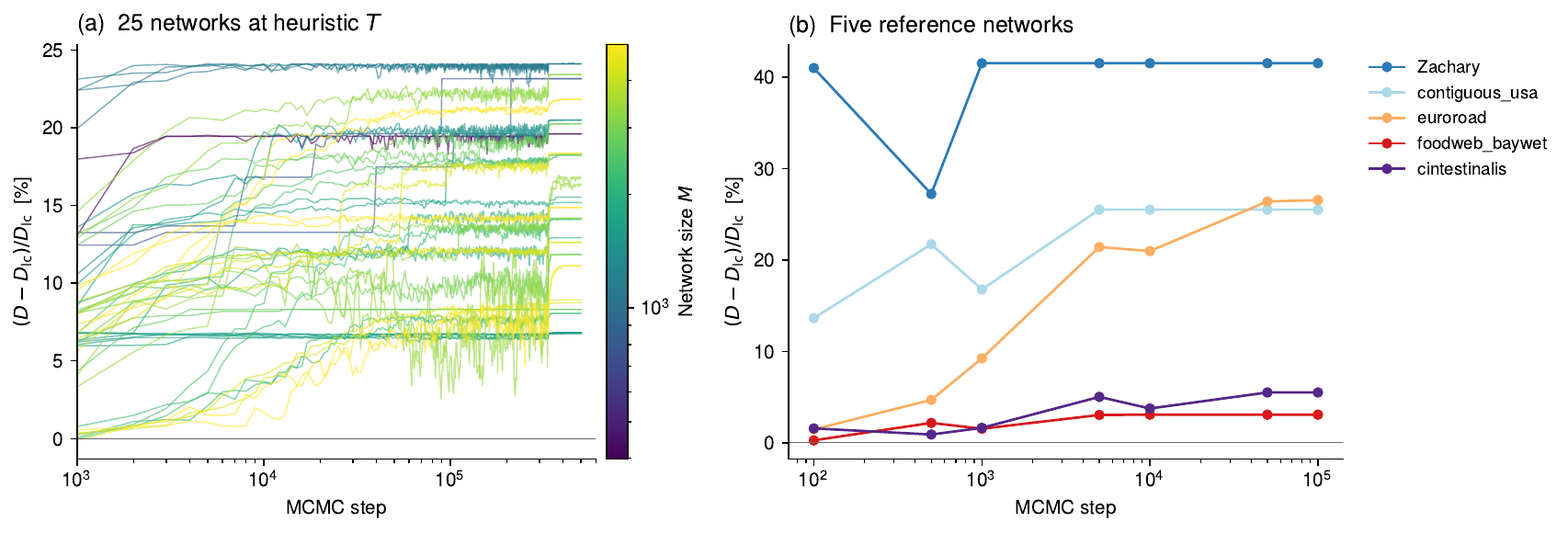}
    \caption{Convergence of $D$ along MCMC chains.
    (a) Fractional improvement vs.\ step for 25 networks in the 5--25\% improvement band at the heuristic temperature, two trials per network.
    (b) Five reference networks on a step grid spanning $10^2$ to $10^5$ steps. All but \texttt{euroroad} reach their $10^5$-step value at $10^4$ steps; \texttt{euroroad} is within $\sim 4\%$ at $10^4$ steps.}
    \label{fig:diag-convergence}
\end{figure}

Figure~\ref{fig:diag-convergence} shows fractional improvement $(D - D_{\text{lc}})/D_{\text{lc}}$ as a function of MCMC step for two complementary experiments. Panel~(a) plots 25 networks in the moderate-improvement band (5--25\% gain), each at the heuristic temperature with two independent seeds, logged every 1{,}000 steps for $5\times 10^5$ steps. Panel~(b) plots five reference networks (Zachary, contiguous\_usa, foodweb\_baywet, cintestinalis, euroroad) at a step-grid ($10^2$ to $10^5$).

Two observations:
\begin{itemize}
    \item For 22 of the 25 networks in panel~(a), the trajectory at the heuristic temperature is within $|\Delta D| < 0.003$ of its running maximum by $2 \times 10^5$ steps, and within $|\Delta D| < 0.001$ by $4 \times 10^5$ steps; the remaining three networks drift inside a band of width $< 0.013$. The chain has mixed below our default $2 \times 10^5$-step budget on this band of networks.
    \item Panel~(b) confirms that for networks with $M < 5{,}000$ edges, $10^4$ steps is sufficient: \texttt{Zachary} and \texttt{contiguous\_usa} reach their plateau by $1{,}000$ steps; \texttt{foodweb\_baywet} and \texttt{cintestinalis} by $5{,}000$ steps; \texttt{euroroad} still creeps slightly between $10^4$ and $10^5$ steps (within $\sim 4\%$ of its $10^5$-step value at $10^4$ steps), the only network in this set that benefits from chains longer than $10^4$ steps.
\end{itemize}

Table~\ref{tab:conv-grid} reports $D$ at three step counts on a grid for the five reference networks of panel (b).

\begin{table}[!htbp]
\centering
\caption{$D$ at $10^3$, $10^4$, and $10^5$ MCMC steps on the five reference networks. All but \texttt{euroroad} reach their $10^5$-step value to within $0.5\%$ by $10^4$ steps (\texttt{euroroad} within $\sim 4\%$); for $M<200$, $10^3$ steps is sufficient.}
\label{tab:conv-grid}
\small
\begin{tabular}{lrrrrr}
\toprule
Network & $M$ & $D_{\mathrm{lc}}$ & $D$ at $10^3$ & $D$ at $10^4$ & $D$ at $10^5$ \\
\midrule
Zachary          &    78 & 0.285 & 0.403 & 0.403 & 0.403 \\
contiguous\_usa  &   107 & 0.277 & 0.324 & 0.348 & 0.348 \\
euroroad         & 1{,}305 & 0.068 & 0.075 & 0.083 & 0.086 \\
foodweb\_baywet  & 2{,}075 & 0.299 & 0.304 & 0.308 & 0.308 \\
cintestinalis   & 2{,}575 & 0.165 & 0.168 & 0.172 & 0.174 \\
\bottomrule
\end{tabular}
\end{table}

\subsection{Robustness to initialisation}
\label{sec:diag-init}

The MCMC is initialised at the single-level cut $\sigma_{\text{lc}}$ in all results of the main text and Supplementary Material. This initialisation is justified because $\sigma_{\text{lc}}$ is the global maximum of $D$ along single-level cuts of the dendrogram, a well-defined high-density starting point in the multi-level cut space.

\begin{figure}[h]
    \centering
    \includegraphics[width=\linewidth]{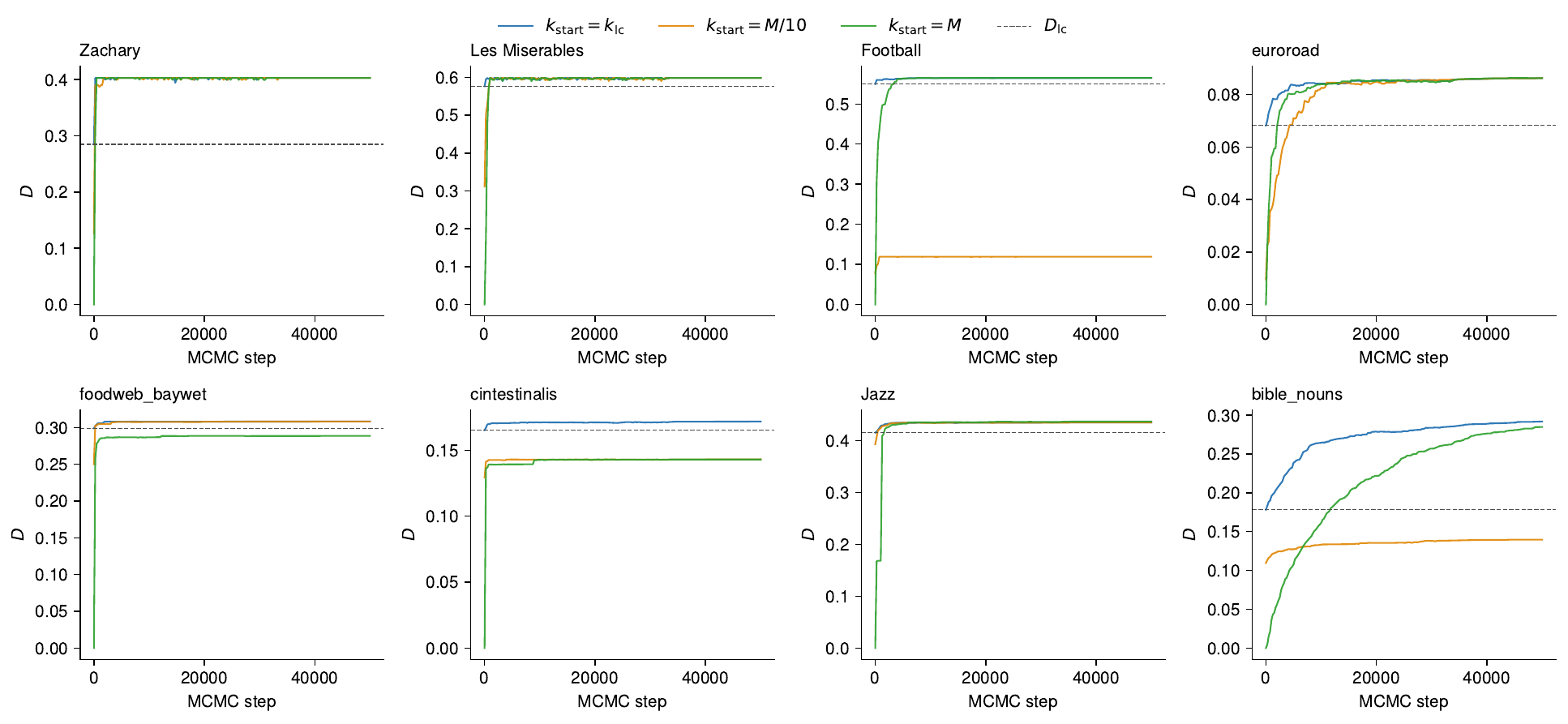}
    \caption{Effect of initialisation on $D$ trajectories. Eight networks (same set as Fig.~\ref{fig:diag-temperature}), three initialisations $k_{\text{start}} \in \{k_{\text{lc}}, M/10, M\}$, $5\times 10^4$ steps, heuristic $T_0$. Dashed line: $D_{\text{lc}}$.}
    \label{fig:diag-init}
\end{figure}

To check that the final $D_{\text{mcmc}}$ is not an artefact of this initialisation, we ran the chain from three starting partitions ($k_{\text{start}} \in \{k_{\text{lc}},\, M/10,\, M\}$) on a fixed eight-network set (Zachary, Les Miserables, Football, euroroad, foodweb\_baywet, cintestinalis, Jazz, bible\_nouns; the same eight are reused for the temperature sweep in Section~\ref{sec:diag-temperature}), for $5\times 10^4$ MCMC steps at the heuristic $T_0$ (Fig.~\ref{fig:diag-init}). Five of the eight networks (\texttt{Zachary}, \texttt{Les Miserables}, \texttt{euroroad}, \texttt{foodweb\_baywet}, \texttt{Jazz}) reach the same plateau from all three initialisations; the chain initialised at $\sigma_{\text{lc}}$ converges first because $\sigma_{\text{lc}}$ already has $D$ close to the plateau. On three networks (\texttt{Football}, \texttt{cintestinalis}, \texttt{bible\_nouns}) the $M/10$ initialisation has not caught up to the $\sigma_{\text{lc}}$ chain within the $5\times 10^4$-step budget. In every case the chain initialised at $\sigma_{\text{lc}}$ matches or exceeds the deeper-init chains, so using $\sigma_{\text{lc}}$ as the standard initialisation never penalises the final $D$ at fixed chain length.

For a stronger test we ran a height-based deep initialisation on \texttt{foodweb\_baywet}, a network with a small improvement under adaptive cut ($D_{\text{lc}} \approx D_{\text{ac}} \approx 0.30$). Starting the chain at a dendrogram cut at similarity height $0.4$ ($k_{\text{start}} \approx 562$, almost $7\times$ deeper than $k_{\text{lc}} = 81$) and running for $10^6$ steps at $T = 10^{-4}$, all three seeds climb back up to within $0.06$ of $D_{\text{lc}}$, with one trial within $0.002$ (Table~\ref{tab:init-stress}). The sampler recovers the single-level-cut plateau on this low-improvement network from a deep initialisation. We still use $\sigma_{\text{lc}}$ in the main results because it converges in a small fraction of the steps and is a well-defined starting point.

Tables~\ref{tab:init-final} and~\ref{tab:init-stress} report the numerical endpoints behind Fig.~\ref{fig:diag-init} and the height-based stress test.

\begin{table}[!htbp]
\centering
\caption{Final $D$ at $5\times 10^4$ steps for each (network, initialisation) pair. $k_{\mathrm{start}}$ in parentheses. Bold marks the highest $D$ among the three initialisations for each network.}
\label{tab:init-final}
\small
\begin{tabular}{lrrlll}
\toprule
Network & $M$ & $D_{\mathrm{lc}}$ & $k_{\mathrm{start}}=k_{\mathrm{lc}}$ & $k_{\mathrm{start}}=M/10$ & $k_{\mathrm{start}}=M$ \\
\midrule
Zachary           &    78 & 0.285 & \textbf{0.403} ($k\!=\!22$) & \textbf{0.403} ($k\!=\!7$) & \textbf{0.403} ($k\!=\!78$) \\
Les Miserables   &   254 & 0.577 & \textbf{0.598} ($k\!=\!52$) & \textbf{0.598} ($k\!=\!25$) & \textbf{0.598} ($k\!=\!254$) \\
Football          &   613 & 0.550 & \textbf{0.565} ($k\!=\!161$) & 0.119 ($k\!=\!61$) & \textbf{0.565} ($k\!=\!613$) \\
euroroad          & 1{,}305 & 0.068 & \textbf{0.086} ($k\!=\!1053$) & \textbf{0.086} ($k\!=\!130$) & \textbf{0.086} ($k\!=\!1305$) \\
foodweb\_baywet   & 2{,}075 & 0.299 & \textbf{0.308} ($k\!=\!81$) & \textbf{0.308} ($k\!=\!207$) & 0.289 ($k\!=\!2075$) \\
cintestinalis    & 2{,}575 & 0.165 & \textbf{0.172} ($k\!=\!663$) & 0.143 ($k\!=\!257$) & 0.143 ($k\!=\!2575$) \\
Jazz              & 2{,}742 & 0.416 & 0.435 ($k\!=\!223$) & 0.435 ($k\!=\!274$) & \textbf{0.437} ($k\!=\!2742$) \\
bible\_nouns      & 9{,}059 & 0.178 & \textbf{0.292} ($k\!=\!3372$) & 0.140 ($k\!=\!905$) & 0.285 ($k\!=\!9059$) \\
\bottomrule
\end{tabular}
\end{table}

\begin{table}[!htbp]
\centering
\caption{Height-based stress test on \texttt{foodweb\_baywet}: chain initialised at a height-0.4 cut ($k_{\mathrm{start}} = 562$, $D = 0.149$ at $t=0$), $10^6$ steps, $T=10^{-4}$, three independent seeds. All trials climb back to within $0.06$ of $D_{\mathrm{lc}} = 0.299$, with one trial within $0.002$.}
\label{tab:init-stress}
\small
\begin{tabular}{lrrr}
\toprule
Trial & $k_{\mathrm{start}}$ & $D$ at $t=0$ & $D$ at $t=10^6$ \\
\midrule
0 & 562 & 0.149 & 0.244 \\
1 & 562 & 0.149 & 0.260 \\
2 & 562 & 0.149 & 0.297 \\
\bottomrule
\end{tabular}
\end{table}

\subsection{Temperature sensitivity}
\label{sec:diag-temperature}

We use a size- and balancedness-adaptive heuristic for the initial temperature,
\begin{equation}
\label{eq:T-heuristic}
T_0 = \frac{c}{M}\, f_B\, f_D,
\quad
f_B = \begin{cases} 3 & B < 0.15 \\ 1 & \text{otherwise} \end{cases},
\quad
f_D = \begin{cases} 0.3 & D_{\text{lc}} > 0.7 \\ 0.01 & D_{\text{lc}} < 0.01 \\ 1 & \text{otherwise} \end{cases},
\end{equation}
with $c = 0.1$ and the result clamped to $[10^{-8}, 10^{-3}]$. The factor $1/M$ keeps the acceptance rate roughly constant as network size grows; $f_B$ raises $T_0$ on very unbalanced dendrograms where the chain must escape narrow basins; $f_D$ lowers $T_0$ on near-saturated or near-zero landscapes where finer probes are needed.

\begin{figure}[h]
    \centering
    \includegraphics[width=\linewidth]{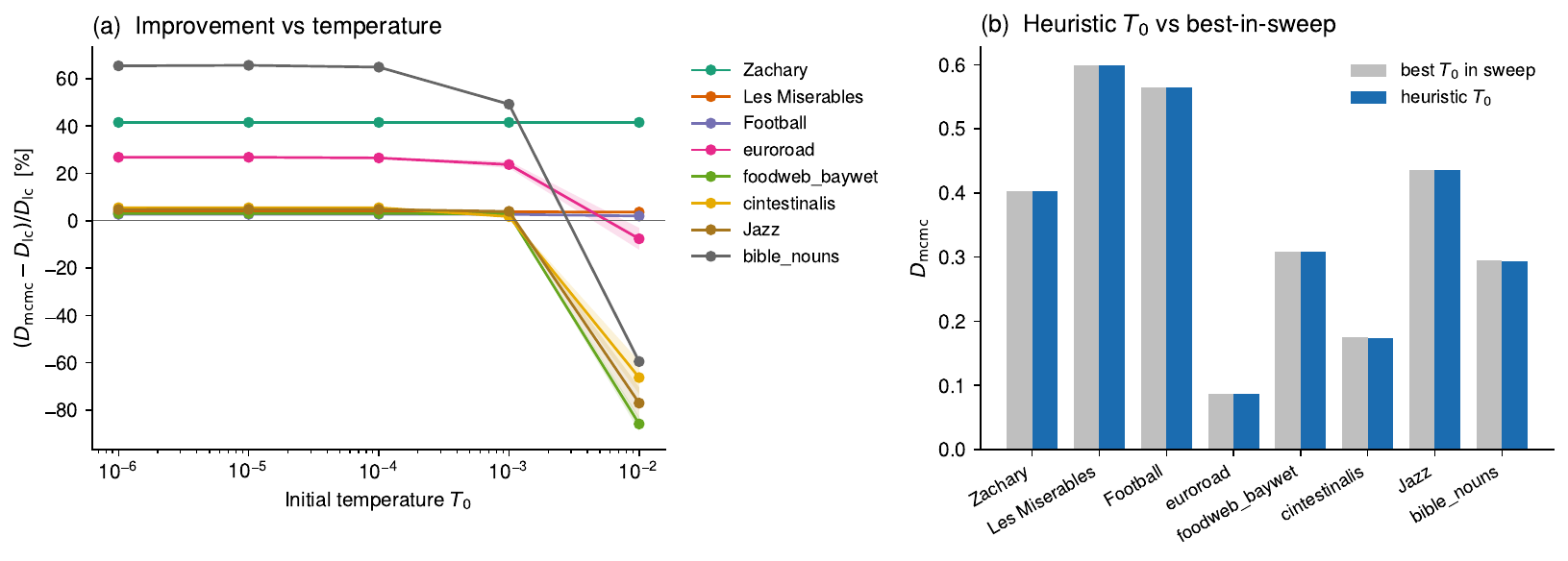}
    \caption{Temperature sensitivity. (a) Fractional improvement $D_{\text{mcmc}}/D_{\text{lc}}$ as a function of initial temperature $T_0$ for eight networks at $2\times 10^5$ steps, mean over three trials. (b) For every network, the heuristic $T_0$ (Eq.~\ref{eq:T-heuristic}, dashed) lies on the plateau of $D(T_0)$, so the heuristic matches the best $T_0$ in the sweep.}
    \label{fig:diag-temperature}
\end{figure}

We validated the heuristic in two ways. First, we swept $T_0$ over five orders of magnitude, $T_0 \in \{10^{-6},10^{-5},10^{-4},10^{-3},10^{-2}\}$, on eight representative networks at $2\times 10^5$ steps with three trials each (Fig.~\ref{fig:diag-temperature}). We also ran a $T_0 = 10^{-7}$ point on each network; the chain matches the $10^{-6}$ values to within noise and the corresponding curves are flat to the left of $10^{-6}$, so we omit $10^{-7}$ from the plot. For every network, the heuristic temperature lies on the plateau of the $D(T_0)$ curve. Hotter values of $T_0$ do not help: at $T_0 = 10^{-2}$ most networks lose ground and $D_{\text{mcmc}}$ falls below $D_{\text{lc}}$ (negative improvement of $50$--$85\%$ on the softer landscapes such as \texttt{foodweb\_baywet}, \texttt{cintestinalis}, \texttt{Jazz}, and \texttt{bible\_nouns}). Only the most ``soft'' landscapes benefit measurably from a colder $T_0$, and even then by less than $1\%$.

Second, we ran a $\{1\times,\, 10\times\}$ comparison around the heuristic $T_0$ on the same 25 networks used for convergence (Section~\ref{sec:diag-convergence}). At $5\times 10^5$ steps, the $1\times$ chain matched or beat the $10\times$ chain on all 25 networks (median gap $0.006$, max gap $0.37$): the chain at the heuristic $T_0$ has already settled while the hotter chain is still wandering. The heuristic is conservative: it is on the right side of the $D(T_0)$ plateau for all networks tested, and a $10\times$ hotter $T_0$ never improves the final $D$ within $5\times 10^5$ steps.

Table~\ref{tab:T-heuristic} reports the heuristic $T_0$ produced by Eq.~\ref{eq:T-heuristic} on the eight networks of Fig.~\ref{fig:diag-temperature}.

\begin{table}[!htbp]
\centering
\caption{Heuristic $T_0$ from Eq.~\ref{eq:T-heuristic} for the eight networks of Fig.~\ref{fig:diag-temperature}.}
\label{tab:T-heuristic}
\small
\begin{tabular}{lrrrr}
\toprule
Network & $M$ & $B$ & $D_{\text{lc}}$ & $T_0$ (heuristic) \\
\midrule
Zachary           &    78 & 0.325 & 0.285 & $1.0\times 10^{-3}$ \\
Les Miserables   &   254 & 0.092 & 0.577 & $1.0\times 10^{-3}$ \\
Football          &   613 & 0.194 & 0.550 & $1.6\times 10^{-4}$ \\
euroroad          & 1{,}305 & 0.238 & 0.068 & $7.7\times 10^{-5}$ \\
foodweb\_baywet   & 2{,}075 & 0.318 & 0.299 & $4.8\times 10^{-5}$ \\
cintestinalis    & 2{,}575 & 0.279 & 0.165 & $3.9\times 10^{-5}$ \\
Jazz              & 2{,}742 & 0.195 & 0.416 & $3.7\times 10^{-5}$ \\
bible\_nouns      & 9{,}059 & 0.148 & 0.178 & $3.3\times 10^{-5}$ \\
\bottomrule
\end{tabular}
\end{table}

\subsection{Complexity and wall-clock timing}
\label{sec:diag-complexity}

Before the MCMC begins, the sampler performs a one-off precomputation in $O(M)$ time: the geometric-cut normalisers $\log Z(v)$ and $\log Z_c(v)$, the leaf sets $\mathrm{leaves}(v)$, the down-move probability tables $p_{\text{stop}}$ / $p_{\text{both}}$ / $p_{\text{split},l/r}$, and the per-subtree partition-density contribution $D(v) = m_v \cdot D_c(m_v, n_v)$.

The per-step cost of the main loop is:
\begin{itemize}
    \item UP-move: $O(k)$, where $k = |\sigma|$ is the current partition size. The dominant cost is the multi-collapse check, which scans current communities and tests disjointness against the sibling's leaf set; each test is amortised $O(\min(|\sigma_i|, |\mathrm{leaves}(\text{sibling})|))$.
    \item DOWN-move: $O(|\sigma'|)$, where $|\sigma'|$ is the size of the sampled cut. Sampling traverses the subtree top-down with constant-time tests against the precomputed probability tables.
    \item Acceptance / update: $O(|\sigma' - \sigma|)$ on accept, $O(1)$ on reject.
\end{itemize}
Because every partition community is by construction $\mathrm{leaves}(v)$ for some node $v$, the partition density is updated incrementally from the cached $D(v)$ values rather than recomputed from the raw edges, which keeps the per-step cost independent of the number of edges.

Table~\ref{tab:wallclock} reports the wall-clock breakdown on ten reference networks spanning two orders of magnitude in $M$, using a fixed seed, $2\times 10^5$ MCMC steps, $T = 10^{-4}$, and a single CPU process (no parallelism). For every network in the table, total wall-clock is under $4$ minutes, and dendrogram construction dominates the larger networks. \texttt{us\_congress-H93} is the largest network in the table and finishes the MCMC in $18\,\mathrm{s}$ because its single-cut partition is already a strong local optimum (the chain stays close to $\sigma_{\text{lc}}$). Across the 202 networks, the wall-clock falls in the brackets: $1$--$10\,\mathrm{s}$ for $M < 1{,}000$, tens of seconds to a few minutes for $1{,}000 \le M < 10{,}000$, and $1$--$10\,\mathrm{min}$ for $M \ge 10{,}000$ (a handful of large, high-improvement networks dominate the upper end).

\begin{table}[!htbp]
\centering
\caption{Single-process wall-clock benchmark of adaptive cut. Networks ordered by $M$. \emph{Dendro.}\ is the time to read the edge list, compute Jaccard similarities, and run single-linkage; \emph{MCMC} is $2\times 10^5$ steps of the adaptive-cut sampler at $T = 10^{-4}$, seed $42$.}
\label{tab:wallclock}
\small
\begin{tabular}{lrrrrr}
\toprule
Network & $M$ & $D_{\text{lc}}$ & $D_{\text{mcmc}}$ & Dendro.\ (s) & MCMC (s) \\
\midrule
Zachary           &    78 & 0.285 & 0.403 &  0.01 &   3.20 \\
contiguous\_usa    &   107 & 0.277 & 0.348 &  0.01 &   3.52 \\
Les\,Miserables    &   254 & 0.577 & 0.598 &  0.02 &   3.75 \\
Football          &   613 & 0.550 & 0.565 &  0.05 &   8.80 \\
euroroad          & 1{,}305 & 0.068 & 0.086 &  0.06 &  16.12 \\
foodweb\_baywet    & 2{,}075 & 0.299 & 0.308 &  1.07 &   7.91 \\
cintestinalis     & 2{,}575 & 0.165 & 0.174 &  1.12 &  24.02 \\
Jazz              & 2{,}742 & 0.416 & 0.435 &  1.21 &  12.84 \\
bible\_nouns       & 9{,}059 & 0.178 & 0.294 &  6.20 & 183.67 \\
us\_congress-H93   & 18{,}083 & 0.179 & 0.179 & 63.68 &  18.00 \\
\bottomrule
\end{tabular}
\end{table}

\section{Balancedness as a Structural Descriptor}
\label{sec:B-structural}

The balancedness $B$ defined in the main text is correlated with adaptive-cut improvement (Fig.~\ref{fig:link_clustering_results}). This section reports two uses of $B$: (i) $B$ as a practical decision criterion for whether to run adaptive cut on a given network, and (ii) $B$ as a structural descriptor that characterises the network through the heterogeneity of its community structure. The manuscript's own VD-SBM example (Fig.~\ref{fig:adaptive-cut}) illustrates the second reading: the plain SBM with uniform community density gives $B = 0.23$, the varying-density SBM with the same number of communities but a density gradient across them gives $B = 0.03$. $B$ tracks the heterogeneity of community sizes and densities, not just the number of communities.

\subsection{Decision criterion}
\label{sec:B-decision}

\begin{figure}[h]
\centering
\includegraphics[width=0.8\linewidth]{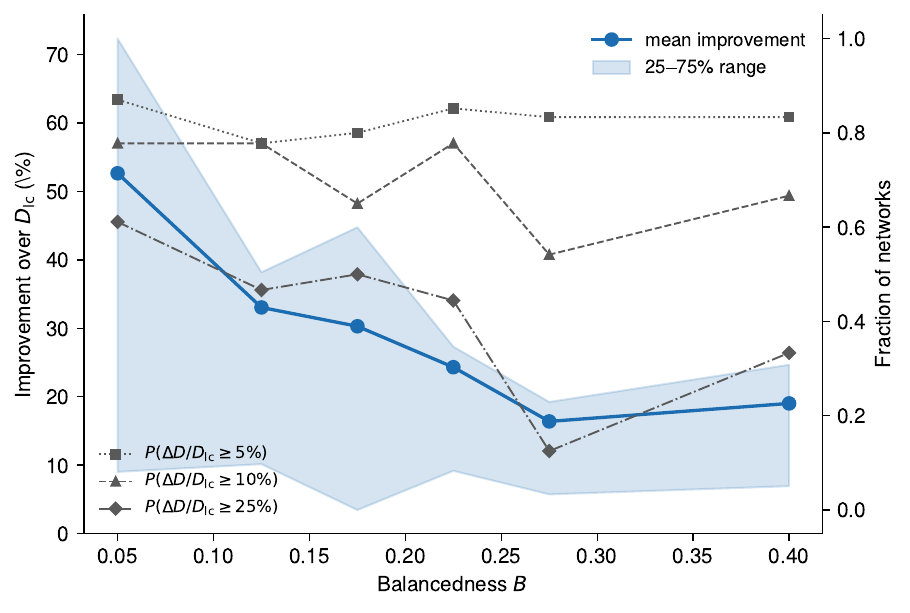}
\caption{Decision criterion. Adaptive-cut improvement $(D_{\text{ac}} - D_{\text{lc}})/D_{\text{lc}}$ vs $B$, binned: solid line is the mean per bin; band is the 25--75\% range across networks. Dashed step lines (right axis) give the fraction of networks with improvement $\geq 5/10/25\%$.}
\label{fig:B-decision}
\end{figure}

We binned the 202 catalogue networks by balancedness and reported, per bin, the mean and median improvement over the single-level cut, and the fraction of networks with improvement of at least $5\%$, $10\%$, and $25\%$ (Table~\ref{tab:B-decision} and Fig.~\ref{fig:B-decision}). The mean improvement falls from $52.6\%$ in the lowest bin ($B < 0.10$) to $16.4\%$ at $B \in [0.25, 0.30)$. The fraction of networks with at least $25\%$ improvement is $0.61$ at $B < 0.10$ and drops to $0.12$ at $B \in [0.25, 0.30)$. Networks with $B < 0.15$ have a $\geq 75\%$ chance of a $\geq 10\%$ improvement under adaptive cut.

\begin{table}[!htbp]
\centering
\caption{Improvement of adaptive cut over single-level cut, binned by balancedness $B$.}
\label{tab:B-decision}
\small
\begin{tabular}{lrrrrrr}
\toprule
$B$ bin & $n$ & mean $\Delta D / D_{\text{lc}}$ (\%) & median (\%) & $P(\Delta D \geq 5\%)$ & $P(\Delta D \geq 10\%)$ & $P(\Delta D \geq 25\%)$ \\
\midrule
$[0.00,\, 0.10)$ & 54 & 52.6 & 56.8 & 0.87 & 0.78 & 0.61 \\
$[0.10,\, 0.15)$ & 45 & 33.0 & 23.2 & 0.78 & 0.78 & 0.47 \\
$[0.15,\, 0.20)$ & 40 & 30.3 & 25.0 & 0.80 & 0.65 & 0.50 \\
$[0.20,\, 0.25)$ & 27 & 24.3 & 17.9 & 0.85 & 0.78 & 0.44 \\
$[0.25,\, 0.30)$ & 24 & 16.4 & 13.2 & 0.83 & 0.54 & 0.12 \\
$[0.30,\, 0.50)$ & 12 & 19.0 & 14.5 & 0.83 & 0.67 & 0.33 \\
\bottomrule
\end{tabular}
\end{table}

\subsection{Partition disagreement, tree balance, and multiscale span}
\label{sec:B-structural-axes}

\begin{figure}[h]
\centering
\includegraphics[width=\linewidth]{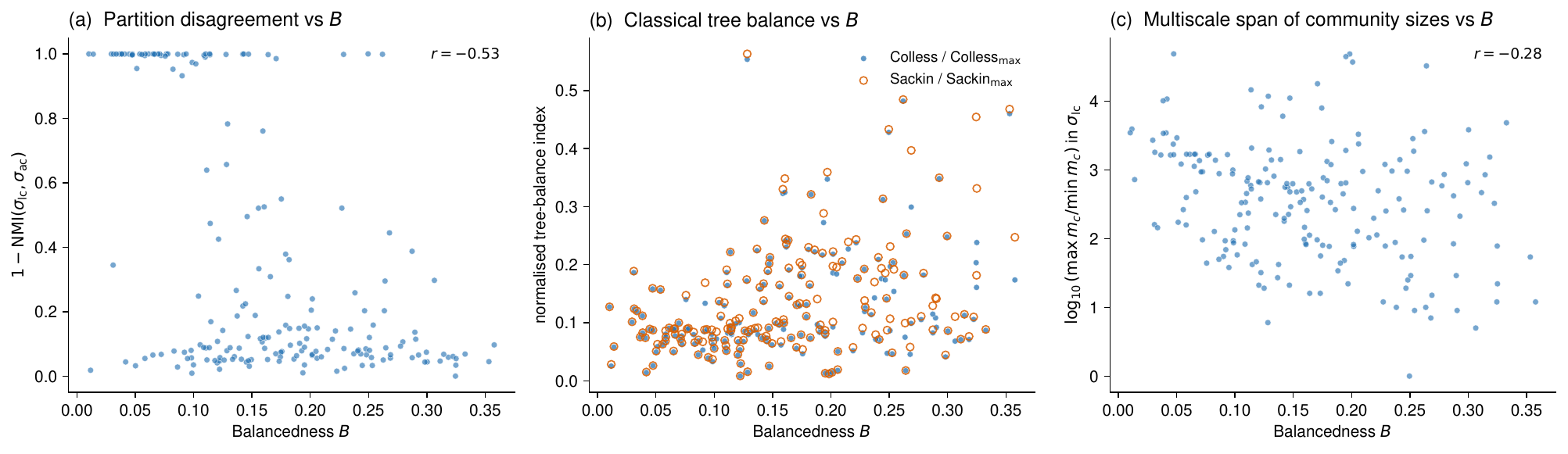}
\caption{Balancedness as a structural descriptor. (a) Partition disagreement $1 - \mathrm{NMI}(\sigma_{\text{lc}}, \sigma_{\text{ac}})$ vs $B$; low $B$ implies single-level and multi-level partitions disagree. (b) Classical leaf-balance indices Colless and Sackin (normalised by their caterpillar maxima) vs $B$; the weak positive correlation shows that $B$ captures dendrogram structure not fully described by topological leaf-balance. (c) $\log_{10}(\max\,m_c / \min\,m_c)$ over $\sigma_{\text{lc}}$ communities vs $B$; low $B$ networks carry communities spanning several orders of magnitude in edge count.}
\label{fig:B-structural}
\end{figure}

Beyond the predictor role, $B$ relates to three structural axes of the dendrogram (Fig.~\ref{fig:B-structural}; Table~\ref{tab:B-correlations}). First, the partition disagreement $1 - \mathrm{NMI}(\sigma_{\text{lc}}, \sigma_{\text{ac}})$ between the single-level cut and the adaptive cut falls as $B$ rises (Pearson $r = -0.53$). Networks with low $B$ admit single-level and multi-level partitions that disagree, supporting the use of $B$ as a diagnostic for clustering reliability: when $B$ is low, the single-level cut is not a stable representation of the dendrogram's community structure.

Second, $B$ shows a weak positive correlation with the classical Colless ($r = +0.31$) and Sackin ($r = +0.37$) indices of leaf-balance. The Colless and Sackin indices are purely topological measures of how subtree sizes split at each internal node, while $B$ aggregates entropy across levels of the dendrogram defined by similarity height. The weak correlation indicates that $B$ captures dendrogram structure that classical leaf-balance indices do not fully describe; classical indices remain low on most link-clustering dendrograms (median Colless$_\text{norm} \approx 0.12$) while $B$ spans the $[0, 0.36]$ range observed across the catalogue, revealing finer level-wise structure.

Third, the log span of community sizes in $\sigma_{\text{lc}}$, $\log_{10}(\max\,m_c / \min\,m_c)$, decreases with $B$ ($r = -0.28$). Low $B$ coincides with $\sigma_{\text{lc}}$ communities that span several orders of magnitude in edge count: the dendrogram contains communities at very different scales that a single horizontal cut cannot simultaneously resolve. This is the dendrogram-level analogue of the resolution-limit problem of single-resolution objectives \cite{fortunato2007resolution, lancichinetti2011limits}. Standard tools for multiscale community detection, such as multiresolution modularity \cite{reichardt2006statistical} and Markov-stability methods \cite{delvenne2010stability}, address the same problem by sweeping a continuous resolution parameter or Markov time. Balancedness $B$ describes the same multiscale property from the geometry of the link-clustering dendrogram, without a resolution sweep, and applies to any hierarchical clustering method that produces a dendrogram.

\begin{table}[!htbp]
\centering
\caption{Pearson and Spearman correlations of structural metrics with balancedness $B$, with $95\%$ bootstrap confidence intervals over $1000$ resamples ($n = 195$ networks where both partitions and the dendrogram are available).}
\label{tab:B-correlations}
\small
\begin{tabular}{lrr}
\toprule
Metric & Pearson $r$ (95\% CI) & Spearman $\rho$ (95\% CI) \\
\midrule
$1 - \mathrm{NMI}(\sigma_{\text{lc}}, \sigma_{\text{ac}})$  & $-0.53$ ($-0.62,\,-0.43$) & $-0.42$ ($-0.55,\,-0.28$) \\
Colless$_\text{norm}$                                       & $+0.31$ ($+0.18,\,+0.44$) & $+0.31$ ($+0.18,\,+0.42$) \\
Sackin$_\text{norm}$                                        & $+0.37$ ($+0.24,\,+0.50$) & $+0.37$ ($+0.24,\,+0.48$) \\
$\log_{10}(\max\,m_c / \min\,m_c)$ in $\sigma_{\text{lc}}$  & $-0.28$ ($-0.41,\,-0.13$) & $-0.27$ ($-0.41,\,-0.12$) \\
\bottomrule
\end{tabular}
\end{table}

\subsection{What \texorpdfstring{$B$}{B} captures: a controlled \texorpdfstring{$2\times 2$}{2x2} design}
\label{sec:B-2x2-design}

The variable $B$ is constructed from the level-wise partition entropy of the edge-similarity dendrogram. To identify what property of the network drives $B$, we crossed two binary axes on a single synthetic family: uniform vs varying intra-community density, and flat vs nested community structure. Each cell of the resulting $2\times 2$ design is generated as a stochastic block model with $4$ leaf communities of $N_c = 20$ nodes ($N = 80$ nodes total). Mean intra-density is fixed at $\bar{p}_{\text{intra}} = 0.35$ in every cell. On the density axis, uniform cells use $p_{\text{intra}}^{(c)} = 0.35$ for every community $c$; varying cells use $p_{\text{intra}}^{(c)} \in \{0.65, 0.45, 0.25, 0.05\}$, the same mean and the same overall edge count $M$. On the nesting axis, flat cells use $p_{\text{inter}} = 0.01$ between every pair of leaves; nested cells use $p_{\text{inter}} = 0.10$ between leaves of the same super-community and $p_{\text{inter}} = 0.005$ between leaves of different super-communities, where leaves $\{L0, L1\}$ form super-$0$ and $\{L2, L3\}$ form super-$1$.

\begin{figure}[h]
\centering
\includegraphics[width=\linewidth]{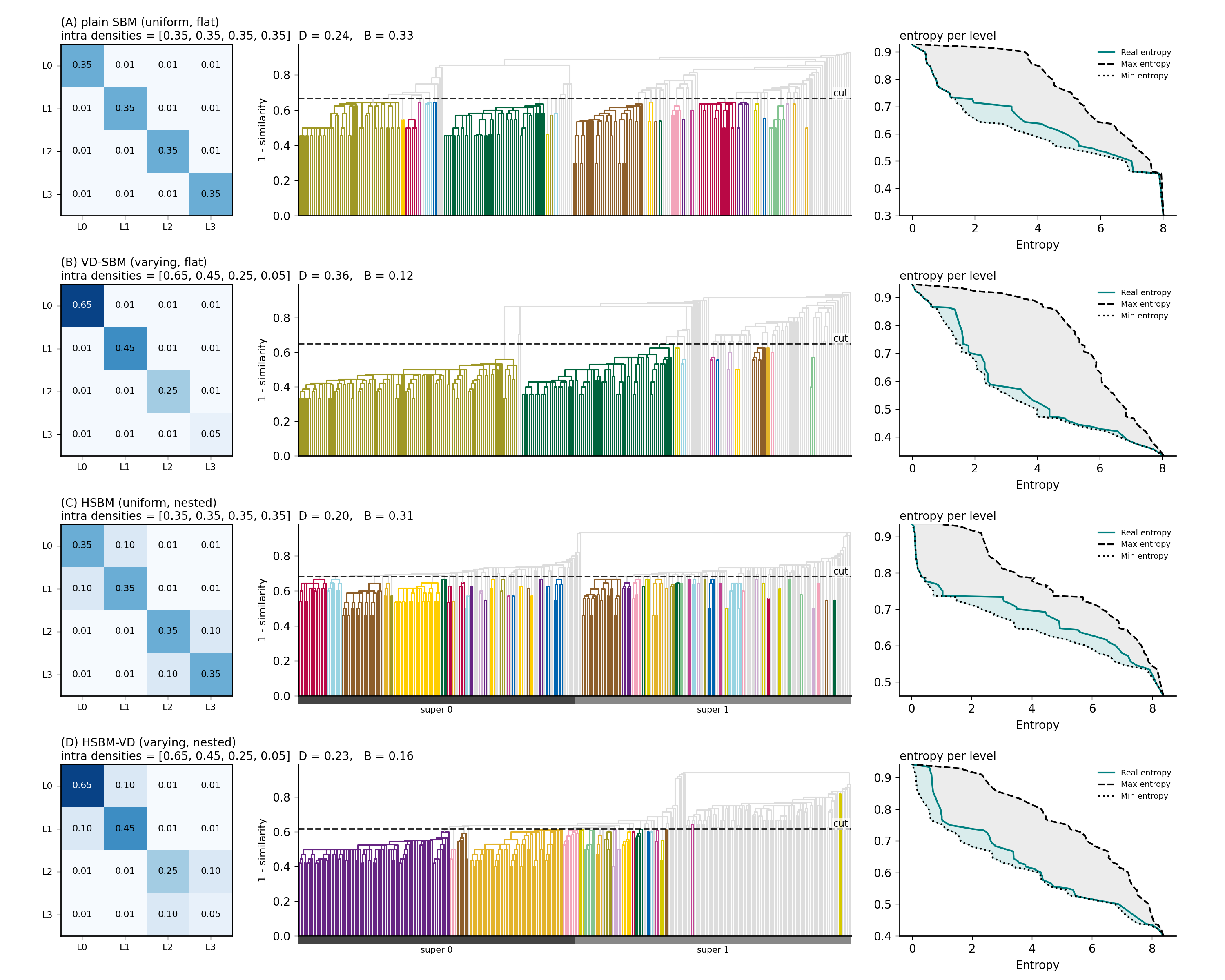}
\caption{$2\times 2$ design crossing uniform vs varying intra-community density (columns) against flat vs $2$-level nested community structure (rows), seed $0$. For each cell: block-density matrix, dendrogram coloured by $\sigma_{\text{lc}}$ with the optimal cut as a dashed line, and entropy curves $H_{\min} \le H_{\text{real}} \le H_{\max}$ vs similarity. The drop in $B$ along the density-variation axis (A$\to$B and C$\to$D) is visible as the real entropy curve pulling down to $H_{\min}$. The nesting axis (A$\to$C and B$\to$D) does not move the entropy curve appreciably.}
\label{fig:B-2x2-design}
\end{figure}

For each cell we draw $5$ independent networks, build the edge-similarity dendrogram with Jaccard similarity and single linkage, and compute $B$ as in the main text. Table~\ref{tab:B-2x2-design} reports the mean and standard deviation of $B$, the number of communities $k_{\text{lc}}$ at the optimal single-level cut, and the network edge count $M$ across the $5$ seeds. Figure~\ref{fig:B-2x2-design} shows, for seed $0$ of each cell, the block-density matrix used to draw the SBM, the resulting dendrogram coloured by $\sigma_{\text{lc}}$ with the optimal cut drawn as a dashed line, and the entropy curves $(H_{\min}, H_{\text{real}}, H_{\max})$ at each similarity level.

\begin{table}[!htbp]
\centering
\caption{Mean $B$, $k_{\text{lc}}$, and edge count $M$ over $5$ seeds in each cell of the $2\times 2$ design.}
\label{tab:B-2x2-design}
\small
\begin{tabular}{lllrrr}
\toprule
cell & density & structure & $B$ (mean $\pm$ std) & $k_{\text{lc}}$ (mean $\pm$ std) & $M$ (mean) \\
\midrule
(A) & uniform  & flat   & $0.237 \pm 0.055$ & $38.8 \pm 20.6$ & $283$ \\
(B) & varying  & flat   & $0.144 \pm 0.036$ & $54.8 \pm 14.0$ & $285$ \\
(C) & uniform  & nested & $0.273 \pm 0.042$ & $36.6 \pm 60.1$ & $356$ \\
(D) & varying  & nested & $0.151 \pm 0.027$ & $97.0 \pm 26.2$ & $358$ \\
\bottomrule
\end{tabular}
\end{table}

The columns of the design (A vs B, C vs D) separate by $0.09$--$0.12$ in $B$; the rows of the design (A vs C, B vs D) shift $B$ by less than $0.04$, within the seed-to-seed standard deviation ($n = 5$ seeds per cell). The density-variation axis is the dominant driver of $B$ in this design; at this sample size we find no evidence for a nesting effect, but cannot rule out a small effect ($\lesssim 0.04$) that a larger seed budget might resolve. The entropy panels of Fig.~\ref{fig:B-2x2-design} show the mechanism: in cells A and C the real entropy curve hugs $H_{\max}$ across most levels (so $B$ is high); in cells B and D the real entropy is pulled down toward $H_{\min}$ across many levels (so $B$ is low). Nesting changes the topology of the dendrogram (the nested cells C and D show two top-level super-branches, marked at the bottom of each dendrogram) but does not move the entropy curve appreciably.

\begin{figure}[h]
\centering
\includegraphics[width=\linewidth]{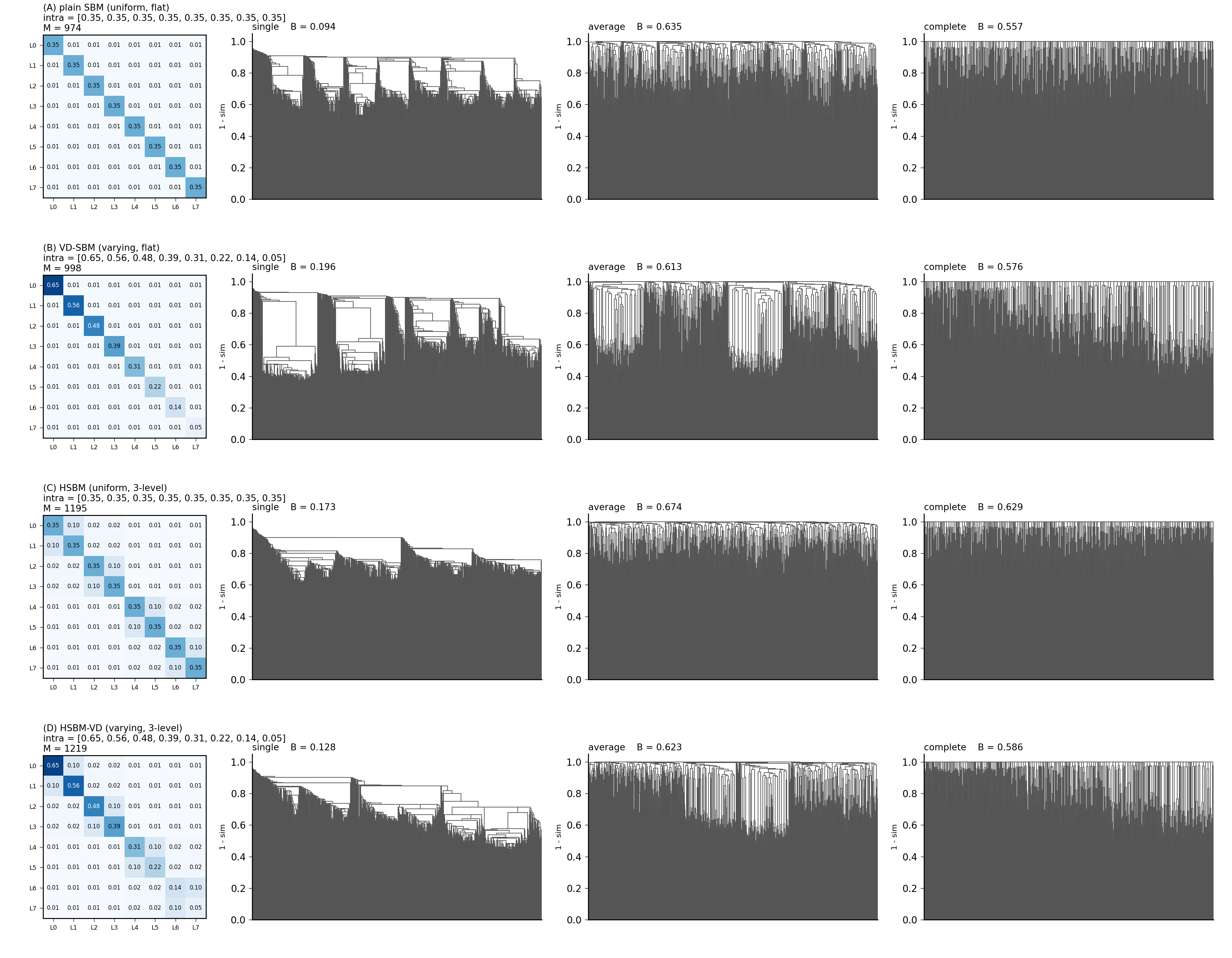}
\caption{Linkage sensitivity of $B$ on the $8$-leaf, $3$-level $2\times 2$ design. Each row is a cell (A, B, C, D); the first column shows the SBM block-density matrix; the remaining three columns show the dendrogram and $B$ value when single, average, or complete linkage is applied to the same Jaccard edge-distance matrix. Single linkage on this larger graph chains and gives $B \in [0.12, 0.16]$ across all cells, masking the cell-to-cell signal. Under both average and complete linkage the uniform-density cells (A, C) score above their density-varying counterparts (B, D) at a higher absolute scale.}
\label{fig:B-linkage-sensitivity}
\end{figure}

\paragraph{Linkage sensitivity.} The absolute value of $B$ depends on the linkage method used to build the dendrogram. To test this we run a separate, larger version of the $2\times 2$ design: eight leaves of $N_c = 25$ nodes (vs four leaves of $N_c = 20$ above) and a three-level nesting in the C and D cells (vs two-level above), with the same uniform/varying density crossing. On the same Jaccard edge-distance matrix per draw we build the dendrogram with single, average, and complete linkage. Across the four cells, single linkage on this larger graph chains the densest leaves together early and gives $B \in [0.12, 0.16]$ with no cell-to-cell signal above seed noise. Average and complete linkage give $B \in [0.57, 0.69]$ with the same cell ordering: density variation lowers $B$ by about $0.05$ under average linkage and about $0.02$ under complete linkage, in both rows (Fig.~\ref{fig:B-linkage-sensitivity}, Table~\ref{tab:B-linkage-sensitivity}). The qualitative claim, low $B$ tracks density variation, holds under average and complete linkage; under single linkage the chaining effect compresses $B$ into the narrow band above and the cell signal is lost in seed noise, so both the absolute scale of $B$ and the resolvability of the density signal are linkage-dependent. The values quoted in the main text (plain SBM $B = 0.23$, VD-SBM $B = 0.03$ in Fig.~\ref{fig:adaptive-cut}) are the single-linkage balancedness of those dendrograms; average and complete linkage would place the same block structures at a higher absolute scale, as this table shows.

\begin{table}[!htbp]
\centering
\caption{Mean $B$ across $5$ seeds in each cell of the $2\times 2$ design ($8$ leaves, $3$-level nesting), under three linkage methods applied to the same Jaccard edge-distance matrix per draw.}
\label{tab:B-linkage-sensitivity}
\small
\begin{tabular}{lllrrr}
\toprule
cell & density & structure & $B_{\text{single}}$ & $B_{\text{average}}$ & $B_{\text{complete}}$ \\
\midrule
(A) & uniform  & flat   & $0.132 \pm 0.033$ & $0.663 \pm 0.017$ & $0.591 \pm 0.022$ \\
(B) & varying  & flat   & $0.155 \pm 0.042$ & $0.606 \pm 0.014$ & $0.568 \pm 0.013$ \\
(C) & uniform  & 3-level nested & $0.131 \pm 0.025$ & $0.693 \pm 0.012$ & $0.624 \pm 0.015$ \\
(D) & varying  & 3-level nested & $0.120 \pm 0.045$ & $0.651 \pm 0.023$ & $0.600 \pm 0.024$ \\
\bottomrule
\end{tabular}
\end{table}

\paragraph{Implications.} Three points follow. (i) $B$ measures dendrogram-level density heterogeneity, not nesting depth or hierarchical depth in any topological sense. (ii) $B$ is related to multiscale community structure in only one direction: low $B$ implies communities at different scales, so a single horizontal cut cannot resolve them all (this is the regime in which adaptive cut delivers most of its improvement). High $B$ does not imply absence of multiscale structure. (iii) The connection to the resolution limit of \cite{fortunato2007resolution} and to multi-resolution / Markov-stability methods \cite{reichardt2006statistical, delvenne2010stability} is indirect: those methods target community size relative to total network scale through an objective-function or diffusion parameter, while $B$ targets density heterogeneity from the geometry of a given dendrogram. The two are complementary diagnostics of the same multiscale phenomenon, computed from different objects.

\subsection{Relation to the Fern{\'a}ndez--G{\'o}mez normalised tree balance}
\label{sec:ntb-comparison}

The closest published index to our balancedness is the Normalised Tree Balance (NTB) of Fern{\'a}ndez \& G{\'o}mez 2020 \cite{fernandez_gomez_2020}, defined (their Eqs.~18--20) on a dendrogram with $M$ leaves as
\begin{equation}
H_I = -\sum_{i \in I} p_i \log_{|I|}(p_i),
\quad p_i = \frac{|X_i|}{|X_I|},
\end{equation}
$H = \operatorname{mean}_{I} H_I$ averaged over the $M-1$ internal nodes, and $\mathrm{NTB} = (H - H_{\min})/(1 - H_{\min})$ where $H_{\min} = \frac{1}{M-1}\!\left[\log_2 M + \sum_{n=2}^{M-1} \frac{\log_2 n}{n+1}\right]$ is the value attained by the caterpillar binary tree of $M$ leaves. For binary single-linkage dendrograms $|I| = 2$ everywhere, so each $H_I \in [0, 1]$ uses $\log_2$.

NTB and our $B$ differ in how they aggregate balance: NTB averages a local Shannon entropy over internal \emph{nodes} (a property of the topology), while $B$ averages a normalised partition entropy over distinct similarity \emph{levels} of the dendrogram. The latter is indexed along the dimension a horizontal cut runs over, and the weighted variant $B_{\text{weighted}}$ further weighs each level by its similarity gap $\Delta s$.

\begin{figure}[h]
\centering
\includegraphics[width=\linewidth]{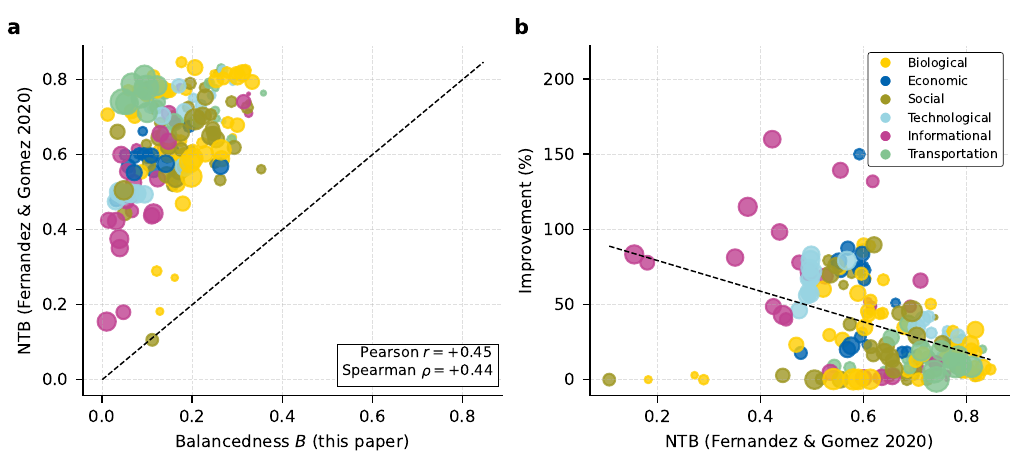}
\caption{Comparison of our balancedness $B$ with the Fern{\'a}ndez--G{\'o}mez 2020 normalised tree balance (NTB). (a) $B$ vs NTB on 202 catalogue networks (Pearson $r = +0.45$, Spearman $\rho = +0.44$); the two indices are correlated but distinct. (b) Adaptive-cut improvement over $D_{\text{lc}}$ vs NTB, reproducing the Fig.~4 scatter with NTB on the $x$-axis. Pearson correlations of improvement with the two balance indices are nearly identical: $r(\text{impr}, B) = -0.38$ and $r(\text{impr}, \mathrm{NTB}) = -0.40$ ($n = 202$).}
\label{fig:B-vs-NTB}
\end{figure}

We computed NTB on the 202 catalogue networks (\texttt{results/ntb\_catalogue.csv}). The two indices are moderately correlated (Pearson $r = +0.45$, Spearman $\rho = +0.44$; Fig.~\ref{fig:B-vs-NTB}a). The two indices capture related but distinct aspects of dendrogram balance. As predictors of adaptive-cut improvement (over $D_{\text{lc}}$) the two indices perform almost identically: Pearson($B$, improv.) $= -0.38$ and Pearson(NTB, improv.) $= -0.40$ (Fig.~\ref{fig:B-vs-NTB}b). NTB is marginally the stronger predictor on this corpus, but the gap (0.02) is within the bootstrap noise. We keep $B$ as the headline predictor because its level-wise indexing matches the object adaptive cut competes against: a single horizontal cut. NTB averages a local entropy over internal nodes regardless of similarity height, so it cannot distinguish a dendrogram whose communities sit at one similarity level from one whose communities are spread across many, and that spread is exactly what adaptive cut exploits.

\begin{figure}[h]
\centering
\includegraphics[width=\linewidth]{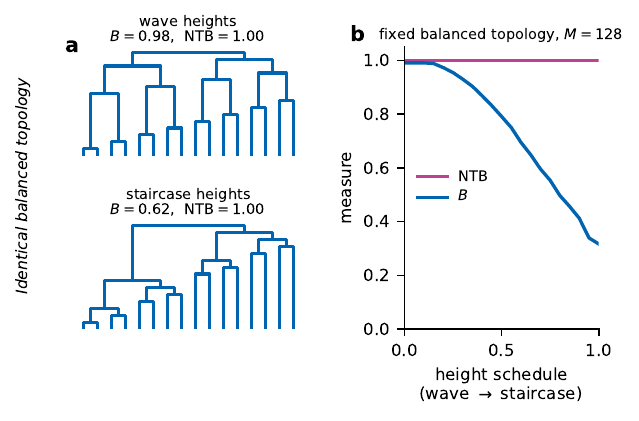}
\caption{Balancedness $B$ and NTB capture contrasting dendrogram properties. (a) Two dendrograms with the \emph{same} balanced binary topology (NTB $= 1$ for both, $M = 16$): merging in synchronised waves keeps every horizontal cut balanced ($B = 0.98$), while a staircase height schedule makes each cut dominated by one growing cluster ($B = 0.62$). (b) Holding the balanced topology fixed at $M = 128$ and interpolating the height schedule from waves to staircase moves $B$ from $0.99$ to $0.32$ while NTB stays at $1$: NTB is blind to merge heights, $B$ is not.}
\label{fig:ntb-divergence}
\end{figure}

\paragraph{Where the two indices diverge.} Because NTB reads only the branching topology and $B$ reads the similarity levels, the two can be driven apart by construction. Fix a perfectly balanced binary topology, for which NTB $= 1$ at every $M$, and vary only the heights at which its merges occur (Fig.~\ref{fig:ntb-divergence}a,b). When the merges form synchronised waves, every horizontal cut splits the leaves into equal blocks and $B \to 1$. When they form a staircase, with one subtree assembling before the next, every cut is dominated by a single growing block and $B$ falls to $0.32$ at $M = 128$, while NTB stays at $1$ throughout. NTB cannot resolve the two cases because it never references merge heights, yet the difference is what decides whether a single horizontal cut suffices.

The opposite corner is unreachable. A caterpillar topology (NTB $\to 0$) forces every level partition to be one large cluster plus singletons, so $B$ is small whenever NTB is small. Balanced topology is necessary but not sufficient for balanced levels, which confines the feasible $(\mathrm{NTB}, B)$ region to $B \lesssim \mathrm{NTB}$; on the 202 catalogue networks NTB $\ge B$ in 201 of 202 cases (the lone exception, an ego network, is a tie within $0.005$; Fig.~\ref{fig:B-vs-NTB}a). The networks farthest into the high-NTB/low-$B$ corner are road networks, trees of life, and the yeast spliceosome (NTB $\approx 0.71$--$0.81$, $B \approx 0.01$--$0.11$): their single-linkage dendrograms split into comparable parts at most nodes (median child-size ratio $\approx 1{:}2$, well away from the $k{:}1$ caterpillar) but collapse onto few distinct similarity levels, so no single cut resolves their communities (us\_roads-VT: $20$ distinct similarity levels across $10^5$ merges, $B = 0.09$, NTB $= 0.81$). These are the dendrograms where adaptive cut gains the most, and the property that flags them, the spread of communities across cut levels, is the one $B$ measures and NTB discards.

\section{Network Catalogue}
\label{sec:network_table}

Table~\ref{tab:SI-networks} lists every network used to evaluate the adaptive cut in Figures~4 and~5 of the main text, with its domain, size $(N, M)$, balancedness $B$, single-cut partition density $D_{\text{lc}}$, adaptive-cut partition density $D_{\text{ac}}$, relative improvement, source, and (where available) a clickable reference to the original publication.

\begingroup
\tiny\let\normalsize\tiny\let\small\tiny\let\footnotesize\tiny
\setlength{\tabcolsep}{1pt}
\renewcommand{\arraystretch}{1.05}
\begin{longtable}{@{}lllrrrrrrll@{}}
\caption{All 202 networks used to generate Figures~4 and 5, with their domain, size $(N,M)$, balancedness $B$, single-cut $D_{\text{lc}}$ and adaptive-cut $D_{\text{ac}}$ partition densities, relative improvement, source, and original publication. Most networks come from the Netzschleuder catalog (\url{https://networks.skewed.de}); a smaller set from the graph-tool built-in collection (\url{https://graph-tool.skewed.de/static/doc/collection.html}); six classical benchmarks are cited individually. Improvement is $(D_{\text{ac}}-D_{\text{lc}})/D_{\text{lc}}\times 100\%$; entries shown as $+0.0$ improve by less than $0.05\%$. At full numerical precision $195$ of the $202$ networks strictly improve ($7$ are exact ties), and the mean ($+34\%$) and median ($+21\%$) improvements are computed at full precision.}\label{tab:SI-networks}\\
\toprule
\textbf{\#} & \textbf{Name} & \textbf{Domain} & \textbf{$N$} & \textbf{$M$} & \textbf{$B$} & \textbf{$D_{\text{lc}}$} & \textbf{$D_{\text{ac}}$} & \textbf{Impr.\%} & \textbf{Source} & \textbf{Reference}\\
\midrule
\endfirsthead
\multicolumn{11}{c}{\tablename\ \thetable{} -- continued from previous page}\\
\toprule
\textbf{\#} & \textbf{Name} & \textbf{Domain} & \textbf{$N$} & \textbf{$M$} & \textbf{$B$} & \textbf{$D_{\text{lc}}$} & \textbf{$D_{\text{ac}}$} & \textbf{Impr.\%} & \textbf{Source} & \textbf{Reference}\\
\midrule
\endhead
\midrule
\multicolumn{11}{r}{\textit{continued on next page}}\\
\endfoot
\bottomrule
\endlastfoot
1 & \parbox[t]{2.6cm}{\raggedright Caenorhabditis elegans} & biological & 453 & 2,025 & 0.11 & 0.14 & 0.22 & +50 & graph-tool & \href{https://doi.org/10.1103/PhysRevE.72.027104}{Duch \& Arenas (2005)}\\
2 & \parbox[t]{2.6cm}{\raggedright Clements \& Long plant-pollinator web} & biological & 275 & 906 & 0.16 & 0.036 & 0.048 & +32 & graph-tool & Clements \& Long (1923)\\
3 & \parbox[t]{2.6cm}{\raggedright Messel Shale food web} & biological & 700 & 6,395 & 0.30 & 0.052 & 0.070 & +35 & graph-tool & \href{https://doi.org/10.1371/journal.pbio.1001987}{Dunne et al. (2014)}\\
4 & \parbox[t]{2.6cm}{\raggedright Yeast spliceosome} & biological & 103 & 4,119 & 0.012 & 0.91 & 0.92 & +0.0 & graph-tool & Korkin et al. (2006)\\
5 & \parbox[t]{2.6cm}{\raggedright budapest\_\allowbreak{}connectome-all\_\allowbreak{}200k} & biological & 1,015 & 104,292 & 0.20 & 0.39 & 0.39 & +0.40 & Netzschleuder & \href{https://doi.org/10.1016/j.neulet.2015.03.071}{Szalkai et al. (2015)}\\
6 & \parbox[t]{2.6cm}{\raggedright budapest\_\allowbreak{}connectome-all\_\allowbreak{}20k} & biological & 1,015 & 70,654 & 0.26 & 0.28 & 0.28 & +0.27 & Netzschleuder & \href{https://doi.org/10.1016/j.neulet.2015.03.071}{Szalkai et al. (2015)}\\
7 & \parbox[t]{2.6cm}{\raggedright budapest\_\allowbreak{}connectome-female\_\allowbreak{}200k} & biological & 1,015 & 95,867 & 0.20 & 0.37 & 0.37 & +0.17 & Netzschleuder & \href{https://doi.org/10.1016/j.neulet.2015.03.071}{Szalkai et al. (2015)}\\
8 & \parbox[t]{2.6cm}{\raggedright budapest\_\allowbreak{}connectome-male\_\allowbreak{}200k} & biological & 1,015 & 79,278 & 0.20 & 0.31 & 0.31 & +0.16 & Netzschleuder & \href{https://doi.org/10.1016/j.neulet.2015.03.071}{Szalkai et al. (2015)}\\
9 & \parbox[t]{2.6cm}{\raggedright celegans\_\allowbreak{}interactomes-wi2007} & biological & 1,108 & 1,500 & 0.11 & 0.013 & 0.021 & +58 & Netzschleuder & \href{https://doi.org/10.1038/nmeth.1279}{Simonis et al. (2009)}\\
10 & \parbox[t]{2.6cm}{\raggedright cintestinalis} & biological & 205 & 2,575 & 0.28 & 0.17 & 0.17 & +5.5 & Netzschleuder & \href{https://doi.org/10.7554/eLife.16962.001}{Ryan (2016)}\\
11 & \parbox[t]{2.6cm}{\raggedright foodweb\_\allowbreak{}baywet} & biological & 128 & 2,075 & 0.32 & 0.30 & 0.31 & +3.1 & Netzschleuder & \href{https://www.cbl.umces.edu/~atlss/FBay701.html}{Ulanowicz et al. (1998)}\\
12 & \parbox[t]{2.6cm}{\raggedright genetic\_\allowbreak{}multiplex-Candida} & biological & 303 & 314 & 0.13 & 0.048 & 0.048 & +0.0 & Netzschleuder & \href{https://doi.org/10.1093/comnet/cnu038}{De Domenico (2014)}\\
13 & \parbox[t]{2.6cm}{\raggedright genetic\_\allowbreak{}multiplex-Gallus} & biological & 205 & 234 & 0.16 & 0.034 & 0.035 & +2.9 & Netzschleuder & \href{https://doi.org/10.1093/comnet/cnu038}{De Domenico (2014)}\\
14 & \parbox[t]{2.6cm}{\raggedright genetic\_\allowbreak{}multiplex-HumanHIV1} & biological & 1,005 & 1,155 & 0.12 & 0.00040 & 0.00040 & +0.0 & Netzschleuder & \href{https://doi.org/10.1093/comnet/cnu038}{De Domenico (2014)}\\
15 & \parbox[t]{2.6cm}{\raggedright genetic\_\allowbreak{}multiplex-Plasmodium} & biological & 1,158 & 2,402 & 0.16 & 0.0098 & 0.016 & +66 & Netzschleuder & \href{https://doi.org/10.1093/comnet/cnu038}{De Domenico (2014)}\\
16 & \parbox[t]{2.6cm}{\raggedright genetic\_\allowbreak{}multiplex-Rattus} & biological & 2,350 & 3,484 & 0.15 & 0.053 & 0.068 & +30 & Netzschleuder & \href{https://doi.org/10.1093/comnet/cnu038}{De Domenico (2014)}\\
17 & \parbox[t]{2.6cm}{\raggedright human\_\allowbreak{}brains-BNU1\_\allowbreak{}0025864\_\allowbreak{}1\_\allowbreak{}DTI\_\allowbreak{}AAL} & biological & 116 & 1,563 & 0.18 & 0.44 & 0.47 & +6.8 & Netzschleuder & \href{https://doi.org/10.1109/GlobalSIP.2013.6736878}{Roncal et al. (2013)}\\
18 & \parbox[t]{2.6cm}{\raggedright human\_\allowbreak{}brains-BNU1\_\allowbreak{}0025864\_\allowbreak{}1\_\allowbreak{}DTI\_\allowbreak{}CPAC200} & biological & 200 & 1,832 & 0.28 & 0.25 & 0.29 & +18 & Netzschleuder & \href{https://doi.org/10.1109/GlobalSIP.2013.6736878}{Roncal et al. (2013)}\\
19 & \parbox[t]{2.6cm}{\raggedright human\_\allowbreak{}brains-BNU1\_\allowbreak{}0025864\_\allowbreak{}1\_\allowbreak{}DTI\_\allowbreak{}DS00583} & biological & 540 & 7,425 & 0.33 & 0.21 & 0.23 & +6.4 & Netzschleuder & \href{https://doi.org/10.1109/GlobalSIP.2013.6736878}{Roncal et al. (2013)}\\
20 & \parbox[t]{2.6cm}{\raggedright human\_\allowbreak{}brains-BNU1\_\allowbreak{}0025864\_\allowbreak{}1\_\allowbreak{}DTI\_\allowbreak{}DS01216} & biological & 1,088 & 17,948 & 0.31 & 0.22 & 0.30 & +33 & Netzschleuder & \href{https://doi.org/10.1109/GlobalSIP.2013.6736878}{Roncal et al. (2013)}\\
21 & \parbox[t]{2.6cm}{\raggedright human\_\allowbreak{}brains-BNU1\_\allowbreak{}0025870\_\allowbreak{}1\_\allowbreak{}DTI\_\allowbreak{}DS01216} & biological & 1,133 & 16,478 & 0.30 & 0.24 & 0.29 & +24 & Netzschleuder & \href{https://doi.org/10.1109/GlobalSIP.2013.6736878}{Roncal et al. (2013)}\\
22 & \parbox[t]{2.6cm}{\raggedright interactome\_\allowbreak{}figeys} & biological & 2,217 & 6,418 & 0.086 & 0.0079 & 0.014 & +75 & Netzschleuder & \href{https://doi.org/10.1038/msb4100134}{Ewing et al. (2007)}\\
23 & \parbox[t]{2.6cm}{\raggedright interactome\_\allowbreak{}pdz} & biological & 161 & 209 & 0.19 & 0.075 & 0.078 & +3.6 & Netzschleuder & \href{https://doi.org/10.1093/bioinformatics/bti098}{Beuming et al. (2005)}\\
24 & \parbox[t]{2.6cm}{\raggedright interactome\_\allowbreak{}stelzl} & biological & 1,615 & 3,106 & 0.14 & 0.010 & 0.015 & +53 & Netzschleuder & \href{https://doi.org/10.1016/j.cell.2005.08.029}{Stelzl et al. (2005)}\\
25 & \parbox[t]{2.6cm}{\raggedright interactome\_\allowbreak{}vidal} & biological & 2,783 & 6,007 & 0.14 & 0.054 & 0.079 & +46 & Netzschleuder & \href{https://doi.org/10.1038/nature04209}{JF. Rual et al. (2005)}\\
26 & \parbox[t]{2.6cm}{\raggedright interactome\_\allowbreak{}yeast} & biological & 1,458 & 1,948 & 0.12 & 0.088 & 0.10 & +15 & Netzschleuder & \href{https://doi.org/10.1098/rspb.2005.3092}{Coulomb et al. (2005)}\\
27 & \parbox[t]{2.6cm}{\raggedright kegg\_\allowbreak{}metabolic-aae} & biological & 880 & 2,296 & 0.18 & 0.045 & 0.076 & +67 & Netzschleuder & \href{https://doi.org/10.1049/iet-syb:20060077}{Huss (2007)}\\
28 & \parbox[t]{2.6cm}{\raggedright kegg\_\allowbreak{}metabolic-afu} & biological & 861 & 2,011 & 0.17 & 0.036 & 0.052 & +44 & Netzschleuder & \href{https://doi.org/10.1049/iet-syb:20060077}{Huss (2007)}\\
29 & \parbox[t]{2.6cm}{\raggedright kegg\_\allowbreak{}metabolic-ana} & biological & 1,314 & 3,552 & 0.15 & 0.030 & 0.051 & +71 & Netzschleuder & \href{https://doi.org/10.1049/iet-syb:20060077}{Huss (2007)}\\
30 & \parbox[t]{2.6cm}{\raggedright kegg\_\allowbreak{}metabolic-ape} & biological & 769 & 1,858 & 0.20 & 0.030 & 0.057 & +89 & Netzschleuder & \href{https://doi.org/10.1049/iet-syb:20060077}{Huss (2007)}\\
31 & \parbox[t]{2.6cm}{\raggedright kegg\_\allowbreak{}metabolic-atc} & biological & 1,538 & 4,315 & 0.16 & 0.028 & 0.052 & +90 & Netzschleuder & \href{https://doi.org/10.1049/iet-syb:20060077}{Huss (2007)}\\
32 & \parbox[t]{2.6cm}{\raggedright kegg\_\allowbreak{}metabolic-ath} & biological & 1,531 & 4,285 & 0.13 & 0.027 & 0.039 & +43 & Netzschleuder & \href{https://doi.org/10.1049/iet-syb:20060077}{Huss (2007)}\\
33 & \parbox[t]{2.6cm}{\raggedright kegg\_\allowbreak{}metabolic-atu} & biological & 1,542 & 4,323 & 0.18 & 0.028 & 0.051 & +86 & Netzschleuder & \href{https://doi.org/10.1049/iet-syb:20060077}{Huss (2007)}\\
34 & \parbox[t]{2.6cm}{\raggedright kegg\_\allowbreak{}metabolic-bas} & biological & 619 & 1,501 & 0.22 & 0.052 & 0.075 & +44 & Netzschleuder & \href{https://doi.org/10.1049/iet-syb:20060077}{Huss (2007)}\\
35 & \parbox[t]{2.6cm}{\raggedright malaria\_\allowbreak{}genes-HVR\_\allowbreak{}1} & biological & 307 & 2,812 & 0.11 & 0.44 & 0.53 & +20 & Netzschleuder & \href{https://doi.org/10.1371/journal.pcbi.1003268}{Larremore et al. (2013)}\\
36 & \parbox[t]{2.6cm}{\raggedright malaria\_\allowbreak{}genes-HVR\_\allowbreak{}5} & biological & 298 & 2,684 & 0.19 & 0.26 & 0.35 & +32 & Netzschleuder & \href{https://doi.org/10.1371/journal.pcbi.1003268}{Larremore et al. (2013)}\\
37 & \parbox[t]{2.6cm}{\raggedright malaria\_\allowbreak{}genes-HVR\_\allowbreak{}6} & biological & 291 & 3,251 & 0.27 & 0.57 & 0.61 & +7.0 & Netzschleuder & \href{https://doi.org/10.1371/journal.pcbi.1003268}{Larremore et al. (2013)}\\
38 & \parbox[t]{2.6cm}{\raggedright malaria\_\allowbreak{}genes-HVR\_\allowbreak{}7} & biological & 306 & 11,688 & 0.21 & 0.59 & 0.64 & +8.2 & Netzschleuder & \href{https://doi.org/10.1371/journal.pcbi.1003268}{Larremore et al. (2013)}\\
39 & \parbox[t]{2.6cm}{\raggedright malaria\_\allowbreak{}genes-HVR\_\allowbreak{}8} & biological & 273 & 3,933 & 0.30 & 0.70 & 0.72 & +3.0 & Netzschleuder & \href{https://doi.org/10.1371/journal.pcbi.1003268}{Larremore et al. (2013)}\\
40 & \parbox[t]{2.6cm}{\raggedright malaria\_\allowbreak{}genes-HVR\_\allowbreak{}9} & biological & 297 & 7,562 & 0.25 & 0.50 & 0.53 & +5.9 & Netzschleuder & \href{https://doi.org/10.1371/journal.pcbi.1003268}{Larremore et al. (2013)}\\
41 & \parbox[t]{2.6cm}{\raggedright mist-genetic\_\allowbreak{}fly} & biological & 4,055 & 13,779 & 0.27 & 0.046 & 0.073 & +58 & Netzschleuder & \href{http://dx.doi.org/10.1093/nar/gkx1116}{Yanhui Hu et al. (2017)}\\
42 & \parbox[t]{2.6cm}{\raggedright mist-genetic\_\allowbreak{}human} & biological & 3,570 & 8,703 & 0.18 & 0.086 & 0.11 & +23 & Netzschleuder & \href{http://dx.doi.org/10.1093/nar/gkx1116}{Yanhui Hu et al. (2017)}\\
43 & \parbox[t]{2.6cm}{\raggedright mist-genetic\_\allowbreak{}mouse} & biological & 214 & 252 & 0.29 & 0.12 & 0.12 & +0.42 & Netzschleuder & \href{http://dx.doi.org/10.1093/nar/gkx1116}{Yanhui Hu et al. (2017)}\\
44 & \parbox[t]{2.6cm}{\raggedright mist-genetic\_\allowbreak{}worm} & biological & 3,037 & 7,914 & 0.18 & 0.056 & 0.071 & +27 & Netzschleuder & \href{http://dx.doi.org/10.1093/nar/gkx1116}{Yanhui Hu et al. (2017)}\\
45 & \parbox[t]{2.6cm}{\raggedright mist-ppi\_\allowbreak{}rat} & biological & 6,803 & 14,636 & 0.17 & 0.023 & 0.037 & +60 & Netzschleuder & \href{http://dx.doi.org/10.1093/nar/gkx1116}{Yanhui Hu et al. (2017)}\\
46 & \parbox[t]{2.6cm}{\raggedright plant\_\allowbreak{}pol\_\allowbreak{}robertson} & biological & 1,882 & 15,254 & 0.23 & 0.017 & 0.023 & +35 & Netzschleuder & \href{https://doi.org/10.5962/bhl.title.11538}{Robertson (1929)}\\
47 & \parbox[t]{2.6cm}{\raggedright tree-of-life-331678} & biological & 531 & 1,848 & 0.058 & 0.44 & 0.52 & +18 & Netzschleuder & \href{https://doi.org/10.1073/pnas.1818013116}{Marinka Zitnik et al. (2019)}\\
48 & \parbox[t]{2.6cm}{\raggedright tree-of-life-333990} & biological & 178 & 282 & 0.25 & 0.26 & 0.29 & +10 & Netzschleuder & \href{https://doi.org/10.1073/pnas.1818013116}{Marinka Zitnik et al. (2019)}\\
49 & \parbox[t]{2.6cm}{\raggedright tree-of-life-335543} & biological & 641 & 4,334 & 0.072 & 0.52 & 0.58 & +13 & Netzschleuder & \href{https://doi.org/10.1073/pnas.1818013116}{Marinka Zitnik et al. (2019)}\\
50 & \parbox[t]{2.6cm}{\raggedright tree-of-life-338966} & biological & 595 & 3,140 & 0.098 & 0.48 & 0.55 & +15 & Netzschleuder & \href{https://doi.org/10.1073/pnas.1818013116}{Marinka Zitnik et al. (2019)}\\
51 & \parbox[t]{2.6cm}{\raggedright tree-of-life-338969} & biological & 879 & 5,873 & 0.083 & 0.46 & 0.55 & +20 & Netzschleuder & \href{https://doi.org/10.1073/pnas.1818013116}{Marinka Zitnik et al. (2019)}\\
52 & \parbox[t]{2.6cm}{\raggedright tree-of-life-339670} & biological & 1,360 & 9,646 & 0.12 & 0.45 & 0.53 & +17 & Netzschleuder & \href{https://doi.org/10.1073/pnas.1818013116}{Marinka Zitnik et al. (2019)}\\
53 & \parbox[t]{2.6cm}{\raggedright tree-of-life-340177} & biological & 434 & 1,477 & 0.11 & 0.46 & 0.55 & +20 & Netzschleuder & \href{https://doi.org/10.1073/pnas.1818013116}{Marinka Zitnik et al. (2019)}\\
54 & \parbox[t]{2.6cm}{\raggedright tree-of-life-365044} & biological & 820 & 4,408 & 0.14 & 0.46 & 0.55 & +18 & Netzschleuder & \href{https://doi.org/10.1073/pnas.1818013116}{Marinka Zitnik et al. (2019)}\\
55 & \parbox[t]{2.6cm}{\raggedright tree-of-life-366602} & biological & 861 & 4,970 & 0.15 & 0.44 & 0.51 & +15 & Netzschleuder & \href{https://doi.org/10.1073/pnas.1818013116}{Marinka Zitnik et al. (2019)}\\
56 & \parbox[t]{2.6cm}{\raggedright tree-of-life-406817} & biological & 714 & 3,283 & 0.14 & 0.41 & 0.51 & +23 & Netzschleuder & \href{https://doi.org/10.1073/pnas.1818013116}{Marinka Zitnik et al. (2019)}\\
57 & \parbox[t]{2.6cm}{\raggedright tree-of-life-452652} & biological & 1,021 & 7,516 & 0.12 & 0.47 & 0.54 & +14 & Netzschleuder & \href{https://doi.org/10.1073/pnas.1818013116}{Marinka Zitnik et al. (2019)}\\
58 & \parbox[t]{2.6cm}{\raggedright tree-of-life-469383} & biological & 838 & 7,232 & 0.065 & 0.47 & 0.54 & +17 & Netzschleuder & \href{https://doi.org/10.1073/pnas.1818013116}{Marinka Zitnik et al. (2019)}\\
59 & \parbox[t]{2.6cm}{\raggedright TrumpWorld} & economic & 2,669 & 3,380 & 0.031 & 0.024 & 0.029 & +18 & graph-tool & \href{https://www.buzzfeednews.com/article/johntemplon/help-us-map-trumpworld}{Buzzfeed News (2017)}\\
60 & \parbox[t]{2.6cm}{\raggedright bitcoin\_\allowbreak{}alpha} & economic & 3,775 & 14,120 & 0.14 & 0.017 & 0.029 & +76 & Netzschleuder & \href{https://doi.org/10.1109/icdm.2016.0033}{Kumar et al. (2016)}\\
61 & \parbox[t]{2.6cm}{\raggedright bitcoin\_\allowbreak{}trust} & economic & 5,875 & 21,489 & 0.16 & 0.023 & 0.039 & +70 & Netzschleuder & \href{https://doi.org/10.1109/icdm.2016.0033}{Kumar et al. (2016)}\\
62 & \parbox[t]{2.6cm}{\raggedright board\_\allowbreak{}directors-net1m\_\allowbreak{}2011-08-01} & economic & 854 & 2,745 & 0.099 & 0.85 & 0.92 & +8.3 & Netzschleuder & \href{https://doi.org/10.1016/j.scaman.2010.10.002}{Seierstad et al. (2011)}\\
63 & \parbox[t]{2.6cm}{\raggedright board\_\allowbreak{}directors-net2m\_\allowbreak{}2011-08-01} & economic & 1,066 & 1,148 & 0.23 & 0.016 & 0.020 & +26 & Netzschleuder & \href{https://doi.org/10.1016/j.scaman.2010.10.002}{Seierstad et al. (2011)}\\
64 & \parbox[t]{2.6cm}{\raggedright eu\_\allowbreak{}procurements\_\allowbreak{}alt-AT\_\allowbreak{}2008} & economic & 1,684 & 1,921 & 0.099 & 0.0016 & 0.0040 & +150 & Netzschleuder & \href{https://doi.org/10.1007/s41060-019-00204-1}{Wachs et al. (2020)}\\
65 & \parbox[t]{2.6cm}{\raggedright eu\_\allowbreak{}procurements\_\allowbreak{}alt-CZ\_\allowbreak{}2008} & economic & 1,777 & 2,110 & 0.082 & 0.0017 & 0.0029 & +73 & Netzschleuder & \href{https://doi.org/10.1007/s41060-019-00204-1}{Wachs et al. (2020)}\\
66 & \parbox[t]{2.6cm}{\raggedright eu\_\allowbreak{}procurements\_\allowbreak{}alt-DK\_\allowbreak{}2011} & economic & 2,219 & 2,904 & 0.072 & 0.0028 & 0.0049 & +75 & Netzschleuder & \href{https://doi.org/10.1007/s41060-019-00204-1}{Wachs et al. (2020)}\\
67 & \parbox[t]{2.6cm}{\raggedright eu\_\allowbreak{}procurements\_\allowbreak{}alt-EE\_\allowbreak{}2008} & economic & 480 & 540 & 0.16 & 0.0065 & 0.0097 & +49 & Netzschleuder & \href{https://doi.org/10.1007/s41060-019-00204-1}{Wachs et al. (2020)}\\
68 & \parbox[t]{2.6cm}{\raggedright eu\_\allowbreak{}procurements\_\allowbreak{}alt-ES\_\allowbreak{}2008} & economic & 8,097 & 13,286 & 0.071 & 0.0019 & 0.0034 & +79 & Netzschleuder & \href{https://doi.org/10.1007/s41060-019-00204-1}{Wachs et al. (2020)}\\
69 & \parbox[t]{2.6cm}{\raggedright eu\_\allowbreak{}procurements\_\allowbreak{}alt-FI\_\allowbreak{}2008} & economic & 2,517 & 3,701 & 0.11 & 0.0047 & 0.0081 & +72 & Netzschleuder & \href{https://doi.org/10.1007/s41060-019-00204-1}{Wachs et al. (2020)}\\
70 & \parbox[t]{2.6cm}{\raggedright eu\_\allowbreak{}procurements\_\allowbreak{}alt-GR\_\allowbreak{}2008} & economic & 1,933 & 2,495 & 0.11 & 0.0015 & 0.0026 & +73 & Netzschleuder & \href{https://doi.org/10.1007/s41060-019-00204-1}{Wachs et al. (2020)}\\
71 & \parbox[t]{2.6cm}{\raggedright eu\_\allowbreak{}procurements\_\allowbreak{}alt-HU\_\allowbreak{}2015} & economic & 2,163 & 2,929 & 0.092 & 0.0062 & 0.010 & +66 & Netzschleuder & \href{https://doi.org/10.1007/s41060-019-00204-1}{Wachs et al. (2020)}\\
72 & \parbox[t]{2.6cm}{\raggedright eu\_\allowbreak{}procurements\_\allowbreak{}alt-IT\_\allowbreak{}2008} & economic & 6,622 & 9,121 & 0.11 & 0.0022 & 0.0040 & +84 & Netzschleuder & \href{https://doi.org/10.1007/s41060-019-00204-1}{Wachs et al. (2020)}\\
73 & \parbox[t]{2.6cm}{\raggedright eu\_\allowbreak{}procurements\_\allowbreak{}alt-LT\_\allowbreak{}2008} & economic & 1,359 & 2,486 & 0.14 & 0.015 & 0.020 & +28 & Netzschleuder & \href{https://doi.org/10.1007/s41060-019-00204-1}{Wachs et al. (2020)}\\
74 & \parbox[t]{2.6cm}{\raggedright eu\_\allowbreak{}procurements\_\allowbreak{}alt-LV\_\allowbreak{}2008} & economic & 1,628 & 2,269 & 0.051 & 0.0060 & 0.011 & +75 & Netzschleuder & \href{https://doi.org/10.1007/s41060-019-00204-1}{Wachs et al. (2020)}\\
75 & \parbox[t]{2.6cm}{\raggedright eu\_\allowbreak{}procurements\_\allowbreak{}alt-PL\_\allowbreak{}2008} & economic & 12,456 & 23,991 & 0.14 & 0.015 & 0.019 & +23 & Netzschleuder & \href{https://doi.org/10.1007/s41060-019-00204-1}{Wachs et al. (2020)}\\
76 & \parbox[t]{2.6cm}{\raggedright eu\_\allowbreak{}procurements\_\allowbreak{}alt-PT\_\allowbreak{}2008} & economic & 692 & 776 & 0.090 & 0.0038 & 0.0057 & +51 & Netzschleuder & \href{https://doi.org/10.1007/s41060-019-00204-1}{Wachs et al. (2020)}\\
77 & \parbox[t]{2.6cm}{\raggedright eu\_\allowbreak{}procurements\_\allowbreak{}alt-SE\_\allowbreak{}2008} & economic & 4,421 & 5,905 & 0.058 & 0.0024 & 0.0045 & +88 & Netzschleuder & \href{https://doi.org/10.1007/s41060-019-00204-1}{Wachs et al. (2020)}\\
78 & \parbox[t]{2.6cm}{\raggedright eu\_\allowbreak{}procurements\_\allowbreak{}alt-SK\_\allowbreak{}2008} & economic & 660 & 773 & 0.10 & 0.0090 & 0.011 & +23 & Netzschleuder & \href{https://doi.org/10.1007/s41060-019-00204-1}{Wachs et al. (2020)}\\
79 & \parbox[t]{2.6cm}{\raggedright product\_\allowbreak{}space-HS} & economic & 866 & 2,532 & 0.19 & 0.23 & 0.27 & +18 & Netzschleuder & \href{https://doi.org/10.1126/science.1144581}{Hausmann et al. (2014)}\\
80 & \parbox[t]{2.6cm}{\raggedright product\_\allowbreak{}space-SITC} & economic & 774 & 1,779 & 0.16 & 0.22 & 0.26 & +19 & Netzschleuder & \href{https://doi.org/10.1126/science.1144581}{Hausmann et al. (2014)}\\
81 & \parbox[t]{2.6cm}{\raggedright us\_\allowbreak{}agencies-alabama} & economic & 1,115 & 4,983 & 0.17 & 0.12 & 0.15 & +22 & Netzschleuder & \href{https://doi.org/10.1073/pnas.1803228115}{Kosack et al. (2018)}\\
82 & \parbox[t]{2.6cm}{\raggedright us\_\allowbreak{}agencies-florida} & economic & 2,010 & 27,570 & 0.17 & 0.34 & 0.36 & +7.0 & Netzschleuder & \href{https://doi.org/10.1073/pnas.1803228115}{Kosack et al. (2018)}\\
83 & \parbox[t]{2.6cm}{\raggedright us\_\allowbreak{}agencies-indiana} & economic & 1,017 & 6,505 & 0.25 & 0.15 & 0.17 & +13 & Netzschleuder & \href{https://doi.org/10.1073/pnas.1803228115}{Kosack et al. (2018)}\\
84 & \parbox[t]{2.6cm}{\raggedright Oregon\_\allowbreak{}0} & informational & 10,670 & 22,002 & 0.045 & 0.0063 & 0.011 & +70 & graph-tool & \href{https://snap.stanford.edu/data/Oregon-1.html}{Leskovec et al. (2005, Oregon AS)}\\
85 & \parbox[t]{2.6cm}{\raggedright Oregon\_\allowbreak{}1} & informational & 10,729 & 21,999 & 0.077 & 0.0060 & 0.010 & +67 & graph-tool & \href{https://snap.stanford.edu/data/Oregon-1.html}{Leskovec et al. (2005, Oregon AS)}\\
86 & \parbox[t]{2.6cm}{\raggedright Oregon\_\allowbreak{}2} & informational & 10,790 & 22,469 & 0.047 & 0.0071 & 0.011 & +56 & graph-tool & \href{https://snap.stanford.edu/data/Oregon-1.html}{Leskovec et al. (2005, Oregon AS)}\\
87 & \parbox[t]{2.6cm}{\raggedright Oregon\_\allowbreak{}3} & informational & 10,859 & 22,747 & 0.094 & 0.0070 & 0.011 & +59 & graph-tool & \href{https://snap.stanford.edu/data/Oregon-1.html}{Leskovec et al. (2005, Oregon AS)}\\
88 & \parbox[t]{2.6cm}{\raggedright Oregon\_\allowbreak{}4} & informational & 10,886 & 22,493 & 0.066 & 0.0060 & 0.011 & +83 & graph-tool & \href{https://snap.stanford.edu/data/Oregon-1.html}{Leskovec et al. (2005, Oregon AS)}\\
89 & \parbox[t]{2.6cm}{\raggedright Oregon\_\allowbreak{}5} & informational & 10,943 & 22,607 & 0.078 & 0.0065 & 0.010 & +57 & graph-tool & \href{https://snap.stanford.edu/data/Oregon-1.html}{Leskovec et al. (2005, Oregon AS)}\\
90 & \parbox[t]{2.6cm}{\raggedright Oregon\_\allowbreak{}6} & informational & 11,011 & 22,677 & 0.061 & 0.0060 & 0.010 & +72 & graph-tool & \href{https://snap.stanford.edu/data/Oregon-1.html}{Leskovec et al. (2005, Oregon AS)}\\
91 & \parbox[t]{2.6cm}{\raggedright Oregon\_\allowbreak{}7} & informational & 11,051 & 22,724 & 0.065 & 0.0059 & 0.010 & +74 & graph-tool & \href{https://snap.stanford.edu/data/Oregon-1.html}{Leskovec et al. (2005, Oregon AS)}\\
92 & \parbox[t]{2.6cm}{\raggedright Political books} & informational & 105 & 441 & 0.33 & 0.29 & 0.31 & +7.1 & V. Krebs & \href{http://www.orgnet.com/divided.html}{Krebs (n.d.)}\\
93 & \parbox[t]{2.6cm}{\raggedright Route views} & informational & 6,474 & 12,572 & 0.070 & 0.0084 & 0.014 & +69 & graph-tool & \href{http://www.routeviews.org/}{University of Oregon Route Views (n.d.)}\\
94 & \parbox[t]{2.6cm}{\raggedright adjnoun} & informational & 112 & 425 & 0.15 & 0.063 & 0.096 & +52 & Netzschleuder & \href{https://doi.org/10.1103/PhysRevE.74.036104}{M.E.J. Newman (2006)}\\
95 & \parbox[t]{2.6cm}{\raggedright bible\_\allowbreak{}nouns} & informational & 1,707 & 9,059 & 0.15 & 0.18 & 0.30 & +66 & Netzschleuder & \href{http://www.chrisharrison.net/index.php/Visualizations/BibleViz}{Römhild}\\
96 & \parbox[t]{2.6cm}{\raggedright caida\_\allowbreak{}as-20040105} & informational & 16,301 & 32,955 & 0.041 & 0.0068 & 0.011 & +68 & Netzschleuder & \href{http://www.caida.org/data/as-relationships/}{The CAIDA AS Relationships Dataset}\\
97 & \parbox[t]{2.6cm}{\raggedright caida\_\allowbreak{}as-20050502} & informational & 19,489 & 39,859 & 0.059 & 0.0059 & 0.011 & +80 & Netzschleuder & \href{http://www.caida.org/data/as-relationships/}{The CAIDA AS Relationships Dataset}\\
98 & \parbox[t]{2.6cm}{\raggedright caida\_\allowbreak{}as-20060116} & informational & 21,232 & 42,872 & 0.037 & 0.0060 & 0.010 & +72 & Netzschleuder & \href{http://www.caida.org/data/as-relationships/}{The CAIDA AS Relationships Dataset}\\
99 & \parbox[t]{2.6cm}{\raggedright caida\_\allowbreak{}as-20070917} & informational & 8,020 & 18,203 & 0.030 & 0.0095 & 0.014 & +46 & Netzschleuder & \href{http://www.caida.org/data/as-relationships/}{The CAIDA AS Relationships Dataset}\\
100 & \parbox[t]{2.6cm}{\raggedright edit\_\allowbreak{}wikibooks-aa} & informational & 23 & 23 & 0.33 & 0.0095 & 0.0095 & +0.0 & Netzschleuder & \href{http://dumps.wikimedia.org/}{Wikimedia Foundation. Wikimedia downloads}\\
101 & \parbox[t]{2.6cm}{\raggedright edit\_\allowbreak{}wikibooks-ar} & informational & 31,797 & 43,896 & 0.010 & 0.00060 & 0.0011 & +83 & Netzschleuder & \href{http://dumps.wikimedia.org/}{Wikimedia Foundation. Wikimedia downloads}\\
102 & \parbox[t]{2.6cm}{\raggedright edit\_\allowbreak{}wikibooks-ca} & informational & 7,767 & 15,421 & 0.11 & 0.0058 & 0.011 & +98 & Netzschleuder & \href{http://dumps.wikimedia.org/}{Wikimedia Foundation. Wikimedia downloads}\\
103 & \parbox[t]{2.6cm}{\raggedright edit\_\allowbreak{}wikibooks-id} & informational & 13,150 & 21,547 & 0.039 & 0.0016 & 0.0029 & +81 & Netzschleuder & \href{http://dumps.wikimedia.org/}{Wikimedia Foundation. Wikimedia downloads}\\
104 & \parbox[t]{2.6cm}{\raggedright edit\_\allowbreak{}wikibooks-sq} & informational & 6,534 & 7,491 & 0.047 & 0.00090 & 0.0016 & +78 & Netzschleuder & \href{http://dumps.wikimedia.org/}{Wikimedia Foundation. Wikimedia downloads}\\
105 & \parbox[t]{2.6cm}{\raggedright edit\_\allowbreak{}wikibooks-tr} & informational & 6,802 & 12,212 & 0.014 & 0.0037 & 0.0055 & +49 & Netzschleuder & \href{http://dumps.wikimedia.org/}{Wikimedia Foundation. Wikimedia downloads}\\
106 & \parbox[t]{2.6cm}{\raggedright edit\_\allowbreak{}wikibooks-uk} & informational & 2,833 & 4,020 & 0.066 & 0.0074 & 0.010 & +40 & Netzschleuder & \href{http://dumps.wikimedia.org/}{Wikimedia Foundation. Wikimedia downloads}\\
107 & \parbox[t]{2.6cm}{\raggedright gnutella-04} & informational & 10,876 & 39,994 & 0.18 & 0.020 & 0.035 & +79 & Netzschleuder & \href{https://doi.org/10.1007/3-540-45748-8_8}{Ripeanu et al. (2002)}\\
108 & \parbox[t]{2.6cm}{\raggedright unicodelang} & informational & 858 & 1,249 & 0.076 & 0.016 & 0.024 & +49 & Netzschleuder & \href{https://doi.org/10.1145/2487788.2488173}{Kunegis (2016)}\\
109 & \parbox[t]{2.6cm}{\raggedright webkb-webkb\_\allowbreak{}cornell\_\allowbreak{}cocite} & informational & 346 & 13,416 & 0.12 & 0.76 & 0.79 & +4.9 & Netzschleuder & \href{https://doi.org/10.1007/bfb0027309}{Slattery (1998)}\\
110 & \parbox[t]{2.6cm}{\raggedright webkb-webkb\_\allowbreak{}texas\_\allowbreak{}cocite} & informational & 334 & 16,494 & 0.042 & 0.78 & 0.79 & +1.1 & Netzschleuder & \href{https://doi.org/10.1007/bfb0027309}{Slattery (1998)}\\
111 & \parbox[t]{2.6cm}{\raggedright webkb-webkb\_\allowbreak{}washington\_\allowbreak{}cocite} & informational & 434 & 15,231 & 0.15 & 0.72 & 0.73 & +1.6 & Netzschleuder & \href{https://doi.org/10.1007/bfb0027309}{Slattery (1998)}\\
112 & \parbox[t]{2.6cm}{\raggedright webkb-webkb\_\allowbreak{}wisconsin\_\allowbreak{}cocite} & informational & 348 & 16,625 & 0.13 & 0.81 & 0.83 & +2.2 & Netzschleuder & \href{https://doi.org/10.1007/bfb0027309}{Slattery (1998)}\\
113 & \parbox[t]{2.6cm}{\raggedright wiki\_\allowbreak{}science} & informational & 677 & 6,517 & 0.31 & 0.36 & 0.39 & +11 & Netzschleuder & \href{https://doi.org/10.6084/m9.figshare.11638932.v5}{Calderone (2020)}\\
114 & \parbox[t]{2.6cm}{\raggedright word\_\allowbreak{}adjacency-darwin} & informational & 7,377 & 44,205 & 0.11 & 0.0060 & 0.0086 & +43 & Netzschleuder & \href{https://doi.org/10.1126/science.1089167}{Milo. et al. (2004)}\\
115 & \parbox[t]{2.6cm}{\raggedright word\_\allowbreak{}adjacency-french} & informational & 8,308 & 23,832 & 0.031 & 0.0010 & 0.0026 & +160 & Netzschleuder & \href{https://doi.org/10.1126/science.1089167}{Milo. et al. (2004)}\\
116 & \parbox[t]{2.6cm}{\raggedright word\_\allowbreak{}adjacency-japanese} & informational & 2,698 & 7,995 & 0.054 & 0.0022 & 0.0039 & +78 & Netzschleuder & \href{https://doi.org/10.1126/science.1089167}{Milo. et al. (2004)}\\
117 & \parbox[t]{2.6cm}{\raggedright word\_\allowbreak{}adjacency-spanish} & informational & 11,558 & 43,050 & 0.038 & 0.0020 & 0.0043 & +115 & Netzschleuder & \href{https://doi.org/10.1126/science.1089167}{Milo. et al. (2004)}\\
118 & \parbox[t]{2.6cm}{\raggedright CS-Aarhus\_\allowbreak{}multiplex} & social & 61 & 353 & 0.24 & 0.42 & 0.44 & +4.6 & graph-tool & \href{https://doi.org/10.1017/nws.2013.5}{Magnani et al. (2013)}\\
119 & \parbox[t]{2.6cm}{\raggedright Ego facebook} & social & 2,888 & 2,981 & 0.11 & 0.00020 & 0.00020 & +0.0 & graph-tool & \href{https://snap.stanford.edu/data/ego-Facebook.html}{McAuley \& Leskovec (2012)}\\
120 & \parbox[t]{2.6cm}{\raggedright Email} & social & 1,133 & 5,451 & 0.29 & 0.14 & 0.16 & +17 & graph-tool & \href{https://doi.org/10.1103/PhysRevE.68.065103}{Guimerà et al. (2003)}\\
121 & \parbox[t]{2.6cm}{\raggedright Epinions} & social & 964 & 14,214 & 0.26 & 0.33 & 0.40 & +20 & graph-tool & \href{https://doi.org/10.1007/978-3-540-39718-2_23}{Richardson et al. (2003)}\\
122 & \parbox[t]{2.6cm}{\raggedright Erdos} & social & 6,927 & 11,850 & 0.056 & 0.0056 & 0.013 & +139 & graph-tool & \href{http://vlado.fmf.uni-lj.si/pub/networks/data/Erdos/Erdos02.net}{Batagelj \& Mrvar (2000)}\\
123 & \parbox[t]{2.6cm}{\raggedright Football} & social & 115 & 613 & 0.19 & 0.55 & 0.56 & +2.7 & Girvan \& Newman (2002) & \href{https://doi.org/10.1073/pnas.122653799}{Girvan \& Newman (2002)}\\
124 & \parbox[t]{2.6cm}{\raggedright Gallery Proximity} & social & 410 & 2,765 & 0.32 & 0.36 & 0.42 & +15 & graph-tool & \href{https://doi.org/10.1016/j.jtbi.2010.11.033}{Isella et al. (2011)}\\
125 & \parbox[t]{2.6cm}{\raggedright Haggle human proximity network} & social & 274 & 2,124 & 0.13 & 0.53 & 0.54 & +2.3 & graph-tool & \href{https://doi.org/10.1109/TMC.2007.1018}{Chaintreau et al. (2007)}\\
126 & \parbox[t]{2.6cm}{\raggedright High school dynamic contacts 1} & social & 126 & 1,709 & 0.16 & 0.45 & 0.48 & +6.8 & graph-tool & \href{https://doi.org/10.1371/journal.pone.0136497}{Mastrandrea et al. (2015)}\\
127 & \parbox[t]{2.6cm}{\raggedright High school dynamic contacts 2} & social & 180 & 2,220 & 0.19 & 0.33 & 0.33 & +2.7 & graph-tool & \href{https://doi.org/10.1371/journal.pone.0107878}{Fournet \& Barrat (2014)}\\
128 & \parbox[t]{2.6cm}{\raggedright Hospital ward dynamic contacts} & social & 75 & 1,139 & 0.098 & 0.47 & 0.47 & +1.2 & graph-tool & \href{https://doi.org/10.1371/journal.pone.0073970}{Vanhems et al. (2013)}\\
129 & \parbox[t]{2.6cm}{\raggedright Hypertext 2009 dynamic contact network} & social & 113 & 2,196 & 0.16 & 0.37 & 0.37 & +0.41 & graph-tool & \href{https://doi.org/10.1016/j.jtbi.2010.11.033}{Isella et al. (2011)}\\
130 & \parbox[t]{2.6cm}{\raggedright Jazz} & social & 198 & 2,742 & 0.20 & 0.42 & 0.44 & +5.2 & graph-tool & \href{https://doi.org/10.1142/S0219525903001067}{Gleiser \& Danon (2003)}\\
131 & \parbox[t]{2.6cm}{\raggedright Les Miserables} & social & 77 & 254 & 0.092 & 0.58 & 0.60 & +3.8 & Knuth (1993) & Knuth (1993)\\
132 & \parbox[t]{2.6cm}{\raggedright Pretty Good Privacy} & social & 10,680 & 24,316 & 0.13 & 0.26 & 0.36 & +35 & graph-tool & \href{https://doi.org/10.1103/PhysRevE.70.056122}{Bogu\textbackslash\{\}\textasciitilde{}\{n\}\textbackslash\{\}'\{a\} et al. (2004)}\\
133 & \parbox[t]{2.6cm}{\raggedright STEM concept networks Researchers} & social & 1,616 & 3,045 & 0.14 & 0.019 & 0.045 & +132 & graph-tool & Koponen \& Nousiainen (2014)\\
134 & \parbox[t]{2.6cm}{\raggedright Song of ice and fire} & social & 796 & 2,823 & 0.16 & 0.12 & 0.18 & +49 & graph-tool & Beveridge \& Shan (2016)\\
135 & \parbox[t]{2.6cm}{\raggedright Star Wars} & social & 109 & 398 & 0.17 & 0.32 & 0.38 & +20 & Gabasova (2016) & \href{https://evelinag.com/blog/2016/01-25-social-network-force-awakens/}{Gabasova (2016)}\\
136 & \parbox[t]{2.6cm}{\raggedright Workplace contacts} & social & 92 & 755 & 0.20 & 0.33 & 0.33 & +2.4 & graph-tool & \href{https://doi.org/10.1017/nws.2015.10}{Génois et al. (2015)}\\
137 & \parbox[t]{2.6cm}{\raggedright Zachary} & social & 34 & 78 & 0.33 & 0.28 & 0.40 & +42 & Zachary (1977) & \href{https://doi.org/10.1086/jar.33.4.3629752}{Zachary (1977)}\\
138 & \parbox[t]{2.6cm}{\raggedright arxiv\_\allowbreak{}authors-GrQc} & social & 4,158 & 13,422 & 0.13 & 0.49 & 0.60 & +23 & Netzschleuder & \href{https://doi.org/10.1145/1217299.1217301}{Leskovec et al. (2007)}\\
139 & \parbox[t]{2.6cm}{\raggedright arxiv\_\allowbreak{}authors-HepTh} & social & 8,638 & 24,806 & 0.22 & 0.29 & 0.37 & +29 & Netzschleuder & \href{https://doi.org/10.1145/1217299.1217301}{Leskovec et al. (2007)}\\
140 & \parbox[t]{2.6cm}{\raggedright arxiv\_\allowbreak{}collab-hep-th-1999} & social & 5,835 & 13,815 & 0.23 & 0.35 & 0.43 & +24 & Netzschleuder & \href{https://doi.org/10.1073/pnas.98.2.404}{Newman (2001)}\\
141 & \parbox[t]{2.6cm}{\raggedright copenhagen-fb\_\allowbreak{}friends} & social & 800 & 6,418 & 0.24 & 0.15 & 0.21 & +44 & Netzschleuder & \href{https://doi.org/10.1038/s41597-019-0325-x}{Sapiezynski et al. (2019)}\\
142 & \parbox[t]{2.6cm}{\raggedright ego\_\allowbreak{}social-facebook\_\allowbreak{}0} & social & 324 & 2,514 & 0.23 & 0.32 & 0.35 & +10 & Netzschleuder & \href{https://doi.org/10.1145/2556612}{McAuley \& Leskovec (2014)}\\
143 & \parbox[t]{2.6cm}{\raggedright ego\_\allowbreak{}social-facebook\_\allowbreak{}348} & social & 224 & 3,192 & 0.20 & 0.36 & 0.37 & +3.3 & Netzschleuder & \href{https://arxiv.org/abs/1210.8182}{McAuley \& Leskovec (2012)}\\
144 & \parbox[t]{2.6cm}{\raggedright facebook\_\allowbreak{}friends} & social & 329 & 1,954 & 0.29 & 0.41 & 0.44 & +5.7 & Netzschleuder & \href{https://arxiv.org/abs/1706.02356}{Maier et al. (2017)}\\
145 & \parbox[t]{2.6cm}{\raggedright facebook\_\allowbreak{}organizations-L1} & social & 5,793 & 30,753 & 0.24 & 0.13 & 0.15 & +15 & Netzschleuder & \href{https://arxiv.org/abs/1303.3741}{Fire et al. (2016)}\\
146 & \parbox[t]{2.6cm}{\raggedright facebook\_\allowbreak{}organizations-M1} & social & 1,429 & 19,357 & 0.25 & 0.13 & 0.15 & +18 & Netzschleuder & \href{https://arxiv.org/abs/1303.3741}{Fire et al. (2016)}\\
147 & \parbox[t]{2.6cm}{\raggedright facebook\_\allowbreak{}organizations-S1} & social & 320 & 2,369 & 0.24 & 0.17 & 0.20 & +13 & Netzschleuder & \href{https://arxiv.org/abs/1303.3741}{Fire et al. (2016)}\\
148 & \parbox[t]{2.6cm}{\raggedright facebook\_\allowbreak{}organizations-S2} & social & 165 & 726 & 0.21 & 0.22 & 0.26 & +17 & Netzschleuder & \href{https://arxiv.org/abs/1303.3741}{Fire et al. (2016)}\\
149 & \parbox[t]{2.6cm}{\raggedright foursquare-NYC\_\allowbreak{}restaurant\_\allowbreak{}checkin} & social & 4,906 & 13,457 & 0.15 & 0.0025 & 0.0047 & +90 & Netzschleuder & \href{https://doi.org/10.1145/2493432.2493464}{Yang et al. (2013)}\\
150 & \parbox[t]{2.6cm}{\raggedright foursquare-NYC\_\allowbreak{}restaurant\_\allowbreak{}tips} & social & 5,372 & 8,852 & 0.034 & 0.00090 & 0.0028 & +211 & Netzschleuder & \href{https://doi.org/10.1145/2493432.2493464}{Yang et al. (2013)}\\
151 & \parbox[t]{2.6cm}{\raggedright marvel\_\allowbreak{}universe} & social & 19,182 & 95,445 & 0.21 & 0.040 & 0.059 & +46 & Netzschleuder & \href{http://arxiv.org/abs/cond-mat/0202174}{Alberich et al. (2002)}\\
152 & \parbox[t]{2.6cm}{\raggedright new\_\allowbreak{}zealand\_\allowbreak{}collab} & social & 1,463 & 4,246 & 0.10 & 0.060 & 0.083 & +37 & Netzschleuder & \href{https://doi.org/10.1145/3167918.3167920}{Aref et al. (2018)}\\
153 & \parbox[t]{2.6cm}{\raggedright physics\_\allowbreak{}collab-pierreAuger} & social & 475 & 6,426 & 0.050 & 0.88 & 0.90 & +2.8 & Netzschleuder & \href{https://arxiv.org/abs/1408.2925}{De Domenico et al. (2015)}\\
154 & \parbox[t]{2.6cm}{\raggedright student\_\allowbreak{}cooperation} & social & 141 & 256 & 0.26 & 0.76 & 0.77 & +1.4 & Netzschleuder & \href{https://doi.org/10.1007/978-3-642-35236-2_59}{Fire et al. (2012)}\\
155 & \parbox[t]{2.6cm}{\raggedright ugandan\_\allowbreak{}village-friendship-1} & social & 202 & 547 & 0.20 & 0.071 & 0.10 & +46 & Netzschleuder & \href{https://doi.org/10.1073/pnas.1700166114}{Chami et al. (2017)}\\
156 & \parbox[t]{2.6cm}{\raggedright ugandan\_\allowbreak{}village-friendship-2} & social & 181 & 688 & 0.35 & 0.081 & 0.13 & +63 & Netzschleuder & \href{https://doi.org/10.1073/pnas.1700166114}{Chami et al. (2017)}\\
157 & \parbox[t]{2.6cm}{\raggedright ugandan\_\allowbreak{}village-friendship-3} & social & 192 & 1,060 & 0.25 & 0.048 & 0.082 & +71 & Netzschleuder & \href{https://doi.org/10.1073/pnas.1700166114}{Chami et al. (2017)}\\
158 & \parbox[t]{2.6cm}{\raggedright ugandan\_\allowbreak{}village-friendship-4} & social & 320 & 2,076 & 0.26 & 0.035 & 0.063 & +79 & Netzschleuder & \href{https://doi.org/10.1073/pnas.1700166114}{Chami et al. (2017)}\\
159 & \parbox[t]{2.6cm}{\raggedright us\_\allowbreak{}congress-H100} & social & 446 & 50,688 & 0.048 & 0.51 & 0.51 & +0.0 & Netzschleuder & \href{https://doi.org/10.1016/j.socnet.2018.07.007}{Neal (1973)}\\
160 & \parbox[t]{2.6cm}{\raggedright us\_\allowbreak{}congress-H93} & social & 446 & 18,083 & 0.17 & 0.18 & 0.18 & +0.0 & Netzschleuder & \href{https://doi.org/10.1016/j.socnet.2018.07.007}{Neal (1973)}\\
161 & \parbox[t]{2.6cm}{\raggedright us\_\allowbreak{}congress-S100} & social & 101 & 2,143 & 0.11 & 0.42 & 0.42 & +0.0 & Netzschleuder & \href{https://doi.org/10.1016/j.socnet.2018.07.007}{Neal (1973)}\\
162 & \parbox[t]{2.6cm}{\raggedright us\_\allowbreak{}congress-S97} & social & 101 & 2,073 & 0.13 & 0.41 & 0.41 & +0.0 & Netzschleuder & \href{https://doi.org/10.1016/j.socnet.2018.07.007}{Neal (1973)}\\
163 & \parbox[t]{2.6cm}{\raggedright Power grid} & technological & 4,941 & 6,594 & 0.17 & 0.14 & 0.18 & +29 & graph-tool & \href{https://doi.org/10.1038/30918}{Watts \& Strogatz (1998)}\\
164 & \parbox[t]{2.6cm}{\raggedright internet\_\allowbreak{}top\_\allowbreak{}pop-Cogentco} & technological & 197 & 243 & 0.31 & 0.065 & 0.073 & +14 & Netzschleuder & \href{https://doi.org/10.1109/JSAC.2011.111002}{Knight et al. (2011)}\\
165 & \parbox[t]{2.6cm}{\raggedright internet\_\allowbreak{}top\_\allowbreak{}pop-Kdl} & technological & 754 & 895 & 0.27 & 0.071 & 0.076 & +6.4 & Netzschleuder & \href{https://doi.org/10.1109/JSAC.2011.111002}{Knight et al. (2011)}\\
166 & \parbox[t]{2.6cm}{\raggedright internet\_\allowbreak{}top\_\allowbreak{}pop-Ntt} & technological & 32 & 63 & 0.27 & 0.24 & 0.28 & +14 & Netzschleuder & \href{https://doi.org/10.1109/JSAC.2011.111002}{Knight et al. (2011)}\\
167 & \parbox[t]{2.6cm}{\raggedright internet\_\allowbreak{}top\_\allowbreak{}pop-TataNld} & technological & 145 & 186 & 0.29 & 0.15 & 0.17 & +14 & Netzschleuder & \href{https://doi.org/10.1109/JSAC.2011.111002}{Knight et al. (2011)}\\
168 & \parbox[t]{2.6cm}{\raggedright software\_\allowbreak{}dependencies-colt} & technological & 504 & 1,139 & 0.20 & 0.14 & 0.17 & +25 & Netzschleuder & \href{https://doi.org/10.1145/2384416.2384418}{Šubelj (2012)}\\
169 & \parbox[t]{2.6cm}{\raggedright software\_\allowbreak{}dependencies-jmail} & technological & 192 & 488 & 0.24 & 0.22 & 0.26 & +18 & Netzschleuder & \href{https://doi.org/10.1145/2384416.2384418}{Šubelj (2012)}\\
170 & \parbox[t]{2.6cm}{\raggedright software\_\allowbreak{}dependencies-jung} & technological & 398 & 943 & 0.12 & 0.16 & 0.21 & +30 & Netzschleuder & \href{https://doi.org/10.1145/2384416.2384418}{Šubelj (2012)}\\
171 & \parbox[t]{2.6cm}{\raggedright software\_\allowbreak{}dependencies-jung-c} & technological & 879 & 2,047 & 0.22 & 0.14 & 0.18 & +31 & Netzschleuder & \href{https://doi.org/10.1145/2384416.2384418}{Šubelj (2012)}\\
172 & \parbox[t]{2.6cm}{\raggedright software\_\allowbreak{}dependencies-org} & technological & 486 & 2,717 & 0.11 & 0.13 & 0.18 & +42 & Netzschleuder & \href{https://doi.org/10.1145/2384416.2384418}{Šubelj (2012)}\\
173 & \parbox[t]{2.6cm}{\raggedright software\_\allowbreak{}dependencies-sjbullet} & technological & 249 & 802 & 0.20 & 0.18 & 0.21 & +15 & Netzschleuder & \href{https://doi.org/10.1145/2384416.2384418}{Šubelj (2012)}\\
174 & \parbox[t]{2.6cm}{\raggedright software\_\allowbreak{}dependencies-sjung} & technological & 399 & 946 & 0.14 & 0.16 & 0.21 & +33 & Netzschleuder & \href{https://doi.org/10.1145/2384416.2384418}{Šubelj (2012)}\\
175 & \parbox[t]{2.6cm}{\raggedright software\_\allowbreak{}dependencies-slucene} & technological & 2,811 & 10,741 & 0.18 & 0.092 & 0.13 & +40 & Netzschleuder & \href{https://doi.org/10.1145/2384416.2384418}{Šubelj (2012)}\\
176 & \parbox[t]{2.6cm}{\raggedright Central Chilean power grid\_\allowbreak{}reduced} & transportation & 218 & 527 & 0.23 & 0.54 & 0.61 & +12 & graph-tool & Kim et al. (2018)\\
177 & \parbox[t]{2.6cm}{\raggedright Chicago road network} & transportation & 12,979 & 20,627 & 0.16 & 0.12 & 0.17 & +37 & graph-tool & Eash et al. (1979)\\
178 & \parbox[t]{2.6cm}{\raggedright EUAirTransportation\_\allowbreak{}multiplex} & transportation & 417 & 2,953 & 0.21 & 0.17 & 0.19 & +8.0 & graph-tool & \href{https://doi.org/10.1038/srep01344}{Cardillo et al. (2013)}\\
179 & \parbox[t]{2.6cm}{\raggedright Train} & transportation & 64 & 243 & 0.25 & 0.56 & 0.63 & +13 & graph-tool & \href{https://doi.org/10.1103/PhysRevX.4.041014}{De Domenico et al. (2014)}\\
180 & \parbox[t]{2.6cm}{\raggedright US airport network} & transportation & 500 & 2,980 & 0.14 & 0.24 & 0.27 & +12 & graph-tool & \href{https://doi.org/10.1073/pnas.0510525103}{Colizza et al. (2006)}\\
181 & \parbox[t]{2.6cm}{\raggedright contiguous\_\allowbreak{}usa} & transportation & 49 & 107 & 0.36 & 0.28 & 0.35 & +25 & Netzschleuder & \href{http://www-cs-faculty.stanford.edu/~uno/sgb.html}{Knuth (1993)}\\
182 & \parbox[t]{2.6cm}{\raggedright euroroad} & transportation & 1,039 & 1,305 & 0.24 & 0.068 & 0.086 & +27 & Netzschleuder & \href{https://doi.org/10.1140/epjb/e2011-10979-2}{Šubelj (2011)}\\
183 & \parbox[t]{2.6cm}{\raggedright london\_\allowbreak{}transport} & transportation & 369 & 430 & 0.26 & 0.092 & 0.11 & +20 & Netzschleuder & \href{https://doi.org/10.1073/pnas.1318469111}{Manlio De Domenico et al. (2014)}\\
184 & \parbox[t]{2.6cm}{\raggedright openstreetmap-01-AL-counties-street\_\allowbreak{}networks:01017\_\allowbreak{}Chambers\_\allowbreak{}County} & transportation & 3,947 & 4,935 & 0.12 & 0.13 & 0.15 & +11 & Netzschleuder & \href{https://doi.org/10.3390/urbansci3010028}{Boeing (2018)}\\
185 & \parbox[t]{2.6cm}{\raggedright openstreetmap-01-AL-counties-street\_\allowbreak{}networks:01031\_\allowbreak{}Coffee\_\allowbreak{}County} & transportation & 3,844 & 5,079 & 0.12 & 0.16 & 0.18 & +14 & Netzschleuder & \href{https://doi.org/10.3390/urbansci3010028}{Boeing (2018)}\\
186 & \parbox[t]{2.6cm}{\raggedright openstreetmap-01-AL-counties-street\_\allowbreak{}networks:01121\_\allowbreak{}Talladega\_\allowbreak{}County} & transportation & 9,877 & 11,843 & 0.095 & 0.11 & 0.13 & +13 & Netzschleuder & \href{https://doi.org/10.3390/urbansci3010028}{Boeing (2018)}\\
187 & \parbox[t]{2.6cm}{\raggedright urban\_\allowbreak{}streets-ahmedabad} & transportation & 2,870 & 4,375 & 0.16 & 0.16 & 0.17 & +9.1 & Netzschleuder & \href{https://doi.org/10.1103/PhysRevE.73.036125}{Crucitti et al. (2006)}\\
188 & \parbox[t]{2.6cm}{\raggedright urban\_\allowbreak{}streets-bologna} & transportation & 541 & 771 & 0.25 & 0.12 & 0.13 & +4.5 & Netzschleuder & \href{https://doi.org/10.1103/PhysRevE.73.036125}{Crucitti et al. (2006)}\\
189 & \parbox[t]{2.6cm}{\raggedright urban\_\allowbreak{}streets-brasilia} & transportation & 179 & 230 & 0.25 & 0.11 & 0.11 & +6.0 & Netzschleuder & \href{https://doi.org/10.1103/PhysRevE.73.036125}{Crucitti et al. (2006)}\\
190 & \parbox[t]{2.6cm}{\raggedright urban\_\allowbreak{}streets-cairo} & transportation & 1,496 & 2,252 & 0.17 & 0.15 & 0.16 & +8.1 & Netzschleuder & \href{https://doi.org/10.1103/PhysRevE.73.036125}{Crucitti et al. (2006)}\\
191 & \parbox[t]{2.6cm}{\raggedright urban\_\allowbreak{}streets-paris} & transportation & 335 & 494 & 0.25 & 0.14 & 0.17 & +24 & Netzschleuder & \href{https://doi.org/10.1103/PhysRevE.73.036125}{Crucitti et al. (2006)}\\
192 & \parbox[t]{2.6cm}{\raggedright urban\_\allowbreak{}streets-venice} & transportation & 1,840 & 2,397 & 0.19 & 0.11 & 0.13 & +14 & Netzschleuder & \href{https://doi.org/10.1103/PhysRevE.73.036125}{Crucitti et al. (2006)}\\
193 & \parbox[t]{2.6cm}{\raggedright urban\_\allowbreak{}streets-vienna} & transportation & 467 & 691 & 0.29 & 0.087 & 0.092 & +5.8 & Netzschleuder & \href{https://doi.org/10.1103/PhysRevE.73.036125}{Crucitti et al. (2006)}\\
194 & \parbox[t]{2.6cm}{\raggedright us\_\allowbreak{}roads-AK} & transportation & 48,560 & 55,014 & 0.12 & 0.046 & 0.053 & +16 & Netzschleuder & \href{http://www.dis.uniroma1.it/challenge9/data/tiger/}{9th DIMACS Implementation Challenge - Shortest Paths}\\
195 & \parbox[t]{2.6cm}{\raggedright us\_\allowbreak{}roads-CT} & transportation & 152,036 & 184,393 & 0.055 & 0.086 & 0.090 & +5.4 & Netzschleuder & \href{http://www.dis.uniroma1.it/challenge9/data/tiger/}{9th DIMACS Implementation Challenge - Shortest Paths}\\
196 & \parbox[t]{2.6cm}{\raggedright us\_\allowbreak{}roads-DC} & transportation & 9,522 & 14,807 & 0.15 & 0.13 & 0.17 & +28 & Netzschleuder & \href{http://www.dis.uniroma1.it/challenge9/data/tiger/}{9th DIMACS Implementation Challenge - Shortest Paths}\\
197 & \parbox[t]{2.6cm}{\raggedright us\_\allowbreak{}roads-DE} & transportation & 48,812 & 59,502 & 0.086 & 0.076 & 0.091 & +19 & Netzschleuder & \href{http://www.dis.uniroma1.it/challenge9/data/tiger/}{9th DIMACS Implementation Challenge - Shortest Paths}\\
198 & \parbox[t]{2.6cm}{\raggedright us\_\allowbreak{}roads-LA} & transportation & 408,161 & 489,985 & 0.046 & 0.058 & 0.058 & +0.17 & Netzschleuder & \href{http://www.dis.uniroma1.it/challenge9/data/tiger/}{9th DIMACS Implementation Challenge - Shortest Paths}\\
199 & \parbox[t]{2.6cm}{\raggedright us\_\allowbreak{}roads-MT} & transportation & 300,809 & 343,444 & 0.098 & 0.052 & 0.057 & +9.4 & Netzschleuder & \href{http://www.dis.uniroma1.it/challenge9/data/tiger/}{9th DIMACS Implementation Challenge - Shortest Paths}\\
200 & \parbox[t]{2.6cm}{\raggedright us\_\allowbreak{}roads-NH} & transportation & 115,055 & 130,197 & 0.066 & 0.063 & 0.069 & +9.9 & Netzschleuder & \href{http://www.dis.uniroma1.it/challenge9/data/tiger/}{9th DIMACS Implementation Challenge - Shortest Paths}\\
201 & \parbox[t]{2.6cm}{\raggedright us\_\allowbreak{}roads-RI} & transportation & 51,642 & 66,650 & 0.10 & 0.088 & 0.11 & +19 & Netzschleuder & \href{http://www.dis.uniroma1.it/challenge9/data/tiger/}{9th DIMACS Implementation Challenge - Shortest Paths}\\
202 & \parbox[t]{2.6cm}{\raggedright us\_\allowbreak{}roads-VT} & transportation & 95,672 & 104,644 & 0.094 & 0.049 & 0.053 & +8.6 & Netzschleuder & \href{http://www.dis.uniroma1.it/challenge9/data/tiger/}{9th DIMACS Implementation Challenge - Shortest Paths}\\
\end{longtable}
\endgroup

\section{Adaptive Cut on the Louvain Dendrogram}
\label{SI:sec:louvain-table}

Table~\ref{tab:SI-louvain-ac} lists, for each of the 202 networks, the modularity of the single-seed Louvain partition $Q_{\text{Louvain}}$ and of the adaptive cut applied to the Louvain dendrogram $Q_{\text{AC}}$, both measured at resolution $\gamma=1$, with the relative improvement plotted in Figure~\ref{fig:louvain_results}.

\begingroup
\footnotesize\let\normalsize\footnotesize\let\small\footnotesize
\setlength{\tabcolsep}{4pt}
\renewcommand{\arraystretch}{1.05}
\begin{longtable}{@{}lrrr@{}}
\caption{Modularity of the single-seed Louvain partition $Q_{\text{Louvain}}$ and of the adaptive cut applied to the Louvain dendrogram $Q_{\text{AC}}$, for the 202 networks of Figure~4, both at resolution $\gamma=1$. Improvement is $(Q_{\text{AC}}-Q_{\text{Louvain}})/Q_{\text{Louvain}}\times100\%$.}\label{tab:SI-louvain-ac}\\
\toprule
\textbf{Network} & \textbf{$Q_{\text{Louvain}}$} & \textbf{$Q_{\text{AC}}$} & \textbf{Impr.\%}\\
\midrule
\endfirsthead
\multicolumn{4}{c}{\tablename\ \thetable{} -- continued from previous page}\\
\toprule
\textbf{Network} & \textbf{$Q_{\text{Louvain}}$} & \textbf{$Q_{\text{AC}}$} & \textbf{Impr.\%}\\
\midrule
\endhead
\midrule
\multicolumn{4}{r}{\textit{continued on next page}}\\
\endfoot
\bottomrule
\endlastfoot
adjnoun & 0.3016 & 0.3016 & +0.0\\
arxiv\_authors-GrQc & 0.8470 & 0.8510 & +0.5\\
arxiv\_authors-HepTh & 0.7555 & 0.7555 & +0.0\\
arxiv\_collab-hep-th-1999 & 0.8249 & 0.8275 & +0.3\\
bible\_nouns & 0.5065 & 0.5251 & +3.7\\
bitcoin\_alpha & 0.4698 & 0.4799 & +2.1\\
bitcoin\_trust & 0.4829 & 0.5065 & +4.9\\
board\_directors-net1m\_2011-08-01 & 0.8795 & 0.8815 & +0.2\\
board\_directors-net2m\_2011-08-01 & 0.9101 & 0.9101 & +0.0\\
budapest\_connectome-all\_200k & 0.4965 & 0.4965 & +0.0\\
budapest\_connectome-all\_20k & 0.5574 & 0.5574 & +0.0\\
budapest\_connectome-female\_200k & 0.5035 & 0.5035 & +0.0\\
budapest\_connectome-male\_200k & 0.5551 & 0.5551 & +0.0\\
Caenorhabditis elegans & 0.4370 & 0.4457 & +2.0\\
caida\_as-20040105 & 0.6512 & 0.6608 & +1.5\\
caida\_as-20050502 & 0.6582 & 0.6695 & +1.7\\
caida\_as-20060116 & 0.6554 & 0.6751 & +3.0\\
caida\_as-20070917 & 0.5479 & 0.5546 & +1.2\\
Central Chilean power grid\_reduced & 0.7785 & 0.7789 & +0.0\\
Chicago road network & 0.9310 & 0.9334 & +0.3\\
cintestinalis & 0.3414 & 0.3419 & +0.1\\
Clements \& Long plant-pollinator web & 0.3113 & 0.3113 & +0.0\\
Contact & 0.1274 & 0.1342 & +5.3\\
contiguous\_usa & 0.5811 & 0.5844 & +0.6\\
copenhagen-fb\_friends & 0.4684 & 0.4762 & +1.7\\
CS-Aarhus\_multiplex & 0.4162 & 0.4162 & +0.0\\
David Copperfield & 0.3008 & 0.3024 & +0.5\\
edit\_wikibooks-ar & 0.6200 & 0.6217 & +0.3\\
edit\_wikibooks-ca & 0.6192 & 0.6326 & +2.2\\
edit\_wikibooks-id & 0.6391 & 0.6465 & +1.2\\
edit\_wikibooks-sq & 0.7419 & 0.7425 & +0.1\\
edit\_wikibooks-tr & 0.6192 & 0.6348 & +2.5\\
edit\_wikibooks-uk & 0.7370 & 0.7375 & +0.1\\
Ego facebook & 0.8087 & 0.8087 & +0.0\\
Email & 0.5715 & 0.5780 & +1.1\\
Email Europe & 0.4107 & 0.4175 & +1.7\\
Epinions & 0.5294 & 0.5358 & +1.2\\
Erdos & 0.6959 & 0.6970 & +0.2\\
eu\_procurements\_alt-AT\_2008 & 0.8622 & 0.8641 & +0.2\\
eu\_procurements\_alt-CZ\_2008 & 0.8407 & 0.8407 & +0.0\\
eu\_procurements\_alt-DK\_2011 & 0.7803 & 0.7849 & +0.6\\
eu\_procurements\_alt-EE\_2008 & 0.8266 & 0.8266 & +0.0\\
eu\_procurements\_alt-ES\_2008 & 0.7271 & 0.7280 & +0.1\\
eu\_procurements\_alt-FI\_2008 & 0.7194 & 0.7194 & +0.0\\
eu\_procurements\_alt-GR\_2008 & 0.8029 & 0.8033 & +0.1\\
eu\_procurements\_alt-HU\_2015 & 0.8029 & 0.8046 & +0.2\\
eu\_procurements\_alt-IT\_2008 & 0.7843 & 0.7920 & +1.0\\
eu\_procurements\_alt-LT\_2008 & 0.6607 & 0.6648 & +0.6\\
eu\_procurements\_alt-LV\_2008 & 0.7924 & 0.7942 & +0.2\\
eu\_procurements\_alt-PL\_2008 & 0.7202 & 0.7235 & +0.5\\
eu\_procurements\_alt-PT\_2008 & 0.8619 & 0.8619 & +0.0\\
eu\_procurements\_alt-SE\_2008 & 0.7961 & 0.7976 & +0.2\\
eu\_procurements\_alt-SK\_2008 & 0.8319 & 0.8359 & +0.5\\
EUAirTransportation\_multiplex & 0.2964 & 0.3079 & +3.9\\
euroroad & 0.8661 & 0.8661 & +0.0\\
facebook\_friends & 0.6998 & 0.7072 & +1.1\\
facebook\_organizations-L1 & 0.7576 & 0.7591 & +0.2\\
facebook\_organizations-M1 & 0.5098 & 0.5134 & +0.7\\
facebook\_organizations-S1 & 0.4467 & 0.4542 & +1.7\\
facebook\_organizations-S2 & 0.5108 & 0.5153 & +0.9\\
foodweb\_baywet & 0.1793 & 0.1820 & +1.5\\
Football & 0.5978 & 0.6046 & +1.1\\
foursquare-NYC\_restaurant\_checkin & 0.4365 & 0.4365 & +0.0\\
foursquare-NYC\_restaurant\_tips & 0.6332 & 0.6356 & +0.4\\
Gallery Proximity & 0.7048 & 0.7121 & +1.0\\
genetic\_multiplex-Candida & 0.4544 & 0.4544 & +0.0\\
genetic\_multiplex-Gallus & 0.4900 & 0.4900 & +0.0\\
genetic\_multiplex-HumanHerpes4 & 0.3216 & 0.3242 & +0.8\\
genetic\_multiplex-HumanHIV1 & 0.7179 & 0.7179 & +0.0\\
genetic\_multiplex-Plasmodium & 0.5122 & 0.5122 & +0.0\\
genetic\_multiplex-Rattus & 0.7150 & 0.7194 & +0.6\\
gnutella-04 & 0.3816 & 0.3840 & +0.6\\
Haggle human proximity network & 0.1274 & 0.1315 & +3.2\\
High school dynamic contacts 1 & 0.4587 & 0.4587 & +0.0\\
High school dynamic contacts 1 copy & 0.4587 & 0.4587 & +0.0\\
High school dynamic contacts 2 & 0.5232 & 0.5232 & +0.0\\
Hospital ward dynamic contacts & 0.0973 & 0.0981 & +0.9\\
human\_brains-BNU1\_0025864\_1\_DTI\_AAL & 0.3766 & 0.3796 & +0.8\\
human\_brains-BNU1\_0025864\_1\_DTI\_CPAC200 & 0.4465 & 0.4518 & +1.2\\
human\_brains-BNU1\_0025864\_1\_DTI\_DS00583 & 0.4544 & 0.4611 & +1.5\\
human\_brains-BNU1\_0025864\_1\_DTI\_DS01216 & 0.5140 & 0.5179 & +0.7\\
human\_brains-BNU1\_0025870\_1\_DTI\_DS01216 & 0.5439 & 0.5572 & +2.5\\
Hypertext 2009 dynamic contact network & 0.1011 & 0.1061 & +5.0\\
interactome\_figeys & 0.4581 & 0.4638 & +1.2\\
interactome\_stelzl & 0.5873 & 0.5873 & +0.0\\
interactome\_vidal & 0.6361 & 0.6361 & +0.0\\
internet\_top\_pop-Cogentco & 0.8047 & 0.8047 & +0.0\\
internet\_top\_pop-Kdl & 0.8812 & 0.8851 & +0.4\\
internet\_top\_pop-Ntt & 0.4519 & 0.4544 & +0.5\\
internet\_top\_pop-TataNld & 0.7599 & 0.7641 & +0.6\\
Jazz & 0.4449 & 0.4451 & +0.1\\
kegg\_metabolic-aae & 0.5150 & 0.5227 & +1.5\\
kegg\_metabolic-afu & 0.5479 & 0.5479 & +0.0\\
kegg\_metabolic-ana & 0.5191 & 0.5191 & +0.0\\
kegg\_metabolic-ape & 0.5344 & 0.5395 & +1.0\\
kegg\_metabolic-atc & 0.5099 & 0.5099 & +0.0\\
kegg\_metabolic-ath & 0.5114 & 0.5114 & +0.0\\
kegg\_metabolic-atu & 0.5038 & 0.5144 & +2.1\\
kegg\_metabolic-bas & 0.5527 & 0.5580 & +1.0\\
Les Miserables & 0.5600 & 0.5600 & +0.0\\
london\_transport & 0.8297 & 0.8297 & +0.0\\
malaria\_genes-HVR\_1 & 0.6366 & 0.6399 & +0.5\\
malaria\_genes-HVR\_5 & 0.4175 & 0.4371 & +4.7\\
malaria\_genes-HVR\_6 & 0.5189 & 0.5298 & +2.1\\
malaria\_genes-HVR\_7 & 0.3748 & 0.3777 & +0.8\\
malaria\_genes-HVR\_8 & 0.5031 & 0.5041 & +0.2\\
malaria\_genes-HVR\_9 & 0.3662 & 0.3706 & +1.2\\
marvel\_universe & 0.6877 & 0.6922 & +0.7\\
Messel Shale food web & 0.3242 & 0.3604 & +11.2\\
mist-genetic\_fly & 0.5051 & 0.5225 & +3.4\\
mist-genetic\_human & 0.7389 & 0.7397 & +0.1\\
mist-genetic\_mouse & 0.8141 & 0.8143 & +0.0\\
mist-genetic\_worm & 0.5846 & 0.6070 & +3.8\\
mist-ppi\_rat & 0.6236 & 0.6353 & +1.9\\
name & 0.4684 & 0.4762 & +1.7\\
new\_zealand\_collab & 0.4619 & 0.4719 & +2.2\\
openstreetmap-01-AL-counties-street\_networks:01017\_Chambers\_County & 0.9429 & 0.9429 & +0.0\\
openstreetmap-01-AL-counties-street\_networks:01031\_Coffee\_County & 0.9391 & 0.9394 & +0.0\\
openstreetmap-01-AL-counties-street\_networks:01121\_Talladega\_County & 0.9615 & 0.9615 & +0.0\\
Oregon\_0 & 0.6278 & 0.6395 & +1.9\\
Oregon\_1 & 0.6271 & 0.6392 & +1.9\\
Oregon\_2 & 0.6231 & 0.6340 & +1.7\\
Oregon\_3 & 0.6201 & 0.6330 & +2.1\\
Oregon\_4 & 0.6209 & 0.6329 & +1.9\\
Oregon\_5 & 0.6224 & 0.6343 & +1.9\\
Oregon\_6 & 0.6255 & 0.6357 & +1.6\\
Oregon\_7 & 0.6169 & 0.6348 & +2.9\\
physics\_collab-pierreAuger & 0.6144 & 0.6147 & +0.0\\
plant\_pol\_robertson & 0.3542 & 0.3565 & +0.6\\
Political books & 0.5268 & 0.5272 & +0.1\\
Power grid & 0.9359 & 0.9382 & +0.3\\
Pretty Good Privacy & 0.8824 & 0.8846 & +0.3\\
product\_space-HS & 0.7153 & 0.7170 & +0.2\\
product\_space-SITC & 0.7529 & 0.7580 & +0.7\\
Route views & 0.6212 & 0.6353 & +2.3\\
software\_dependencies-colt & 0.6399 & 0.6436 & +0.6\\
software\_dependencies-jmail & 0.6053 & 0.6121 & +1.1\\
software\_dependencies-jung & 0.7425 & 0.7436 & +0.1\\
software\_dependencies-jung-c & 0.7625 & 0.7674 & +0.6\\
software\_dependencies-org & 0.4279 & 0.4320 & +1.0\\
software\_dependencies-scolt & 0.6400 & 0.6410 & +0.2\\
software\_dependencies-sjbullet & 0.5725 & 0.5750 & +0.4\\
software\_dependencies-sjung & 0.7373 & 0.7373 & +0.0\\
software\_dependencies-slucene & 0.7127 & 0.7194 & +0.9\\
Song of ice and fire & 0.5899 & 0.6114 & +3.6\\
Star Wars & 0.4274 & 0.4274 & +0.0\\
STEM concept networks Researchers & 0.5847 & 0.5895 & +0.8\\
student\_cooperation & 0.8064 & 0.8064 & +0.0\\
Train & 0.4382 & 0.4421 & +0.9\\
tree-of-life-331678 & 0.7491 & 0.7508 & +0.2\\
tree-of-life-333990 & 0.8244 & 0.8249 & +0.1\\
tree-of-life-335543 & 0.5880 & 0.5890 & +0.2\\
tree-of-life-338966 & 0.6405 & 0.6523 & +1.8\\
tree-of-life-338969 & 0.6638 & 0.6673 & +0.5\\
tree-of-life-339670 & 0.7326 & 0.7396 & +0.9\\
tree-of-life-340177 & 0.7504 & 0.7504 & -0.0\\
tree-of-life-365044 & 0.7607 & 0.7619 & +0.2\\
tree-of-life-366602 & 0.7274 & 0.7319 & +0.6\\
tree-of-life-406817 & 0.7327 & 0.7372 & +0.6\\
tree-of-life-452652 & 0.7356 & 0.7381 & +0.3\\
tree-of-life-469383 & 0.6577 & 0.6588 & +0.2\\
TrumpWorld & 0.8018 & 0.8027 & +0.1\\
ugandan\_village-friendship-1 & 0.4219 & 0.4219 & +0.0\\
ugandan\_village-friendship-2 & 0.3293 & 0.3293 & +0.0\\
ugandan\_village-friendship-3 & 0.2574 & 0.2574 & +0.0\\
ugandan\_village-friendship-4 & 0.2848 & 0.2936 & +3.1\\
unicodelang & 0.7191 & 0.7191 & +0.0\\
urban\_streets-ahmedabad & 0.9074 & 0.9082 & +0.1\\
urban\_streets-bologna & 0.8327 & 0.8327 & +0.0\\
urban\_streets-brasilia & 0.7898 & 0.7898 & +0.0\\
urban\_streets-cairo & 0.8859 & 0.8859 & +0.0\\
urban\_streets-paris & 0.7983 & 0.7983 & +0.0\\
urban\_streets-venice & 0.9189 & 0.9222 & +0.4\\
urban\_streets-vienna & 0.8107 & 0.8126 & +0.2\\
US airport network & 0.3559 & 0.3663 & +2.9\\
us\_agencies-alabama & 0.5059 & 0.5167 & +2.1\\
us\_agencies-florida & 0.5413 & 0.5427 & +0.3\\
us\_agencies-indiana & 0.4101 & 0.4150 & +1.2\\
us\_congress-H100 & 0.0921 & 0.0934 & +1.4\\
us\_congress-H93 & 0.1449 & 0.1489 & +2.8\\
us\_congress-S100 & 0.1160 & 0.1287 & +11.0\\
us\_congress-S97 & 0.1213 & 0.1233 & +1.6\\
us\_roads-AK & 0.9856 & 0.9856 & +0.0\\
us\_roads-CT & 0.9852 & 0.9852 & +0.0\\
us\_roads-DC & 0.9402 & 0.9402 & +0.0\\
us\_roads-DE & 0.9793 & 0.9793 & +0.0\\
us\_roads-LA & 0.9917 & 0.9917 & +0.0\\
us\_roads-MT & 0.9921 & 0.9921 & +0.0\\
us\_roads-NH & 0.9864 & 0.9867 & +0.0\\
us\_roads-RI & 0.9801 & 0.9805 & +0.0\\
us\_roads-VT & 0.9851 & 0.9853 & +0.0\\
webkb-webkb\_cornell\_cocite & 0.2503 & 0.2569 & +2.6\\
webkb-webkb\_texas\_cocite & 0.2184 & 0.2188 & +0.2\\
webkb-webkb\_washington\_cocite & 0.2633 & 0.2659 & +1.0\\
webkb-webkb\_wisconsin\_cocite & 0.2239 & 0.2257 & +0.8\\
wiki\_science & 0.6199 & 0.6213 & +0.2\\
word\_adjacency-darwin & 0.2393 & 0.2393 & +0.0\\
word\_adjacency-french & 0.4041 & 0.4082 & +1.0\\
word\_adjacency-japanese & 0.3670 & 0.3674 & +0.1\\
word\_adjacency-spanish & 0.3179 & 0.3457 & +8.7\\
Workplace contacts & 0.3857 & 0.3857 & +0.0\\
Zachary & 0.4198 & 0.4198 & +0.0\\
\end{longtable}
\endgroup

\bibliography{refs}

\end{document}